\documentclass[longauth]{aa}  

\usepackage{graphicx}
\usepackage{txfonts}
\usepackage{hyperref}
\usepackage{xcolor}
\usepackage{lipsum}
\usepackage{gensymb}
\usepackage{multirow}
\usepackage{siunitx}
\usepackage[capitalise]{cleveref}
\usepackage{mhchem}
\newcommand{\agkelp}{0.171^{+0.066}_{-0.068}}
\newcommand{\aghelios}{0.133\pm0.062}
\newcommand{\agbb}{0.127\pm0.063}
\newcommand{\asphere}{0.266^{+0.097}_{-0.100}}

\newcommand{\cowanAB}{0.139^{+0.074}_{-0.100}}
\newcommand{\morrisAB}{0.057^{+0.083}_{-0.101}}

\begin{document}

   \title{Spi-OPS: \textit{Spitzer} and \textit{CHEOPS} confirm the near-polar orbit of MASCARA-1\,b and reveal a hint of dayside reflection\thanks{The photometric time series data are only available in electronic form at the CDS via anonymous ftp to cdsarc.u-strasbg.fr (130.79.128.5) or via \url{http://cdsweb.u-strasbg.fr/cgi-bin/qcat?J/A+A/658/A75}}}
   \titlerunning{Spi-OPS: a near-polar orbit and dayside reflection for MASCARA-1\,b} 

   \subtitle{}

   \author{M. J. Hooton$^{1}$\thanks{matthew.hooton@unibe.ch}, 
S. Hoyer$^{2}$, 
D. Kitzmann$^{3}$, 
B. M. Morris$^{3}$, 
A. M. S. Smith$^{4}$, 
A. Collier Cameron$^{5}$, 
D. Futyan$^{6}$, 
P. F. L. Maxted$^{7}$, 
D. Queloz$^{6,8}$, 
B.-O. Demory$^{3}$, 
K. Heng$^{3,9}$, 
M. Lendl$^{6}$, 
J. Cabrera$^{4}$, 
Sz. Csizmadia$^{4}$, 
A. Deline$^{6}$,
H. Parviainen$^{10}$, 
S. Salmon$^{6}$, 
S. Sulis$^{2}$, 
T. G. Wilson$^{5}$, 
A. Bonfanti$^{11}$, 
A. Brandeker$^{12}$,
O. D. S. Demangeon$^{13,14}$, 
M. Oshagh$^{10}$, 
C. M. Persson$^{15}$, 
G. Scandariato$^{16}$, 
Y. Alibert$^{1}$, 
R. Alonso$^{10,17}$, 
G. Anglada Escudé$^{18,19}$, 
T. Bárczy$^{20}$, 
D. Barrado$^{21}$, 
S. C. C. Barros$^{13,14}$, 
W. Baumjohann$^{11}$, 
M. Beck$^{6}$, 
T. Beck$^{1}$, 
W. Benz$^{1,3}$, 
N. Billot$^{6}$, 
X. Bonfils$^{22}$, 
V. Bourrier$^{6}$, 
C. Broeg$^{1,3}$, 
M.-D. Busch$^{1}$, 
S. Charnoz$^{23}$, 
M. B. Davies$^{24}$, 
M. Deleuil$^{2}$,
L. Delrez$^{25,26,6}$, 
D. Ehrenreich$^{6}$, 
A. Erikson$^{4}$, 
J. Farinato$^{27}$, 
A. Fortier$^{1,3}$, 
L. Fossati$^{11}$, 
M. Fridlund$^{28,29}$, 
D. Gandolfi$^{30}$, 
M. Gillon$^{25}$, 
M. Güdel$^{31}$, 
K. G. Isaak$^{32}$, 
K. Jones$^{3}$, 
L. Kiss$^{33,34,35}$, 
J. Laskar$^{36}$, 
A. Lecavelier des Etangs$^{37}$, 
C. Lovis$^{6}$, 
A. Luntzer$^{38}$, 
D. Magrin$^{39}$, 
V. Nascimbeni$^{39}$, 
G. Olofsson$^{12}$, 
R. Ottensamer$^{38}$, 
I. Pagano$^{16}$, 
E. Pallé$^{10}$, 
G. Peter$^{40}$, 
G. Piotto$^{39,41}$, 
D. Pollacco$^{9}$, 
R. Ragazzoni$^{39,41}$, 
N. Rando$^{42}$, 
F. Ratti$^{42}$, 
H. Rauer$^{4,43,44}$, 
I. Ribas$^{18,19}$, 
N. C. Santos$^{13,14}$, 
D. Ségransan$^{6}$, 
A. E. Simon$^{1}$, 
S. G. Sousa$^{13}$, 
M. Steller$^{11}$, 
Gy. M. Szabó$^{45,46}$, 
N. Thomas$^{1}$, 
S. Udry$^{6}$, 
B. Ulmer$^{47}$, 
V. Van Grootel$^{26}$, 
N. A. Walton$^{48}$}
    \authorrunning{M. J. Hooton et al.}

   \institute{$^{1}$ Physikalisches Institut, University of Bern, Gesellsschaftstrasse 6, 3012 Bern, Switzerland\\
$^{2}$ Aix Marseille Univ, CNRS, CNES, LAM, 38 rue Frédéric Joliot-Curie, 13388 Marseille, France\\
$^{3}$ Center for Space and Habitability, Gesellsschaftstrasse 6, 3012 Bern, Switzerland\\
$^{4}$ Institute of Planetary Research, German Aerospace Center (DLR), Rutherfordstrasse 2, 12489 Berlin, Germany\\
$^{5}$ Centre for Exoplanet Science, SUPA School of Physics and Astronomy, University of St Andrews, North Haugh, St Andrews KY16 9SS, UK\\
$^{6}$ Observatoire Astronomique de l'Université de Genève, Chemin Pegasi 51, Versoix, Switzerland\\
$^{7}$ Astrophysics Group, Keele University, Staffordshire, ST5 5BG, United Kingdom\\
$^{8}$ Cavendish Laboratory, JJ Thomson Avenue, Cambridge CB3 0HE, UK\\
$^{9}$ Department of Physics, University of Warwick, Gibbet Hill Road, Coventry CV4 7AL, United Kingdom\\
$^{10}$ Instituto de Astrofisica de Canarias, 38200 La Laguna, Tenerife, Spain\\
$^{11}$ Space Research Institute, Austrian Academy of Sciences, Schmiedlstrasse 6, A-8042 Graz, Austria\\
$^{12}$ Department of Astronomy, Stockholm University, AlbaNova University Center, 10691 Stockholm, Sweden\\
$^{13}$ Instituto de Astrofisica e Ciencias do Espaco, Universidade do Porto, CAUP, Rua das Estrelas, 4150-762 Porto, Portugal\\
$^{14}$ Departamento de Fisica e Astronomia, Faculdade de Ciencias, Universidade do Porto, Rua do Campo Alegre, 4169-007 Porto, Portugal\\
$^{15}$ Department of Space, Earth and Environment, Onsala Space Observatory, Chalmers University of Technology, 439 92  Onsala, Sweden\\
$^{16}$ INAF, Osservatorio Astrofisico di Catania, Via S. Sofia 78, 95123 Catania, Italy\\
$^{17}$ Departamento de Astrofisica, Universidad de La Laguna, 38206 La Laguna, Tenerife, Spain\\
$^{18}$ Institut de Ciencies de l'Espai (ICE, CSIC), Campus UAB, Can Magrans s/n, 08193 Bellaterra, Spain\\
$^{19}$ Institut d'Estudis Espacials de Catalunya (IEEC), 08034 Barcelona, Spain\\
$^{20}$ Admatis, 5. Kandó Kálmán Street, 3534 Miskolc, Hungary\\
$^{21}$ Depto. de Astrofisica, Centro de Astrobiologia (CSIC-INTA), ESAC campus, 28692 Villanueva de la Cañada (Madrid), Spain\\
$^{22}$ Université Grenoble Alpes, CNRS, IPAG, 38000 Grenoble, France\\
$^{23}$ Université de Paris, Institut de physique du globe de Paris, CNRS, F-75005 Paris, France\\
$^{24}$ Centre for Mathematical Sciences, Lund University, Box 118, 22100 Lund, Sweden\\
$^{25}$ Astrobiology Research Unit, Université de Liège, Allée du 6 Août 19C, B-4000 Liège, Belgium\\
$^{26}$ Space sciences, Technologies and Astrophysics Research (STAR) Institute, Université de Liège, Allée du 6 Août 19C, 4000 Liège, Belgium\\
$^{27}$ INAF, Osservatorio Astronomico di Padova,Vicolo dell’Osservatorio 5, 35122 Padova, Italy\\
$^{28}$ Leiden Observatory, University of Leiden, PO Box 9513, 2300 RA Leiden, The Netherlands\\
$^{29}$ Department of Space, Earth and Environment, Onsala space observatory, Chalmers University of Technology, 439 92  Onsala, Sweden\\
$^{30}$ Dipartimento di Fisica, Universita degli Studi di Torino, via Pietro Giuria 1, I-10125, Torino, Italy\\
$^{31}$ University of Vienna, Department of Astrophysics, Türkenschanzstrasse 17, 1180 Vienna, Austria\\
$^{32}$ Science and Operations Department - Science Division (SCI-SC), Directorate of Science, European Space Agency (ESA), European Space Research and Technology Centre (ESTEC), Keplerlaan 1, 2201-AZ Noordwijk, The Netherlands\\
$^{33}$ Konkoly Observatory, Research Centre for Astronomy and Earth Sciences, 1121 Budapest, Konkoly Thege Miklós út 15-17, Hungary\\
$^{34}$ ELTE Eötvös Loránd University, Institute of Physics, Pázmány Péter sétány 1/A, 1117 Budapest, Hungary\\
$^{35}$ Sydney Institute for Astronomy, School of Physics A29, University of Sydney, NSW 2006, Australia\\
$^{36}$ IMCCE, UMR8028 CNRS, Observatoire de Paris, PSL Univ., Sorbonne Univ., 77 av. Denfert-Rochereau, 75014 Paris, France\\
$^{37}$ Institut d'astrophysique de Paris, UMR7095 CNRS, Université Pierre \& Marie Curie, 98bis blvd. Arago, 75014 Paris, France\\
$^{38}$ Department of Astrophysics, University of Vienna, Tuerkenschanzstrasse 17, 1180 Vienna, Austria\\
$^{39}$ INAF, Osservatorio Astronomico di Padova, Vicolo dell'Osservatorio 5, 35122 Padova, Italy\\
$^{40}$ Institute of Optical Sensor Systems, German Aerospace Center (DLR), Rutherfordstrasse 2, 12489 Berlin, Germany\\
$^{41}$ Dipartimento di Fisica e Astronomia ``Galileo Galilei'', Universita degli Studi di Padova, Vicolo dell'Osservatorio 3, 35122 Padova, Italy\\
$^{42}$ ESTEC, European Space Agency, 2201AZ, Noordwijk, NL\\
$^{43}$ Center for Astronomy and Astrophysics, Technical University Berlin, Hardenberstrasse 36, 10623 Berlin, Germany\\
$^{44}$ Institut für Geologische Wissenschaften, Freie UniversitÃ¤t Berlin, 12249 Berlin, Germany\\
$^{45}$ ELTE Eötvös Loránd University, Gothard Astrophysical Observatory, 9700 Szombathely, Szent Imre h. u. 112, Hungary\\
$^{46}$ MTA-ELTE Exoplanet Research Group, 9700 Szombathely, Szent Imre h. u. 112, Hungary\\
$^{47}$ Ingenieurbüro Ulmer, Im Technologiepark 1, 15236 Frankfurt/Oder, Germany\\
$^{48}$ Institute of Astronomy, University of Cambridge, Madingley Road, Cambridge, CB3 0HA, United Kingdom\\}

   \date{Received 28 June 2021; accepted 10 September 2021}

 
  \abstract
   {The light curves of tidally locked hot Jupiters transiting fast-rotating, early-type stars are a rich source of information about both the planet and star, with full-phase coverage enabling a detailed atmospheric characterisation of the planet. Although it is possible to determine the true spin-orbit angle $\Psi$ -- a notoriously difficult parameter to measure -- from any transit asymmetry resulting from gravity darkening induced by the stellar rotation, the correlations that exist between the transit parameters have led to large disagreements in published values of $\Psi$ for some systems.}
   {We aimed to study these phenomena in the light curves of the ultra-hot Jupiter MASCARA-1\,b, which is characteristically similar to well-studied contemporaries such as KELT-9\,b and WASP-33\,b. }
   {We obtained optical \textit{CHaracterising ExOPlanet Satellite} (\textit{CHEOPS}) transit and occultation light curves of MASCARA-1\,b, and analysed them jointly with a \textit{Spitzer}/IRAC \SI{4.5}{\micro\meter} full-phase curve to model the asymmetric transits, occultations, and phase-dependent flux modulation. For the latter, we employed a novel physics-driven approach to jointly fit the phase modulation by generating a single 2D temperature map and integrating it over the two bandpasses as a function of phase to account for the differing planet-star flux contrasts. The reflected light component was modelled using the general ab initio solution for a semi-infinite atmosphere.}
   {When fitting the \textit{CHEOPS} and \textit{Spitzer} transits together, the degeneracies are greatly diminished and return results consistent with previously published Doppler tomography. Placing priors informed by the tomography achieves even better precision, allowing a determination of $\Psi=72.1^{+2.5}_{-2.4}$ deg. From the occultations and phase variations, we derived dayside and nightside temperatures of $3062^{+66}_{-68}$ K and $1720\pm330$ K, respectively. Our retrieval suggests that the dayside emission spectrum closely follows that of a blackbody. As the \textit{CHEOPS} occultation is too deep to be attributed to blackbody flux alone, we could separately derive geometric albedo $A_\mathrm{g}=\agkelp$ and spherical albedo $A_\mathrm{s}=\asphere$ from the \textit{CHEOPS} data, and Bond albedo $A_\mathrm{B}=\morrisAB$ from the \textit{Spitzer} phase curve. Although small, the $A_\mathrm{g}$ and $A_\mathrm{s}$ indicate that MASCARA-1\,b is more reflective than most other ultra-hot Jupiters, where \ce{H-} absorption is expected to dominate.}
   {Where possible, priors informed by Doppler tomography should be used when fitting transits of fast-rotating stars, though multi-colour photometry may also unlock an accurate measurement of $\Psi$. Our approach to modelling the phase variations at different wavelengths provides a template for how to separate thermal emission from reflected light in spectrally resolved \textit{James Webb Space Telescope} phase curve data.}


   \maketitle
%

\section{Introduction}
\label{sec:intro}

Stars with types earlier than $\sim$F6 usually rotate rapidly, with O-type stars being observed to reach line of sight projected rotational velocities ($v\sin{i_\star}$) as high as 610 km s$^{-1}$ \citep{VFTS102}. Unlike the convective zones present in the outer layer of later types, the radiative outer layers of these stars lack the magnetic activity that dissipates their angular momentum with time, with the rapid rotation causing the radius at the equator to bulge with respect to the poles. \citet{vonZeipel} predicted that an effect of this oblateness would be that the flux emitted from the surface of the star would vary with local effective gravity. This results in both the brightness and temperature decreasing from the poles to the equator: a phenomenon known as gravity darkening.

The combined effect of the oblateness and surface temperature variations in the star causes the light curves of any transiting planet to deviate from the characteristic transit signatures of cooler hosts significantly \citep[see][]{Barnes09}. Fitting for these deviations can reveal a wealth of new information about the system including the spin-orbit alignment, which provides an insight into the formation and evolution of the system.

Spectroscopy of exoplanets during transit has proven to be a highly successful method of measuring their spin-orbit alignment. Observations of the Rossiter-McLaughlin (RM) effect \citep{Rossiter,McLaughlin} have been the most productive source of these measurements to date for slow-rotating stars \citep[e.g.][]{QuelozRM,BrothwellRM,T1RM}. Although these observations facilitate a measurement of the sky-projected spin-orbit angle $\lambda$, the fact that the stellar inclination $i_\star$ is degenerate with the stellar rotational velocity $v$ means that a precise measurement of the true, three-dimensional spin-obit angle $\Psi$ is not possible for the majority of cases where $v$ is difficult to measure. Doppler tomography observations can yield a similar, reliable measurement for faster rotators with few spectral lines \citep[e.g.][]{WASP33disc,KELT9Gaudi,MASCARA4disc}, but studies to date have not been able to resolve $i_\star$. A common method of breaking the aforementioned degeneracy is the use of asteroseismology to estimate $i_\star$ \citep[e.g.][]{Asteroseis_k50,Asteroseis_hp7,Asteroseis_G9}, although analytical criteria derived by \citet{Asteroseis_inc} suggest that this method is only reliable for the range of $20\degree<i_\star<80\degree$. Another method is the reloaded RM technique defined in \citet{Reloaded_def}, although the requirements of radial velocity measurements and a high-precision transit light curve has so far limited its application to a small number of systems \citep[e.g.][]{Reloaded_GJ436,Reloaded_HD3167,Reloaded_HD209}.

Unlike most spectroscopic methods, the transit light curve of a gravity-darkened host simultaneously encodes information about $v$, $i_\star$, and $\lambda$. This facilitates a measurement of $\Psi$ on the basis of time-series photometry alone, making it a powerful tool to study the history of planets orbiting early-type stars. Due to the high photometric precision required, studies have been limited until recently to planets and low-mass stellar companions orbiting a handful of the brightest stars in the \textit{Kepler} field \citep{Barnesk13,Barnesk2138,Szabo11,Szabo20,Masudak13,ZhouHuang13,Ahlersk368,Ahlersk89,Ahlersk976,Howarthk13,Hermank13}. Observations from the \textit{Transiting Exoplanet Survey Satellite} \citep[\textit{TESS};][]{TESS} have the benefit of observing many more of the systems producing the most pronounced gravity darkening signatures than \textit{Kepler}. However, its smaller collecting area, shorter observing baselines, and redder and narrower bandpass limit spin-orbit measurements to particularly favourable systems \citep[e.g.][]{Szabo20,AhlersM4,Ahlersk9}. To date, studies of this nature have only been conducted at optical wavelengths, and all but one \citep{Szabo20} have only used data taken at one wavelength band.

The main drawback of the gravity darkening method is the extent to which many of the observables from the transit light curve are correlated with each other. This makes accurate measurement of the full range of observables very challenging, and results are often highly sensitive to the choice of fixed parameters and the placement of priors within the model. The clearest example of this is the case of Kepler-13A\,b, where measurements of $\lambda$ range from a slightly misaligned value of $23\pm4$ deg \citep{Barnesk13} to a near-polar value of $59.20\pm0.05$ deg \citep{Howarthk13}: the latter of which is in close agreement with a measurement from Doppler tomography \citep{Johnsonk13}.

Roughly half an orbital period after the transit, a smaller drop in brightness is observed for the duration that the dayside of the exoplanet is occulted by its host. Optical wavelengths are an important window in which to observe the occultations of hot Jupiters, as the sharp drop off in thermal emission in the optical allows the reflective properties of their daysides to be studied. With the exception of Kepler-7b \citep[e.g.][]{Demoryk7}, the results from the \textit{Kepler} and \textit{K2} missions suggested that the population of observed hot Jupiters reflect a tiny fraction of the starlight incident upon their daysides \citep[systematic studies of \textit{Kepler} phase curves are presented in][]{Angerhausen,Esteves}. Results from \textit{TESS} satellite support this, though \citet{WongSouth} and \citet{WongNorth} observe a tentative trend of increasing $A_g$ with planetary equilibrium temperature. \citet{evanshd189} use \textit{Hubble}/STIS to report a significant detection of the occultation of HD 189733\,b at blue wavelengths -- in contrast to a non-detection at redder wavelengths -- perhaps a signature of Rayleigh scattering. The limited results from ground-based occultation observations in the optical and near-ultraviolet include similar non-detections \citep[e.g.][]{Chenw46,Hootonk9}, but also disagreeing results that are challenging to interpret \citep[a full discussion is contained in][]{Hootonw12}. Precise occultation observations of a greater number of hot Jupiters at multiple wavelengths are required to further understand the atmospheric processes shaping the reflective properties of these extreme planets.

The \textit{CHaracterising ExOPlanet Satellite} \citep[\textit{CHEOPS;}][]{BenzCHEOPS} was launched in December 2019, and has been acquiring high-precision optical photometry since going into full scientific operation in April 2020. Unlike the observing strategies of \textit{Kepler} and \textit{TESS}, \textit{CHEOPS} is a mission designed primarily to observe individual targets already known to host transiting planets, with the first results being presented in literature in recent months \citep{bonfanti21,BorsatoTTV,Delrez_nu2lupi,TOI178,Morris55cnce,SwayneEBLM,Szabo_aumic,VanGrootelCHEOPS}. With photometric precision and a wide optical bandpass\footnote{The \textit{CHEOPS} bandpass is almost identical to \textit{Gaia} $G_\mathrm{mag}$. More information can be accessed at \url{https://www.cosmos.esa.int/web/cheops/performances-bandpass}} both comparable to that of \textit{Kepler}, this strategy allows \textit{CHEOPS} to extend photometric measurements of $\Psi$ and occultation depth to the most amenable targets across most of the night sky. Demonstrating this capability, \citet{Lendlw189} reveal the asymmetry in the transit light curve of WASP-189\,b at high precision, confirming the polar $\lambda$ reported by \citep{w189} and measuring $\Psi=86.4^{+2.9}_{-4.4}$ deg. In addition, the occultation depth of $87.9\pm4.3$ ppm measured from four \textit{CHEOPS} occultation light curves suggests a geometric albedo $A_\mathrm{g}$ consistent with 0, although with no existing occultation data at longer wavelengths to disentangle thermal and reflected components of flux, an $A_g$ as high as $\sim0.3$ is also possible.

The Jupiter-sized MASCARA-1\,b \citep[T17 hereafter]{Talens} orbits its A8 type host star HD 201585 (MASCARA-1 hereafter) with a period of $\sim$2.14 d. The resulting planetary equilibrium temperature in excess of 2500 K places it squarely amongst the hottest, most highly irradiated known exoplanets. Doppler tomography of a MASCARA-1\,b transit also in T17 reveals both that the star is a rapid rotator ($v\sin{i_\star}=109\pm4$ km s$^{-1}$), and that its planetary companion orbits in a near-polar orbit ($\lambda=69.5\pm3.0 \degree$). With \textit{Gaia} $G_\mathrm{mag}=8.25$, it is one of the brightest stars known to host a hot Jupiter, making it particularly amenable to classification using high-precision space-based photometry. These aspects are all analogous to the well-studied ultra-hot Jupiters KELT-9\,b \citep{KELT9Gaudi} and WASP-33\,b \citep{WASP33disc}, albeit with a slightly larger separation from its host in each case. Recently, \citet{BellM1} present the analysis of a \textit{Spitzer}/IRAC 4.5 \si{\micro\meter} full-phase curve of MASCARA-1\,b acquired during its final year of operation as part of a population survey of \textit{Spitzer} phase curves. However, MASCARA-1 eluded observation during the \textit{Kepler} and \textit{K2} missions, and has not yet been observed by \textit{TESS} due to its declination close to the ecliptic plane.

In this paper we present Spi-OPS: the first joint analysis of \textit{Spitzer}/IRAC and \textit{CHEOPS} photometry. In it we present new \textit{CHEOPS} observations of the asymmetric transit and occultation of MASCARA-1\,b, which we analysed jointly with the \textit{Spitzer} phase curve. Using this, we characterised the oblate star, the misaligned planetary orbit, and the atmosphere of the highly irradiated exoplanet. We also include an investigation into the effect of using transit photometry at different wavelengths in measuring properties such as stellar rotation, stellar inclination and spin-orbit alignment. We present an independent derivation of the stellar parameters and discuss the possibility that the star is pulsating in \cref{sec:star}; we describe the observations and data reduction in \cref{sec:obs}; we describe how the light curves were jointly modelled \cref{sec:analysis}; we investigate the effect of the placement of different priors when fitting \textit{CHEOPS} and \textit{Spitzer} transits in \cref{sec:transit_fit}; we search for evidence of orbital precession in \cref{sec:precession}; we discuss the results of the joint model and present modelling of the atmosphere of the dayside in \cref{sec:discussion}, and conclude in \cref{sec:conclusion}.



\section{Stellar classification}
\label{sec:star}

\subsection{Derivation of stellar parameters}

In the following section, we describe our independent derivation of various stellar parameters including effective temperature $T_\mathrm{eff}$, average radius $R_\star$, mass $M_\star$, and age $t_\star$.

We used a HARPS high resolution spectra  with a signal-to-noise ratio (S/N) of $\sim1000$ at the central wavelength of 550 nm to derive $T_\mathrm{eff}$
from the broad line wings of H$\alpha$ with  
the  publicly available spectral analysis  
package Spectroscopy Made Easy  \citep[\href{http://www.stsci.edu/~valenti/sme.html}{\texttt{SME}};][]{vp96, pv2017}. 
We used   {\tt {SME}} version 5.2.2, 
selected the Atlas12 \citep{2013ascl.soft03024K} stellar atmosphere grids, and  
retrieved the line data from \href{http://vald.astro.uu.se}{VALD} \citep{Ryabchikova2015}. 
We fixed the projected stellar rotation velocity to 109~km~s$^{-1}$ from T17 
and obtained $T_\mathrm{eff} = 7490\pm 150$~K, which is in agreement with \citet{2012MNRAS.427..343M} and T17. 
The iron abundance relative to hydrogen was modelled from iron lines in the range 6200\,--\,6800~\AA, and was found to be [Fe/H]\,=\,$0.15 \pm 0.15$. 

We determined $R_\star$ using the infrared flux method \citep[IRFM][]{Blackwell1977} via relationships between the bolometric flux, the stellar angular diameter, the effective temperature, and the parallax. Using a Markov-Chain Monte Carlo (MCMC) algorithm, we used the stellar spectral parameters derived above as priors to build spectral energy distributions from the {\sc atlas} Catalogues \citep{Castelli2003}. We subsequently computed synthetic fluxes by convolving these models with the throughput of the considered photometric bands that were compared with the observed fluxes in these bandpasses; {\it Gaia} $G$, $G_{\rm BP}$, and $G_{\rm RP}$ \citep{GaiaCollaboration2020}; 2MASS $J$, $H$, and $K$ \citep{Skrutskie2006}; and {\it WISE} $W1$ and $W2$ \citep{Wright2010} to obtain the stellar bolometric flux and hence the angular diameter. Combined with the offset-corrected {\it Gaia} EDR3 parallax \citep{Lindegren2020}, we determined the stellar radius to be $2.072\pm0.022$\,$R_\odot$.

We inferred $M_{\star}$ and $t_{\star}$ from two different sets of stellar evolutionary models, namely PARSEC v1.2S \citep{marigo17} and CL\'ES \citep{scuflaire08}, by adopting $T_{\mathrm{eff}}$, [Fe/H], and $R_{\star}$ as input parameters. The output values from PARSEC v1.2S were computed interpolating within the grids of models by applying the Isochrone placement algorithm presented in \citet{bonfanti15,bonfanti16}. In the case of CL\'ES, a direct fit to the stellar models was instead performed. After checking the consistency of the two pairs of parameters through the validation procedure described in detail in \citet{bonfanti21}, we combined the probability distributions of both $M_{\star}$ and $t_{\star}$ to retrieve the corresponding medians and standard deviation as our reference values. A selection of parameters derived through the processes described in this section are shown in \cref{tab:star}.

\begin{table}
\caption{Properties of MASCARA-1 and the methods employed to derive them.}
\label{tab:star}
\centering
\begin{tabular}{lll}
\hline\hline
\multicolumn{3}{c}{MASCARA-1} \\    
\hline                        
\multirow{3}{2 cm}{Alternative names} & \multicolumn{2}{l}{HD\,201585} \\
 & \multicolumn{2}{l}{HIP 104513} \\
 & \multicolumn{2}{l}{\textit{Gaia} DR2 1744911763437512064} \\
\hline
Parameter & Value & Method \\
\hline
   $V$ [mag] & 8.27 & Simbad \\
   $G$ [mag] & 8.2424 & \textit{Gaia} EDR3 \\
   $J$ [mag] & 7.819 & 2MASS \\
   $T_{\mathrm{eff}}$ [K] & $7490\pm150$ & spectroscopy \\\relax
   [Fe/H] [dex] & $0.15\pm0.15$ & spectroscopy \\\relax
   $R_{\star}$ [$R_{\odot}$] & $2.072\pm0.022$ & IRFM \\
   $M_{\star}$ [$M_{\odot}$] & $1.825_{-0.095}^{+0.097}$ & isochrones \\
   $t_{\star}$ [Gyr]        & $0.8\pm0.2$ & isochrones \\
   $\rho_{\star}$ [$\rho_{\odot}$] & $0.205\pm0.013$ & from $R_{\star}$ and $M_{\star}$\\
   $L_{\star}$ [$L_{\odot}$] & $12.1\pm1.0$ & from $R_{\star}$ and $T_{\mathrm{eff}}$\\
    
\hline
   $R_{\star}$ [$R_{\odot}$] & $2.1\pm0.2$ & T17 \\
   $M_{\star}$ [$M_{\odot}$] & $1.72\pm0.07$ & T17 \\
   $\rho_{\star}$ [$\rho_{\odot}$] & $0.23_{-0.01}^{+0.03}$ & T17 \\
\hline
\end{tabular}
\end{table}

\subsection{Possible pulsations}
\label{sec:pulse}

The fundamental parameters of MASCARA-1 mean that it is likely to exhibit one of or both $\delta$ Scuti- and $\gamma$ Dor-type stellar pulsations. The former class corresponds to pulsations with periods typically between half an hour to several hours, with the latter between one and three days. The brightness variations induced by these pulsations are mostly the consequence of variations in the effective temperature. Assuming a black-body radiation spectrum, the largest brightness-variation amplitudes would be detected at wavelengths of $\sim$ 400~nm (covered at the blue end of the \textit{CHEOPS} bandpass) for a star like MASCARA-1, while no detectable variation is expected in the infrared domain of \textit{Spitzer}. 

We checked which pulsation modes are predicted to be unstable for the MASCARA-1 stellar model computed with the CL\'ES code. We used the non-adiabatic oscillation code MAD \citep{Dupret01} and found that only radial and non-radial modes of $\delta$ Scuti type are predicted to be excited by the $\kappa$ mechanism \citep{Kappa} in this model: the predicted pulsation periods range from $\sim$0.50 to 2.53 hours. 

We did not observe strong evidence for sizable $\delta$ Scuti-like pulsations in the \textit{CHEOPS} light curves described in \cref{sec:CHEOPS}, although at the high end of the period range each light curve only covers a few periods of the expected signal. Moreover, the 98.77 minutes duration of an orbit of the \textit{CHEOPS} satellite falls within this range, meaning that known instrumental signals occur on this timescale. Therefore, we made no attempt to fit for any low-amplitude $\delta$ Scuti pulsations in the \textit{CHEOPS} light curve, and observations with a much longer baseline would be needed to reveal any pulsations emanating from MASCARA-1\,b. We did observe time-correlated trends in some of the \textit{CHEOPS} light curves that were not removed by conventional methods for detrending \textit{CHEOPS} data, albeit over much longer timescales than the predicted periods of  $\delta$ Scuti pulsations. We discuss how we modelled these in \cref{sec:instrumental_model}.



\section{Observations and data reduction}
\label{sec:obs}

\subsection{\textit{CHEOPS}}
\label{sec:CHEOPS}

\begin{table*}
\caption{Details of each of the five \textit{CHEOPS} observations of MASCARA-1.}             
\label{table:obs}      
\centering          
\begin{tabular}{c c c c c c c c}     
\hline\hline       
Filekey\tablefootmark{(a)} & Start time & End time & $N_\mathrm{exp}/N_\mathrm{int}$\tablefootmark{(b)} & $t_\mathrm{int}$ & Type & $N_\mathrm{frames}$ & Efficiency\\ 
& \multicolumn{2}{c}{(UTC)} &  & (s) &  &  & (\%)\\
\hline                    
PR100016\_TG008501\_V0200 & 2020-07-01 12:34 & 2020-07-01 22:20 & 1 & 23.0 & Occultation & 897 & 58.7 \\
PR100020\_TG000401\_V0200 & 2020-07-09 00:24 & 2020-07-09 12:36 & 2 & 43.6 & Transit & 577 & 57.2 \\
PR100016\_TG008502\_V0200 & 2020-07-14 10:48 & 2020-07-14 20:17 & 1 & 23.0 & Occultation & 919 & 61.9 \\
PR100016\_TG008503\_V0200 & 2020-07-23 01:08 & 2020-07-23 10:54 & 1 & 23.0 & Occultation & 1201 & 78.5 \\
PR100020\_TG001101\_V0200 & 2020-07-24 01:27 & 2020-07-24 13:58 & 2 & 43.6 & Transit & 860 & 83.3 \\
\hline                  
\end{tabular}
\tablefoot{\tablefoottext{a}{The filekey is the unique identifier associated with each dataset processed by the DRP.} \tablefoottext{b}{$N_\mathrm{exp}/N_\mathrm{int}$ denotes the image stacking order.}}
\end{table*}

The \textit{CHEOPS} satellite hosts an f/8 Ritchey-Chr\'etien on-axis telescope with an effective diameter of $\sim$30 cm, with a single frame-transfer back-side illuminated charge-coupled device (CCD) detector. \textit{CHEOPS} observed two transits and three occultations of MASCARA-1\,b in July 2020 as part of the Guaranteed Time Observation (GTO) programme, which are summarised in \cref{table:obs}. Due to the limited downlink bandwidth of the \textit{CHEOPS} satellite, observations with exposure time $t_\mathrm{exp}< 22.65$ s have their frames stacked prior to downlinking\footnote{See the \href{https://www.cosmos.esa.int/documents/1416855/4841242/CHEOPS-UGE-PSO-MAN-001_i2.0-Observers+Manual.pdf/ae8c9725-51d6-97bd-6efe-0bb4c2d4f2c5?t=1604331831450}{\textit{CHEOPS} observers manual.}
}. The transit and occultation observations were acquired as part of two different GTO sub-programmes. Although very similar and chosen to achieve $\sim$90 \% of pixel full-well capacity, the difference in requested $t_\mathrm{exp}$ led to them coincidentally falling either side of the 22.65 s threshold, resulting in the transit observations having a stacking order of two and the occultations a stacking order of one. The result of this is a large difference in integration time $t_\mathrm{int}$ (the total $t_\mathrm{exp}$ per downlinked frame), visible in \cref{table:obs}. As \textit{CHEOPS} occupies a low-Earth orbit with a period of $\sim98.77$ min, the listed efficiencies refer to the proportion of the allocated time that the target was observable. Interruptions are primarily caused by Earth occultations, the satellite was crossing the South Atlantic Anomaly (SAA), and the detected stray light exceeding a predefined threshold computed by the instrument team, which is designed to keep the \textit{CHEOPS} noise budget within requirements. 

\begin{figure}
    \includegraphics{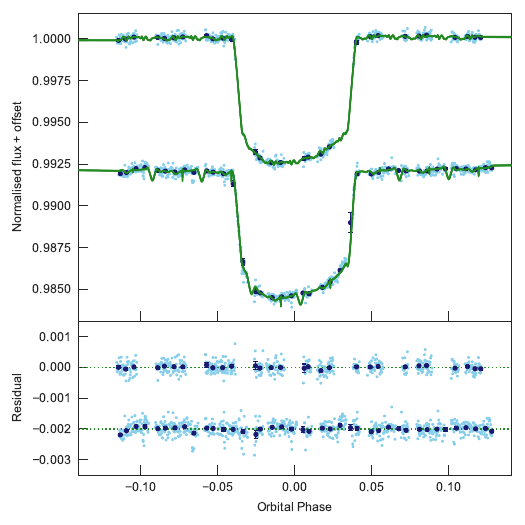}
    \caption{Normalised flux associated with the two \textit{CHEOPS} transit observations as a function of orbital phase, with an arbitrary offset applied for display purposes. The raw data are shown in light blue points, and data in bins of 12 minutes are shown in dark blue points with error bars. The best fitting model is shown with a green line, and the 32 models evenly spaced in the chain are shown in fainter green lines. The top panel shows the raw flux, and the bottom panel shows the residual flux. The asymmetry visible in the transits is caused by stellar rotation-induced gravity darkening. The root mean square (RMS) of the residuals is 365 ppm.}
    \label{fig:transit_DRP}
\end{figure}

\begin{figure}
    \includegraphics{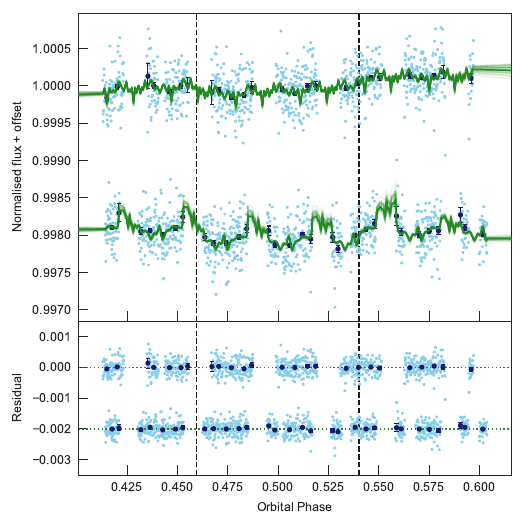}
    \caption{Similar to Fig. \ref{fig:transit_DRP}, but displaying the second and third \textit{CHEOPS} occultation observations. The dashed lines mark the phases of the beginning and end of the occultation. The RMS of the residuals is 226 ppm.}
    \label{fig:eclipse_DRP}
\end{figure}

Following downlink, the data were reduced using version 13 of the \textit{CHEOPS} Data Reduction Pipeline \citep[DRP; described in full in][]{DRPHoyer}. In short, each frame is calibrated (by applying corrections for bias, gain, non-linearity, dark current and flat fielding), corrected for environmental effects (cosmic rays, smearing trails from nearby stars, and background), and then aperture photometry is performed for three fixed aperture sizes, along with a fourth that is performed at a size that minimises scatter due to contamination from background sources. The DRP estimates the contamination from nearby objects by simulating the \textit{CHEOPS} field of view for each frame in the observation based on the \textit{Gaia} DR2 star catalogue \citep{GaiaCollaboration2020} and a template of the extended \textit{CHEOPS} point spread function (PSF). Due to the brightness of MASCARA-1, the mean contamination through all of the \textit{CHEOPS} observations was estimated to be $<$ 0.05 \%. We selected the DEFAULT aperture of 25 pixels (25$^{\prime\prime}$) in all cases, which minimised the route mean square (RMS) of each of the light curves, and discarded data points where the flux fell $>5\sigma$ from the mean of the ten closest data points. The light curves are shown for the transits -- visibly asymmetric due to the stellar rotation-induced gravity darkening -- in \cref{fig:transit_DRP} and for the second and third occultations in \cref{fig:eclipse_DRP}, with the models described in \cref{sec:analysis} also shown in each case. Collectively, the \textit{CHEOPS} light curves cover 43.5\% of full-phase coverage.

\begin{figure}
    \includegraphics{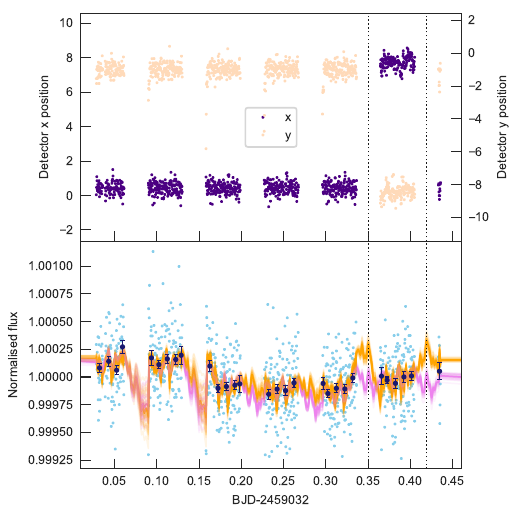}
    \caption{Top: $x$ (indigo) and $y$ (beige) positions of the PSF centroid on the detector. For the orbit that falls between the dotted lines, the PSF occupies a position roughly ten pixels away compared to the other orbits. Bottom: similar to the top panel of \cref{fig:eclipse_DRP}, but displaying the first \textit{CHEOPS} occultation observation. The orange lines display models including $y$ and $y^2$ detector position in the baseline model. The violet lines display models instead including time in the baseline model.}
    \label{fig:eclipse1}
\end{figure}

During one \textit{CHEOPS} orbit of the first occultation observation, a problem with the guiding caused the target PSF to occupy a position on the detector roughly ten pixels away from its position for the other orbits (see top panel of \cref{fig:eclipse1}). For bright targets like MASCARA-1, the \textit{CHEOPS} payload is in-the-loop for improved pointing performances (jitter of order 1-2 arcsec). For the orbit in question, the centroiding algorithm could not lock the pointing on the target and the payload was consequently not in the loop. This issue occurred right after an interruption where the target was not visible for 40 minutes. Updates to the centroiding algorithm have improved its performance since then, and this issue has not been observed again.

When fitting an occultation model to this dataset alone, the measured occultation depth was very sensitive to the combination of auxiliary observing parameters used in the baseline model (see \cref{sec:instrumental_model}). This is because the erroneous orbit fell very close to the end of the observation, with the ten integrations in the final orbit proving insufficient to discriminate between a range of baseline models. This is demonstrated in the bottom panel of \cref{fig:eclipse1}, which shows the raw data for the first occultation. The flux predicted by models that include detector position (orange) and time (violet) in the baseline model significantly diverge during the final few orbits, with the respective occultation depths measured to be $232\pm19$ and $164\pm18$ ppm. Not only do these significantly disagree with each other, but they are much deeper than the depths measured for the other two occultations, which are both relatively insensitive to the choice of linear basis vectors used. For this reason we excluded the first occultation from the joint analysis described in the following section. However, we cannot rule out the possibility of part of the increase in depth materialising due to an increase in the brightness from the MASCARA-1\,b dayside.

\subsection{\textit{Spitzer}}

\begin{figure*}
    \includegraphics{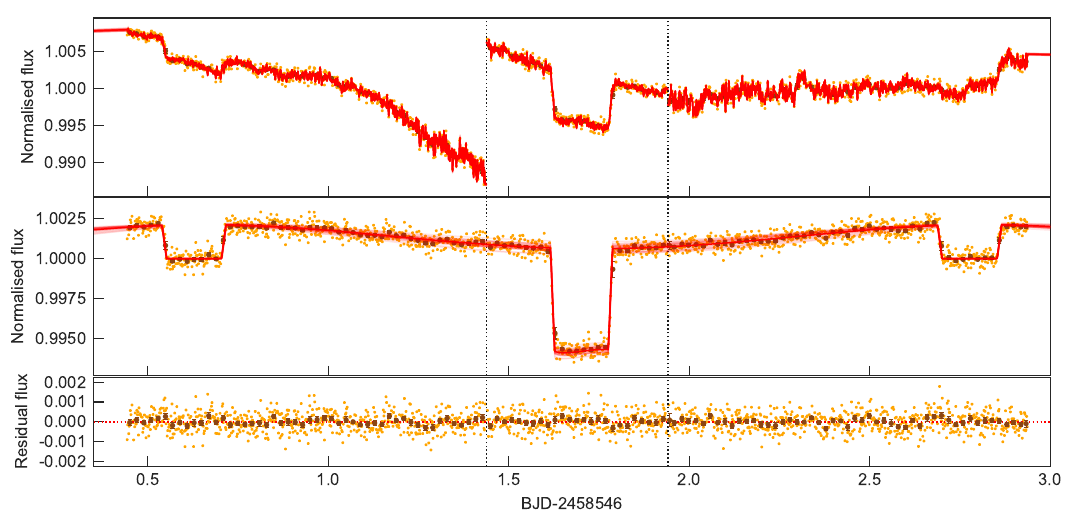}
    \caption{Normalised flux associated with the \textit{Spitzer} phase curve observation. The raw data are shown in orange points, and data in bins of 30 minutes are shown in brown points with error bars. The best fitting model is shown with a red line, and the 32 models evenly spaced in the chain are shown in fainter red lines. The times when a new AOR begins are marked with dashed vertical lines. The top panel shows the raw flux, the middle panel shows the flux corrected for the instrumental effects with the transit, occultations and phase variations, and the bottom panel shows the residual flux. The mean absolute deviation of the residuals is 244 ppm.}
    \label{fig:spitzer}
\end{figure*}

A full phase curve of MASCARA-1\,b \citep[which was previously presented in][]{BellM1} was observed by \textit{Spitzer}/IRAC using the channel 2 \SI{4.5}{\micro\meter} bandpass as part of programme 14059 with PI Jacob Bean, which we downloaded from the Spitzer Heritage Archive \footnote{\url{http://sha.ipac.caltech.edu}}. The observation began at 2019-03-03 22:32 UTC shortly before the ingress of a MASCARA-1\,b occultation, and ended after 3.50 days of continuous observation and shortly after the egress of the following occultation. In this time, 106496 32x32 pixel images (with a 39$^{\prime\prime}$x39$^{\prime\prime}$ field of view) were acquired with an exposure time of 1.92 s, which were packaged in 1664 data cubes of 64 frames and acquired across three Astronomical Observing Requests (AORs). A full description of the method used to reduce the \textit{Spitzer} data is given in \citet{Demory55cnce}, which produces the raw light curve shown in the top panel of \cref{fig:spitzer}, with changes in AORs marked with dotted black lines.


\section{Light curve fitting}
\label{sec:analysis}

A good model for jointly fitting the \textit{CHEOPS} and \textit{Spitzer} data has to take account of a number of signals present: the asymmetric transits of the oblate star, the flux deficit during the occultations, the phase-dependent modulation as the visible hemisphere of the tidally locked planet rotates with respect to the observer, the instrumental trends that are routinely present in \textit{CHEOPS} photometry \citep[e.g.][]{Lendlw189,bonfanti21}, and any other time-correlated trends. To incorporate each of these signals, we created a custom model to simultaneously fit the light curves based on the existing tools available in \texttt{pycheops}\footnote{\url{https://github.com/pmaxted/pycheops}} \citep[see][]{Maxted_classics}, a package that allows simultaneous modelling of transiting and occulting planets or eclipsing binaries along with the most common systematic trends present in \textit{CHEOPS} datasets.

The model we used to fit the light curves is the sum of flux contributions from the star, planet, and instrumental systematics, where each term is normalised by the mean stellar flux when MASCARA-1\,b is fully occulted by its host. We describe each of these terms in the following sections.

\subsection{Stellar model}\label{sec:stellar_model}

The flux from the star decreases during the transit of MASCARA-1\,b, and is assumed to be constant out of transit. We modelled the gravity-darkened transits using the GravityDarkenedModel in \texttt{PyTransit}\footnote{\url{https://github.com/hpparvi/PyTransit}} \citep{pytransit}, which implements the widely used model first presented in \citet{Barnes09}\footnote{We note an error in \citet{Barnes09}, which is that all instances of $(1-f^2)$ should be replaced with $(1-f)^2$ in their Eq. 14. The current equation biases the coordinate of the oblate photosphere along the axis parallel to the line of sight.}. This will be described in full in \citet{2022arXiv220104518D}, but we describe the main steps here. 

As described in \cref{sec:intro}, the surface temperature of a rapidly rotating star is well-described by the \citet{Maeder} adaptation of von Zeipel's theorem \citep{vonZeipel}, which links the temperature and gravity at any point on the stellar surface by the equation. This is given by
\begin{equation}\label{eq:vonzeipel}
    \frac{T(\vartheta)}{T_\mathrm{pole}}=\left (\frac{g(\vartheta)}{g_\mathrm{pole}} \right )^{\beta},
\end{equation}
where $\vartheta$ is the stellar latitude, $g$ is local surface gravity, $g$ is polar surface gravity, and $\beta$ is the gravity darkening coefficient. \texttt{PyTransit} assumes that such a rapidly rotating star is well-approximated by a Roche model. The mass is concentrated at the centre of the star, the inner layers are spherical, and only the outermost layers of the star exhibit distortion due to the rotation. \texttt{PyTransit} computes the flux from the star in a grid of points on the surface projected onto the plane of the sky, and generates the light curve by evaluating the grid points that are hidden during a planetary transit as a function of time.

Changes in the light curve due to both the local temperature differences and oblateness induced by the rapid rotation are taken into account. Local temperature can be derived using \cref{eq:vonzeipel}, with $g$ determined by equating the gravitational and centrifugal forces:
\begin{equation}
	\overrightarrow{g}\!\left(\vartheta\right)=-\frac{G\,M_\star}{r_\vartheta^2}\,\hat{u_r}+\left(\frac{2\pi}{P_\mathrm{rot}}\right)^2R_\star\sin\!\left(\vartheta\right)\,\hat{u_{\bot}}.
\end{equation}
Here, $G$ is the gravitational constant, $M_\star$ is the stellar mass, $P_\mathrm{rot}$ is the stellar rotational period, $R_\star$ is the stellar equatorial radius, and $\hat{u_r}$ and $\hat{u_{\bot}}$ are unit vectors radially from the centre and perpendicular to the stellar spin axis, respectively. $r_\vartheta$ is the latitute-dependent stellar radius accounting for the stellar oblateness $f_\star$, and is equal to $R_\star$ at the equator and $R_\star(1-f_\star)$ at the poles. In turn, $f_\star$ is given by
\begin{equation}
    f_\star=\frac{3\pi}{2G\rho_\star P_\mathrm{rot}^2},
\end{equation}
where $\rho_\star$ is the stellar density. The local flux is then derived from the local temperature by integrating the temperature-dependent stellar spectrum over the filter transmission function, taking into account the effects of limb darkening. As the effect of the gravity darkening varies as a function of wavelength, the light curve is computed by integrating the stellar flux as a function of time over the filter response function. With $T(\vartheta)$ as an input \texttt{PyTransit} allows the use of a Planck function or a PHOENIX spectrum \citep{PHOENIX} to represent the stellar spectrum for each point on the surface. The PHOENIX spectrum is particularly desirable for the \textit{CHEOPS} observations due to the large deviation in the spectra of early-type stars from that of a Planck function. As the spectra only extend to the reddest wavelengths in the \textit{CHEOPS} bandpass, we used a Planck function for the \textit{Spitzer}/IRAC \SI{4.5}{\micro\meter} bandpass, which is much more representative of the star at infrared than at optical wavelengths. For the sake of computational efficiency, the stellar spectra were interpolated onto 15 wavelength bins for each response function. In addition to the model presented here, we obtained consistent results using a fit of only the \textit{CHEOPS} transit light curves using \texttt{TLCM}\footnote{\url{http://www.transits.hu/}} \citep{TLCM} following the same methodology as \citet{Lendlw189}. 

The following terms in our parameterisation of the transit light curve are compatible with a symmetric transit, such as that of \citet{mandelagol}. They are:
\begin{itemize}
    \item the orbital period $P$,
    \item time of inferior conjunction $t_\mathrm{0}$,
    \item transit depth $\delta_\mathrm{trans}=(R_\mathrm{p}/R_\star)^2$,
    \item stellar density $\rho_\star$,
    \item impact parameter $b$,
    \item eccentricity $e$ and argument of peristron $\omega_\mathrm{peri}$, parameterised as $e\cos{\omega_\mathrm{peri}}$ and $e\sin{\omega_\mathrm{peri}}$,
    \item and the quadratic limb darkening parameters $q_\mathrm{1}$ and $q_\mathrm{2}$ \citep{limbdark}.
\end{itemize}
A variety of different parameterisations of gravity-darkened light curves have been presented in literature, which has on occasion led to common names and symbols having different definitions between different papers. We use the same definitions and notation that are used by \citet{Masudak13}, with the light curve asymmetry parameterised by the following additional terms.
\begin{itemize}
    \item stellar equatorial radius $R_\star$.
    \item sky-projected spin-orbit angle $\lambda$: the sky-projected angle between the orbital and stellar equatorial  planes. This term is used interchangeably with (sky-projected) stellar obliquity by \citet{Masudak13} and by much of the related spectroscopy literature. We measure orbital plane anti-clockwise with respect to the stellar equatorial plane (sometimes termed $\alpha$), as opposed to the reverse direction (sometimes termed $\beta$).
    \item stellar inclination $i_\star$: the angle between the stellar rotation axis and the observer's line of sight. This is related to the \citet{Barnes09} definition of stellar obliquity $\phi$---the angle between the equatorial plane and the observer's line of sight---by the equation $i_\star=90$ deg $-\phi$.
    \item stellar rotational velocity projected onto the observer's line of sight $v\sin{i_\star}$,
    \item gravity darkening coefficient $\beta$,
    \item and stellar polar temperature $T_\mathrm{pole}$.
\end{itemize}

This parameterisation differs from that of \cite{Masudak13} by the use of $v\sin{i_\star}$ instead of the stellar rotation frequency $f_\mathrm{rot}$, and $\rho_\star$ instead of $M_\star$: all of which are related by the equation
\begin{equation}
    \frac{v\sin{i_\star}}{2\pi f_\mathrm{rot}}=\left (\frac{3M_\star}{4\pi\rho_\star}\right )^{\frac{1}{3}} \sin{i_\star}.
\end{equation}
Our choice of input parameters was informed by the desire to test the effect of placing Gaussian priors on observables from the Doppler tomography and our stellar characterisation (see \cref{sec:star}).

As the Doppler tomography observation presented in T17 measures a prograde orbit at high significance, we placed a prior of $-90<\lambda<90$ deg to exclude retrograde solutions. With this taken into account, the combination of the transit light curve and the Doppler tomography still cannot distinguish between solutions with $\{b,i_\star,\lambda\}$ and $\{-b,-i_\star,-\lambda\}$. By applying the constraint of $b>0$ and letting $i_\star$ vary in the full range of -90 deg $<i_\star<$ 90 deg\footnote{As for all of our runs without a Gaussian prior on $b$ we saw that $b\rightarrow0$ in the sampling, but the Doppler tomography suggests that $b$ is nonzero with high significance, for the remainder of the study we assume that the solution with positive $b$ exists for $0<i_\star<90$ deg. However, our results cannot distinguish between results with negative $b$ for any given solution.}, our model will sample the only solution that exists within these limits. This is explained in greater detail in \cref{sec:tomography}.


As described in \cref{sec:intro}, large degeneracies exist between the measurements of many of the parameters listed above, and previous studies \citep[e.g.][]{Barnesk13,Masudak13} have demonstrated that large disagreements in the measured asymmetric parameters can arise depending on the choice of free and fixed parameters within the model. We present a detailed investigation of this in \cref{sec:transit_fit}, where we test the effect of fixing and varying different parameters when fitting the \textit{CHEOPS} and \textit{Spitzer} light curves separately and together.

\subsection{Planetary model}\label{sec:planetary_model}

The flux from the planet was modelled as the product of models to account for phase-dependent flux modulation and the occultation. We modelled the phase-dependent flux by self-consistently generating the components of flux due to thermal emission and reflected starlight. Although numerous phase curves of hot Jupiters have also exhibited signals associated with Doppler boosting and tidal ellipsoidal distortion \citep[e.g.][]{Shporerk13,Wongk9,Owens_w12}, relations presented by \citet{Morris_ellipsoid1}, \citet{Morris_ellipsoid2}, and \citet{Shporer_w18} suggest that the amplitude of these signals for MASCARA-1 would be roughly 10 ppm and 25 ppm, respectively. As we would be unable to significantly detect signals of these size in the \textit{Spitzer} phase curve and due to the limited phase coverage of the \textit{CHEOPS} data, we did not fit for these signals in our model.

To model the thermal phase variations in the \textit{CHEOPS} and \textit{Spitzer} light curves, we used a set of mathematical basis functions known as parabolic cylinder functions, which were derived by \citet{HengWorkman} to describe a heated fluid layer on a rotating sphere with frictional forces. \citet{Morris2021_hml} (M21 hereafter) demonstrate that these basis functions naturally describe the chevron-shaped feature in hot Jovian photospheres \citep{ShowmanGCM,Shami2014} and are able to fit a sample of eight \textit{Spitzer} phase curves with a small number of physically motivated parameters. A distinguishing feature of this approach is that it generates a single 2D temperature map, which is used to compute the phase-dependent flux in multiple bandpasses.


The full method is described in M21, but we give a brief summary of the steps here. Firstly, we generated a 2D temperature map as a function of planetary latitude $\theta$ and longitude $\phi$. This is given by the equation (1) of M21,
\begin{equation}\label{eq:T}
    T\left (\theta,\phi\right )=\bar{T}\left ( 1+\sum_{m,\ell}^{\ell_\mathrm{max}}h_{m,\ell}(\theta,\phi)\right ),
\end{equation}
where $\bar{T}$ is the mean background temperature. The $h_{m,\ell}$ basis functions describe perturbations to the mean background temperature and are given by equation (258) of \citet{HengWorkman},
\begin{multline}\label{eq:hml}
    h_{m,\ell}=\frac{C_{m,\ell}}{\omega_\mathrm{drag}^{2}\alpha^4+m^2}e^{-\tilde{\mu}^2/2}[\mu mH_\ell\cos{(m\phi)}\\+\alpha\omega_\mathrm{drag} \left (2lH_{\ell-1}-\tilde{\mu}H_\ell \right ) \sin{(m\phi)}]
\end{multline}
where $\alpha$ is a dimensionless fluid number that depends on the Rossby and Prandtl numbers and $\omega_{\rm drag}$ is the dimensionless drag frequency. M21 demonstrated that synthetic temperature maps generated by general circulation models (GCMs) are well-fitted by $\alpha=0.6$ and $\omega_{\rm drag}=4.5$; we hold these parameter values fixed as they control the latitudinal distribution of temperature, which is not constrained by real phase curves. Each $h_{m,\ell}$ mode is described by a pair of wavenumbers $(m,l)$, which are analogous to quantum numbers in the quantum harmonic oscillator; the power in each mode is quantified by $C_{m,\ell}$.  Other quantities include $\mu=\cos{\theta}$ and $\tilde{\mu}=\alpha\mu$. The physicist's Hermite polynomials are represented by $H_\ell$. M21 further showed that one needs to specify a hot spot offset $\Delta\phi$ within the fitting procedure that translates the entire temperature map in longitude.


After generating the temperature map, we then integrated the ratio of the planetary and stellar flux as a function of orbital phase $\xi$ (normalised such that transits occur at $\xi=\pm\pi$) according to \citet{Cowanphasecurve}:
\begin{equation}\label{eq:fluxratio}
    \frac{F_\mathrm{p}}{F_\star}=\frac{1}{\pi I_\star}\left (\frac{R_\mathrm{p}}{R_\star}\right )^2\int_{0}^{\pi}\int_{-\xi-\pi/2}^{-\xi+\pi/2}I_\mathrm{p}(\theta,\phi)\cos{(\phi+\xi)}\sin{^2(\theta)}d\phi d\theta,
\end{equation}
with planetary intensity $I_\mathrm{p}$ given by
\begin{equation}\label{eq:intensity}
    \int\lambda\mathcal{F}_\lambda \mathcal{B}_\lambda (T(\theta,\phi))d\lambda
\end{equation}
for a filter with photon flux response function $\mathcal{F}_\lambda$ in units of electrons per photon, where $\mathcal{B}_\lambda$ is a Planck function. For $I_\star$, we used a PHOENIX spectrum interpolated according to the values of $T_\mathrm{eff}$, $\log g$, and [Fe/H] listed in \cref{tab:star}.

\begin{figure*}
    \includegraphics{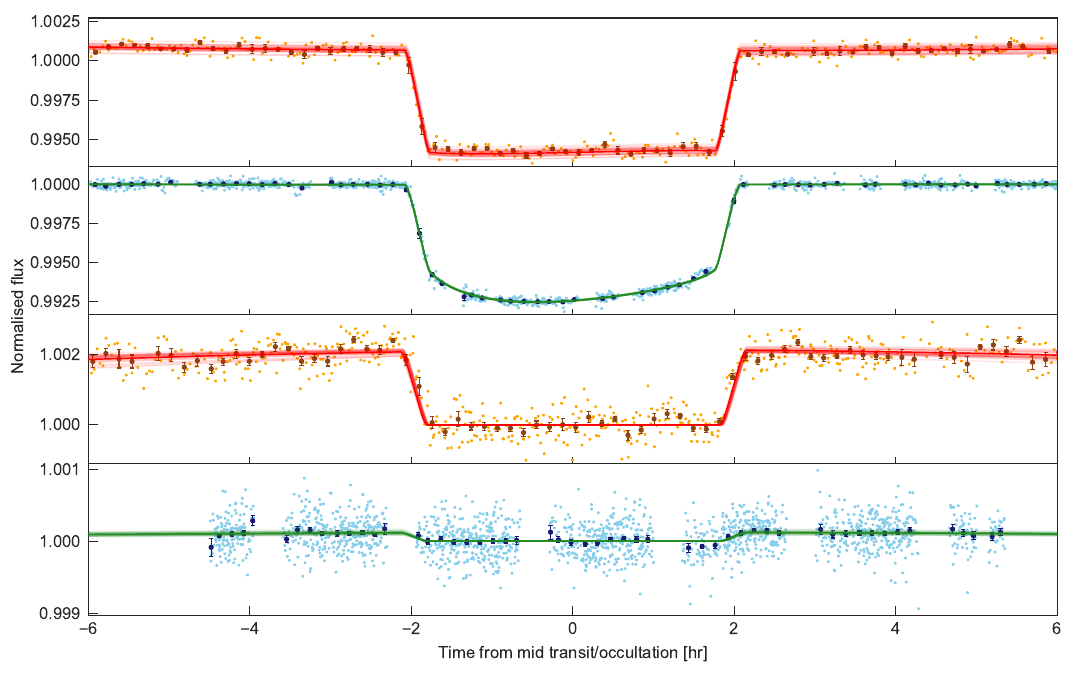}
    \caption{Phase-folded \textit{Spitzer} and \textit{CHEOPS} photometry centred on the transits and occultations, using the same format as the previous light curves. Point with error bars show data in bins of 12 minutes. From top to bottom, the panels display the \textit{Spitzer} transit, the \textit{CHEOPS} transit, the \textit{Spitzer} occultation, and the \textit{CHEOPS} occultation, respectively.}
    \label{fig:phasefold}
\end{figure*}

Alongside other parameters already defined in \cref{sec:stellar_model}, for $\ell_\mathrm{max}=1$\footnote{M21 showed that setting $\ell_\mathrm{max}=1$ provides a good approximation for the analysis of \textit{Spitzer} phase curves. Here, only the Hermite polynomials $H_\mathrm{0}=1$ and $H_\mathrm{1}=2\tilde{\mu}$ in \cref{eq:hml} are used.} the thermal phase modulation can be modelled using three parameters: $C_\mathrm{1,1}$\footnote{For $\ell_\mathrm{max}=1$, $C_\mathrm{1,1}$ is the only non-zero $C_\mathrm{m,\ell}$ power coefficient, as $C_\mathrm{m,0}\equiv0$ and the perturbation arising from $C_\mathrm{m,\ell}$ would be equal and opposite to that created by $-C_\mathrm{m,-\ell}$ occupying the same value.},  $\Delta\phi$, and $\bar{T}$. Conceptually, these three parameters can be compared in an analogy to a sinusoidal phase curve model where $C_\mathrm{1,1}$ is related to the semi-amplitude of the sinusoid, $\Delta\phi$ is a constant phase offset of the sinusoid, and $\bar{T}$ denotes the DC constant offset term from zero-flux. We allowed $\Delta\phi$ and $\bar{T}$ to vary uniformly, and $C_\mathrm{1,1}$ to vary logarithmically. Given the temperature map yielded using these parameters, one may derive the Bond albedo $A_{\rm B}$ and redistribution efficiency $\varepsilon$, which we describe in \cref{sec:budget}.



For the reflected light component of the phase variation, we used the general ab initio solutions for a semi-infinite atmosphere derived by \citet{Heng_reflected}; we emphasise that `semi-infinite' refers to the optical depth and not the spatial extent of the atmosphere. Generally, the amplitude of a reflected-light phase curve is described by the geometric albedo $A_g$, while its shape is quantified by the integral phase function $\Psi_\mathrm{ph}$. These quantities are in turn functions of the single-scattering albedo $\omega_\mathrm{scat}$ and the scattering asymmetry factor $g$, which are fundamental scattering parameters. Forward, reverse, and isotropic scattering correspond to $g=1$, -1 and 0, respectively. The widely used Henyey-Greenstein scattering phase function or reflection law \citep{HenyeyGreenstein} is employed. It is worth noting that this is a more general approach than assuming a Lambertian sphere, which is an artificial construct where an object is equally bright in reflected light regardless of viewing angle; see \citet{Dyudina_phasecurves} for multiple examples of how the reflected light phase curves of Solar System planets and moons are not well described by the Lambertian reflection law. Since no phase offset is ultimately detected, it is sufficient to use equations (8) and (9) of \citet{Heng_reflected}, which respectively describe $A_g$ and $\Psi_\mathrm{ph}$ for a homogeneously reflective sphere. Upon obtaining $\Psi_\mathrm{ph}$, it may be integrated over phase angle to obtain the phase integral $q$ \citep{PhaseIntegral}. The spherical albedo is then $A_s = q A_g$. Previously, the homogeneous-sphere solution was employed, in tandem with a double Henyey-Greenstein scattering phase function, to successfully fit the Cassini phase curves of Jupiter \citep{HengCassini}.

We used a single set of $\{C_\mathrm{1,1}, \bar{T}, \Delta\phi, \omega_\mathrm{scat}, g\}$ to model phase variation in the light curves of both \textit{CHEOPS} and \textit{Spitzer}. The \textit{CHEOPS} observations described in \cref{sec:CHEOPS} cover well under half of an orbit of MASCARA-1\,b across five separate visits, and thus only impose loose upper and lower limits on some parameters and do not meaningfully constrain others. The effect of fitting the phase curve model to the \textit{CHEOPS} and \textit{Spitzer} data simultaneously is to use the precise constraints availed by the full phase coverage of the \textit{Spitzer} light curve to estimate the phase modulation in the \textit{CHEOPS} light curves. Although some level of wavelength-dependence is expected for each of the terms, we assumed that the values for the atmospheric layer probed by the \textit{Spitzer} observation are a good approximation for the layer probed by \textit{CHEOPS}. This also requires us to neglect the role that chemical emission and absorption features may have on the thermal emission. Spectrally resolved observations in the future will be far better placed to detect such deviations than the light curves acquired using two broadband features that we are presenting.


We modelled the occultations using a Mandel and Agol transit model \citep{mandelagol} with the limb darkening parameters fixed to 0. The model uses the same input parameters as the symmetric transit, along with $R_\star$ to compute a delay in the occultation due to the difference in light travel time between the star and planet (roughly 20 seconds in the case of MASCARA-1). Occultation depth is not an input for the model, and is derived by recording the sum of the fluxes of the thermal and reflected components described above at the phase of the occultation centre.

\subsection{Instrumental model}
\label{sec:instrumental_model}

We modelled systematic trends in the \textit{CHEOPS} light curves using a set of linear basis vectors of auxiliary observation parameters that are output by the DRP, such as sky background, contamination, detector smearing correction, detector $xy$-coordinates, and a `glint' function (see Maxted et al. \textit{in review}) created in \texttt{pycheops} by fitting a spline function to the out-of-transit or occultation flux as a function of telescope roll angle. We found that linear decorrelation only using a glint function per \textit{CHEOPS} dataset and no other parameters optimised the Bayes factor \citep[see the following section]{BayesFactor}, with the exception of the final \textit{CHEOPS} occultation where sky background was also included.

As discussed in \cref{sec:pulse}, we observed low-frequency time-correlated trends in the \textit{CHEOPS} light curves that we are unable to attribute to astronomical or instrumental sources. For this reason, we modelled them using a Gaussian process (GP) with a simple harmonic oscillator kernel using the \texttt{celerite2}\footnote{\url{https://github.com/exoplanet-dev/celerite2}} package \citep{celerite1,celerite2}. Following its introduction by \citet{GibsonGP}, GPs offer a flexible method to model correlated noise and accurately determine the parameters in transit light curves \citep{GibsonGP2}. The GP is parameterised by the frequency of the undamped oscillator $\omega_\mathrm{0}$, $S_\mathrm{0}$ is proportional to the power in the power spectral density at $\omega=\omega_\mathrm{0}$, and the quality factor of the oscillator $Q$. We allowed the first two to vary on a logarithmic scale and fixed the latter to 10.


For each \textit{Spitzer} AOR, we performed the same process using polynomial functions of $x$ and $y$ centroid position and $x$ and $y$ FWHM of the point response function up to the second order, including $xy$ cross terms. The full set of linear basis vectors used for both the \textit{CHEOPS} and \textit{Spitzer} detrending is shown in \cref{tab:noise}.


\subsection{Posterior sampling}\label{sec:sampling}

We fitted the trends present in the \textit{CHEOPS} light curves using a \texttt{pycheops} Multivisit object. To derive the joint posterior distribution, we used the Dynamic Nested Sampler \citep[see][]{ns1,ns2,dns1} in the \texttt{dynesty} package \citep{dynesty}. Unlike a Markov Chain Monte Carlo (MCMC) algorithm \citep[implemented in \texttt{pycheops} using the \texttt{emcee} package][]{emcee}, Nested Sampling estimates the Bayesian evidence $\mathcal{Z}$ of the model, allowing for robust model comparison. We ran the fit with numerous combinations of the linear basis vectors described above, and selected the model that returned the highest estimate of $\mathcal{Z}$. 

Due to residual time-correlated noise in the \textit{Spitzer} phase curve, for each of our fits we adopted the time-averaging method described in \citet{Winn_timeave}. This involved determining the mean of the scaling parameter $\beta$ (not to be confused with the gravity darkening coefficient $\beta$) for bin sizes of 1-100 data points, and then running the sampling again with the uncertainties for each of the \textit{Spitzer} data points boosted by a factor of $\beta$.


\section{Accurate spin-orbit angle determination}
\label{sec:transit_fit}

Simultaneously acquiring accurate measurements of the parameters listed in \cref{sec:stellar_model} from a transit light curve presents a major challenge. \citet{Barnes09} described how the light curves spin-orbit aligned planets orbiting oblate stars mimic those of spherical stars at high impact parameters, which introduces a bias into the measurement of the transit depth. Additionally, correlations exist between so many of these parameters that an incorrect assumption about one can lead to biases in the measurements of several others.

The different reported values of sky-projected spin-orbit angle $\lambda$ for Kepler-13A\,b -- where planet and host are comparable to MASCARA-1\,b -- provide a demonstration of this. Doppler tomography of the transit \citep{Johnsonk13} measures $\lambda=58.6\pm2.0$ deg. For measurements using the \textit{Kepler} photometry, this agrees well with \citet{Howarthk13} ($59.20\pm0.05$ deg), but poorly with \citet{Barnesk13} ($23\pm4$ deg) and \citet{Hermank13} ($27.9\pm1.1$ deg\footnote{The published value is $-27.9\pm1.1$ deg. We assume that the disagreement arises due to a definition of $\lambda$ in the opposite direction to the other studies.}). A common feature of the latter two studies is that neither use priors from the Doppler tomography. In addition, they either fix the limb darkening coefficients or do not fit both components in the quadratic law. \citet{Masudak13} perform a detailed analysis using different combinations of priors and found that fitting both quadratic law limb darkening components could produce a result consistent with the tomography, and suggested using a Gaussian prior on $\lambda$ informed by the Doppler tomography.


Our goal in this section is to determine the best combination of uniform priors, Gaussian priors, and fixed parameters to allow accurate and informative measurements of certain parameters of interest. Of particular importance is the true spin-orbit angle $\Psi$, which as described in \cref{sec:intro} is difficult to measure using other techniques. This is given by the relation:
\begin{equation}\label{eq:spinorbit}
    \cos{\Psi}=\cos{i_\star}\cos{i_\mathrm{p}}+\sin{i_\star}\sin{i_\mathrm{p}}\cos{\lambda},
\end{equation}
where $i_\mathrm{p}$ is the orbital inclination. The asymmetric transit light curves of rapidly rotating stars contain information simultaneously about $i_\star$, $i_\mathrm{p}$, and $\lambda$. Thus, an accurate measurement of $\Psi$ will be possible in the event that these three parameters can be accurately determined.

\subsection{Choice of priors}

First of all, we considered what information was available for the MASCARA-1 system from which we could potentially construct relevant priors, whether these came from previous observations or simulations of this system and similar host stars. The stellar classification presented in \cref{sec:star} precisely estimates $R_\star$, $M_\star$ (and therefore $\rho_\star$) and $T_\mathrm{eff}$. We can infer the values of $q_1$ and $q_2$ from previous studies based on the stellar type and wavelength dependency \citep{Claret_Spitzer,Claret_CHEOPS}. Similarly we can infer the value of $\beta$ \citep{Claret_GD}. 

Finally, T17 have already observed the system during transit using Doppler tomography, which facilitated a precise measurement of the impact parameter $b$, sky-projected rotational velocity $v\sin{i_\star}$, and $\lambda$. As these parameters are derived effectively by measuring the geometric path of the planet across its host during the observations, we consider these measurements to be particularly robust against systematic biases. Further, the comparison of our estimations of $b$, $v\sin{i_\star}$, and $\lambda$ when they are left free to vary in the fit with those obtained from tomography is a useful indicator of accuracy. Placing a prior on $v\sin{i_\star}$ is relatively common practice when fitting gravity-darkened light curves \citep[e.g.][]{Masudak13,AhlersM4}. The same strategy has been adopted less commonly for $\lambda$ \citep[e.g.][]{Masudak13}, but there are no examples of this strategy being adopted for $b$ (or equivalently $i_\mathrm{p}$).


To investigate how the results extracted from fitting the asymmetric transit compared to those from the Doppler tomography, we performed a series of different fits to the \textit{CHEOPS} and \textit{Spitzer} transits separately and jointly. In our initial tests where we let all of the transit parameters vary uniformly, a particularly strong degeneracy existed between $i_\star$ and $\beta$. This is because as $i_\star\rightarrow90$ deg, a smaller proportion of the hotter, polar regions of the star that give rise to the most pronounced transit asymmetry are visible, which mandates a larger value of $\beta$ to fit any asymmetry, and vice versa. Even when trying to mitigate against this by placing a Gaussian prior on $\beta$, the fits regularly returned results with $\beta$ at an unphysically high value. For this reason, we fixed $\beta=0.199$ according to \citet{Claret_GD} for later runs. 

We also placed a Gaussian prior $R_\star$ according the IRFM determination described in \cref{sec:star}, as this value is poorly constrained by the model. In addition, we went one step further by fixing $T_\mathrm{pole}=7490$ K according to the spectral classification in \cref{sec:star}, as it is also poorly constrained by the model, and is important in determining the phase variations as described in \cref{sec:planetary_model}. We experimented with placing a prior on $\rho_\star$ according to our determinations of $R_\star$ and $M_\star$, but we let this vary uniformly as the light curves constrained it more strongly. Following the recommendation of \citet{Masudak13}, we let the limb darkening coefficients $q_1$ and $q_2$ vary uniformly. Finally, we also let $\delta_\mathrm{trans}$ and $i_\star$ vary uniformly within wide limits, and $P$ and $t_\mathrm{0}$ within tight limits set by an initial fit to all of the transits. This choice of priors and fixed parameters are consistent for all of the fits presented in this paper, apart from the search for orbital precession performed in \cref{sec:precession}.

We then tested the effect of fitting different combinations of the \textit{CHEOPS} and \textit{Spitzer} transits, both with and without Gaussian priors informed by the tomography placed on $b$, $v\sin{i_\star}$, and $\lambda$. For each instrument combination we performed two fits: one with Gaussian priors on these three parameters, and another where they can vary uniformly within the limits described in \cref{sec:stellar_model}. The results of fits for a selection of parameters are shown in \cref{tab:GDtransit}, and corner plots displaying the correlations for the fit to the \textit{CHEOPS}-only and \textit{CHEOPS}+\textit{Spitzer} light curves both with and without the Gaussian priors is shown in \cref{fig:corner_cheops_noprior,fig:corner_both_noprior,fig:corner_cheops_prior,fig:corner_both_prior}.

\begin{table}
\caption[]{Measurements of a selection of parameters for different combinations of transit light curves both with and without the placement of Gaussian priors on $b$, $v\sin{i_\star}$ and $\lambda$ informed by the Doppler tomography. The correlations between parameters for the \textit{CHEOPS}-only and  \textit{CHEOPS}+\textit{Spitzer} fits are shown in \cref{fig:corner_cheops_noprior,fig:corner_both_noprior,fig:corner_cheops_prior,fig:corner_both_prior}.
}
\label{tab:GDtransit}
\centering
\begin{tabular}{lccc}
\hline \hline
Parameter & Unit & no priors & Tomography priors \\ 
\hline 
\multicolumn{4}{l}{\textit{\textit{CHEOPS} only}} \\ 
$\delta_\mathrm{trans}$ & \% & $0.6396^{+0.0043}_{-0.0053}$ &  $0.6192^{+0.004}_{-0.0048}$\\ 
$\rho_\star$ & g cm$^{-3}$ & $0.3198^{+0.0025}_{-0.004}$ & $0.2943^{+0.0028}_{-0.0031}$ \\ 
$b$ &  & $0.028^{+0.05}_{-0.022}$ & $0.113\pm0.012$ \\ 
$q_1$ &  & $0.153^{+0.034}_{-0.032}$ & $0.227^{+0.026}_{-0.024}$ \\ 
$q_2$ &  & $0.126^{+0.088}_{-0.074}$ & $0.45^{+0.054}_{-0.052}$ \\ 
$i_\mathrm{\star}$ & deg & $29.2^{+12.0}_{-8.2}$ & $58.2\pm2.7$ \\ 
$\lambda$ & deg & $-54\pm26$ & $-72.0^{+3.1}_{-2.9}$ \\ 
$v\sin{i_\mathrm{\star}}$ & km s$^{-1}$ & $68^{+37}_{-16}$ & $105.7^{+3.6}_{-3.4}$ \\ 
\hline 
\multicolumn{4}{l}{\textit{\textit{Spitzer} only}} \\ 
$\delta_\mathrm{trans}$ & \% & $0.601^{+0.013}_{-0.021}$ & $0.5928^{+0.0058}_{-0.006}$ \\ 
$\rho_\star$ & g cm$^{-3}$ & $0.303^{+0.01}_{-0.018}$ & $0.2821^{+0.0033}_{-0.0034}$ \\ 
$b$ &  & $0.062^{+0.081}_{-0.046}$ & $0.1192^{+0.0084}_{-0.0095}$ \\ 
$q_1$ &  & $0.0208^{+0.0051}_{-0.005}$ & $0.0211^{+0.0048}_{-0.0046}$ \\ 
$q_2$ &  & $0.148^{+0.048}_{-0.049}$ & $0.161\pm0.05$ \\ 
$i_\mathrm{\star}$ & deg & $73^{+11}_{-20}$ & $73.6^{+9.0}_{-9.3}$ \\  
$\lambda$ & deg & $-36^{+52}_{-40}$ & $-70.2\pm3.0$ \\ 
$v\sin{i_\mathrm{\star}}$ & km s$^{-1}$ & $90^{+50}_{-61}$ & $108.7^{+3.8}_{-3.9}$ \\ 
\hline 
\multicolumn{4}{l}{\textit{\textit{CHEOPS}+\textit{Spitzer}}} \\ 
$\delta_\mathrm{trans}$ & \% & $0.605^{+0.011}_{-0.01}$ & $0.6149^{+0.0035}_{-0.0043}$ \\ 
$\rho_\star$ & g cm$^{-3}$ & $0.2874^{+0.0086}_{-0.0087}$ & $0.2889^{+0.0028}_{-0.0036}$ \\ 
$b$ &  & $0.017^{+0.03}_{-0.013}$ & $0.110\pm0.012$ \\ 
$q_{1,\mathrm{C}}$ &  & $0.257^{+0.028}_{-0.025}$ & $0.29^{+0.027}_{-0.026}$ \\ 
$q_{2,\mathrm{C}}$ &  & $0.514\pm0.086$ & $0.383^{+0.044}_{-0.04}$ \\ 
$q_{1,\mathrm{S}}$ &  & $0.0018^{+0.0034}_{-0.0013}$ & $0.0119\pm0.0053$ \\ 
$q_{2,\mathrm{S}}$ &  & $0.46^{+0.35}_{-0.31}$ & $0.148^{+0.05}_{-0.052}$ \\
$i_\mathrm{\star}$ & deg & $68.3^{+4.3}_{-6.1}$ & $59.8^{+2.6}_{-2.3}$ \\ 
$\lambda$ & deg & $-69^{+13}_{-12}$ & $-72.7^{+2.8}_{-3.0}$ \\ 
$v\sin{i_\mathrm{\star}}$ & km s$^{-1}$ & $125\pm13$ & $107.9^{+4.5}_{-3.3}$ \\ 
\hline 
\end{tabular}
\end{table} 


Some of the differences between the values for \textit{CHEOPS} transits and \textit{Spitzer} light curves with no priors from tomography are stark. $\delta_\mathrm{trans}$ for \textit{CHEOPS} is $>5\%$ larger than that of \textit{Spitzer}, which is an order of magnitude greater than is expected from wavelength-dependent atmospheric absorption for planets with large scale heights. This disagreement is likely driven by the more significant asymmetry in the \textit{CHEOPS} light curves, which in turn favours a lower $i_\star$ where more of the hotter polar regions are visible to the observer. For the \textit{CHEOPS}-only and \textit{Spitzer}-only fits, the measured values are also highly sensitive to whether or not the priors are placed.

In contrast, our tests show that there are two approaches that result in good agreement between the sampled parameters. The first is the placement of the priors from tomography, where the results for the \textit{CHEOPS}-only and \textit{CHEOPS}+\textit{Spitzer} fits are all within 2$\sigma$ of each other. A larger disagreement is observed with the \textit{Spitzer}-only fit, but this may be explained by the lower S/N for this transit. The second approach is to jointly fit the \textit{CHEOPS} and the \textit{Spitzer} transits together. In this case the transits are forced to share the same transit parameters, which in turn greatly diminishes the degeneracies existing between then. While less precise, they are also in good agreement with those from the \textit{CHEOPS}-only and \textit{CHEOPS}+\textit{Spitzer} fits where priors from tomography were used. \Cref{fig:corner_cheops_noprior,fig:corner_both_noprior,fig:corner_cheops_prior,fig:corner_both_prior} display corner plots from both of the \textit{CHEOPS}-only and \textit{CHEOPS}+\textit{Spitzer} fits, which demonstrates the effect of both of these approaches singularly and together. We discuss these two approaches in the following subsections.

\subsection{Constraints from Doppler tomography}\label{sec:tomography}

\begin{figure*}
    \centering
    \includegraphics{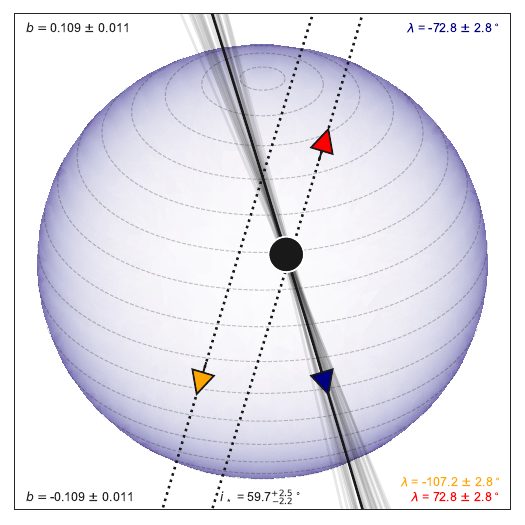}
    \includegraphics{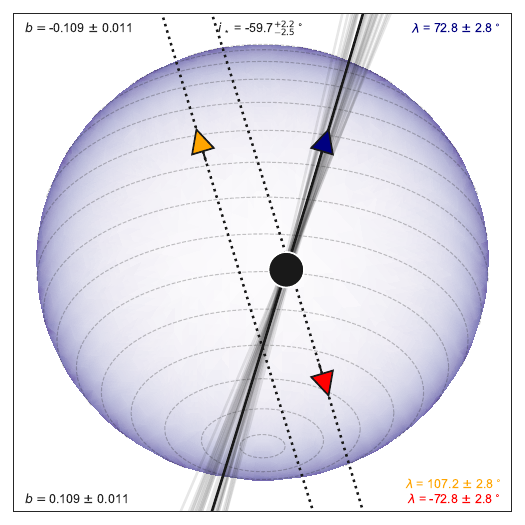}
    \caption{Visualisation of the path of the MASCARA-1\,b transit across the stellar disk using the \textit{CHEOPS}+\textit{Spitzer} fit with priors from Doppler tomography. The star rotates anti-clockwise when viewed from the top, and the majority of the hemisphere in view moves from left to right. Constant stellar latitudes are marked with curved dashed lines. The colour scale is proportional to the flux emitted by the star, with brighter areas marked with lighter colours. For a given $i_\star$, all possible transit paths of the planet across the star can be described using different combinations of $b$ and $\lambda$. The $\{b,\lambda\}$ pair associated with the median path transited by the planet is shown with a solid black line with a blue arrow indicating the direction, and with 32 samples drawn from the posterior shown in grey. The dashed lines with red and orange arrows indicate alternative $\{b,\lambda\}$ pairs that also describe the signals in the Doppler tomography and transit light curves, respectively. However, when considering the tomography and the transit light curve together, only the $\{b,\lambda\}$ pair described by the blue arrow can successfully describe them both. Assuming stellar symmetry about the equatorial plane, there are two indistinguishable solutions for each arrow in the range $-90<i_\star<90$ deg: one with $\{b,i_\star,\lambda\}$ and another with $\{-b,-i_\star,-\lambda\}$. The $\{b,\lambda\}$ pairs with positive $i_\star$ (north pole pointing towards the observer) are shown in the left panel, and the $\{b,\lambda\}$ pairs with negative $i_\star$ (north pole pointing away from the observer) are shown in the right panel. The blue arrow in the left panel is an example of a solution fitting the light curve with $b>0$, and $-90<\lambda<90$ deg, and corresponding solutions are found for all of the transit fits presented in this paper.}

    \label{fig:visualise}
\end{figure*}

The joint consideration of time-series photometry and Doppler tomography has two major advantages. Firstly, they each rule out different orbital configurations that could otherwise be possible. This is demonstrated in the visualisation in the left panel of \cref{fig:visualise}, which displays four different transit paths of MASCARA-1\,b across its gravity-darkened host for the \textit{CHEOPS}+\textit{Spitzer} fit with tomographic priors. The shape of the light curve shows that MASCARA-1\,b transits the hotter polar regions of the star in the first half of the transit (i.e. $-180<\lambda<0$ deg), which are described by the prograde solution marked in blue and the equivalent retrograde solution marked in yellow. The tomography confirms both that the orbit is prograde (i.e. $-90<\lambda<90$ deg), and that the majority of its path across the star is spent occulting red-shifted sections of the stellar disk (rotating away from the observer, the right half of the stellar disk), which are described by the solutions marked in blue and red. Only the blue solution collectively describes both the light curve and the tomography. For completeness, the right panel shows the equivalent transits for the solutions with $-i_\star$ due to the symmetry of the star about the equatorial plane.

Secondly, in most circumstances Doppler tomography facilitates accurate determination of $b$, $v\sin{i_\star}$, and $\lambda$, but not $\Psi$ due to the lack of information about $i_\star$. Transit light curves of fast-rotating stars encode information about all of these parameters, but their accurate determination is more challenging. By placing priors informed by the tomography on these parameters, this allows a robust measurement of $i_\star$, which enables an accurate derivation of $\Psi$. We adopted this strategy for our joint fit to the full set of light curves in \cref{sec:discussion}.

\subsection{Constraints from multi-colour photometry}

Our fits also suggest that in the absence of any Doppler tomography, accurate determination of transit parameters can also be achieved by the use of the \textit{CHEOPS} and \textit{Spitzer} data together. This allows us to report a measurement of $\Psi=70.8^{+10.8}_{-12.1}$ deg independently from the Doppler tomography. Although the precision is much worse than when priors from tomography are used, the result is in close agreement with the measurement using priors and confidently rules out orbits without significant misalignment. It is possible that the use of multi-colour photometry in general -- rather than the specific use of optical and mid-IR photometry -- would also facilitate a similarly accurate result, but further study should be undertaken to confirm this.

To the best knowledge of the authors, this is the only study to date to attempt to use an IR transit light curve to measure the spin-orbit angle of a system. Due to the lack of clear gravity darkening-induced asymmetry in the \textit{Spitzer} transit light curve, we tested the effect of decoupling the gravity darkening coefficient $\beta$ for each instrument and fixing it to 0 for \textit{Spitzer}. This enabled us to test the performance of the model from \citet{Barnes09} compared to a symmetric model. The Bayes factor $\mathcal{B}=\frac{\mathcal{Z}_\mathrm{Barnes}}{\mathcal{Z}_\mathrm{sym}}$ for the two fits is 1.09, indicating that both fits perform similarly well but that the fit with the Barnes \textit{Spitzer} transit is very marginally preferred. On this basis, we cannot claim to significantly detect gravity darkening-induced asymmetry in the \textit{Spitzer} transit. 

The biggest differences in the median models occur unsurprisingly during the transit, with a maximum difference of 250 ppm and an in-transit mean absolute deviation of 51 ppm: both well within the uncertainties of the \textit{Spitzer} photometry. Additionally, with the exception of the \textit{Spitzer} limb darkening coefficients, the fits returned values that were within 1$\sigma$ of each other. As we have no compelling physically or data-driven reasons to suggest that Barnes model does not hold at IR wavelengths, we proceeded using a single value of $\beta$ to describe both the \textit{CHEOPS} and \textit{Spitzer} transits in the joint model presented in \cref{sec:discussion}.

\section{Orbital precession}\label{sec:precession}

We tested the possibility of using the transits to search for evidence of orbital precession, which would allow us to measure the stellar quadrupole moment $J_\mathrm{2}$ of MASCARA-1 \citep{Szabo12}. Although the only previous measurements using photometry \citep{Szabo12,Masudak13} leverage the four years of precise data acquired by \textit{Kepler} \citep[coupled with \textit{TESS} data in the case of][]{Szabo20}, the intervening 16 months between the \textit{CHEOPS} and \textit{Spitzer} transit observations presented an opportunity to see if significant precession had occurred in that time. In addition, we used the NITES $r^\prime$-band transit from 2016 Sep 17 presented in T17\footnote{Although another NITES transit acquired on 2016 Jul 23 was also presented, we did not include this as the observation ended at the beginning of egress, making the detrending more susceptible to systematic bias.}, increasing the time spanned by the observations to 46 months. 

Precession manifests itself in transit light curves as a change in the impact parameter $b$ (alternatively $i_\mathrm{p}$ or transit duration $t_\mathrm{14}$) and $\lambda$. Previous measurements of $J_2$ for other hot Jupiters \citep[e.g.][]{Szabo12,Masudak13,Watanabew33,Borsaw33} are of the order $10^{-4}$. Adopting the same value for MASCARA-1, we would anticipate observing d$b$/d$t\sim$ 0.002 and d$\lambda$/d$t\sim$ 0.34 deg/yr. The precision of the data will prevent us to detect variations at this low level, but $J_2$ is a poorly understood quantity and we cannot rule out the possibility that it is much larger for MASCARA-1. 

In our first search, we fixed all transit parameters apart from $b$ and $\lambda$ to median values in the \textit{CHEOPS}+\textit{Spitzer} fit with prior informed by the Doppler tomography as described in the previous section, and let $b$, $\lambda$ and the baseline parameters be sampled uniformly. However, due to the lower significance of the asymmetry in the NITES and \textit{Spitzer} transits, it was not possible to meaningfully constrain the corresponding measurement of $\lambda$ from these light curves. As all previous reports of observed precession have been detected on the basis of changing $b$ alone, our next search adopted the same strategy using a two step process. Firstly, we fitted the transit light curves together. We fixed $b$, $v\sin{i_\star}$, $\lambda$ to the values measured by the Doppler tomography in T17, and let $P$, $t_\mathrm{0}$, $\delta_\mathrm{trans}$, and $\rho_\star$ and the limb darkening parameters per instrument ($q_1$ and $q_2$) vary uniformly. Secondly, we fitted each transit light curve individually, fixing the varying parameters to the median values in the previous step and allowing only $b$ and the baseline parameters to vary uniformly. The results are presented in \cref{tab:precession}.

\begin{table}
\centering
\caption[]{Measured impact parameter $b$ and transit duration $t_\mathrm{14}$ for the individual fits to four transit light curves, as described in \cref{sec:precession}.}
\label{tab:precession}
\begin{tabular}{lcc}
\hline \hline
Transit & $b$ & $t_\mathrm{14}$ [hr]  \\ 
\hline 
NITES & $0.22^{+0.12}_{-0.14}$ & $4.191^{+0.08}_{-0.13}$ \\ 
\textit{Spitzer} & $0.092^{+0.022}_{-0.028}$ & $4.2698^{+0.008}_{-0.0084}$ \\ 
\textit{CHEOPS} 1 & $0.1595^{+0.0087}_{-0.0089}$ & $4.2381^{+0.0052}_{-0.0053}$ \\ 
\textit{CHEOPS} 2 & $0.1036^{+0.0093}_{-0.011}$ & $4.2654^{+0.0039}_{-0.0037}$ \\ 
\hline 
\end{tabular} 
\end{table} 

The ground-based NITES transit favours a larger value of $b$ than the others, but was detected at relatively low significance and is within 1$\sigma$ of the measurement of $b=0.122\pm0.012$ \citep{Talens_corr} from the Doppler tomography acquired two weeks later. Although the measurements of $b$ from \textit{Spitzer} and the second \textit{CHEOPS} transit are in good agreement (and in turn in good agreement with the Doppler tomography), the first \textit{CHEOPS} transit is $>2\sigma$ higher. This measurement would also appear unreliable, as we would not expect significant precession to occur in the two weeks that separated the two \textit{CHEOPS} transit observations. As the ingress and egress were poorly sampled in the first \textit{CHEOPS} transit due to data gaps in the observation, this resulted in the fit favouring a shorter transit duration with an associated larger $b$.

With the measurements from the NITES and first \textit{CHEOPS} transits deemed unreliable, we found no evidence of the occurrence of significant orbital precession in the time between the $\textit{Spitzer}$ and second \textit{CHEOPS} transit. We discuss strategies for a possible future study to explore this in greater depth in \cref{sec:conclusion}.

\section{Results and discussion}
\label{sec:discussion}

The \textit{CHEOPS} transits, \textit{CHEOPS} occultations, \textit{Spitzer} phase curve, and detrended and phase-folded transits and occultations are displayed in \cref{fig:spitzer,fig:transit_DRP,fig:eclipse_DRP,fig:phasefold}, respectively, along with 32 models drawn from the posterior in each case. The results of the fit are displayed in \cref{tab:results,tab:noise}.

\begin{table*}[!htbp]
\caption[]{Priors and best-fitting values for the transit, occultation, and phase curve model parameters described in \cref{sec:analysis}, along with a selection of derived parameters. The baseline and noise parameters in the model are displayed in \cref{tab:noise}.}
\label{tab:results}
\begin{tabular}{lcccc}
\hline \hline
\multicolumn{2}{c}{Parameter} & Unit & Prior & Best fit \\ 
\hline 
\multicolumn{5}{l}{\textit{Transit, occultation and phase-curve model parameters}} \\ 
Period & $P$ & d & $\mathcal{U}$(2.1487, 2.1488) & $2.14877381^{+0.00000088}_{-0.00000087}$\\ 
Time of inferior conjunction & $t_\mathrm{0}$ & BJD$_\mathrm{TDB}-2458833$ & $\mathcal{U}$(0.487, 0.489) & $0.488151^{+0.000091}_{-0.000092}$\\ 
Transit depth & $\delta_\mathrm{trans}$ & \% & $\mathcal{U}$(0.5, 0.7) & $0.6216^{+0.0035}_{-0.0033}$\\ 
Stellar density & $\rho_\star$ & g cm$^{-3}$ & $\mathcal{U}$(0.2, 0.4) & $0.2966^{+0.0027}_{-0.0024}$\\ 
Impact parameter & $b$ &  & $\mathcal{U}$(0.0, 1.0) & $0.113\pm0.012$\\ 
Cosine eccentric component & $\sqrt{e}\cos{\omega_\mathrm{peri}}$ &  & $\mathcal{U}$(-0.15, 0.15) & $0.001\pm0.015$\\ 
Sine eccentric component & $\sqrt{e}\sin{\omega_\mathrm{peri}}$ &  & $\mathcal{U}$(-0.15, 0.15) & $0.000^{+0.016}_{-0.015}$\\ 
\textit{CHEOPS} Quadratic limb darkening & $q_\mathrm{1,C}$ &  & $\mathcal{U}$(0.0, 1.0) & $0.234^{+0.026}_{-0.022}$\\ 
\textit{CHEOPS} Quadratic limb darkening & $q_\mathrm{2,C}$ &  & $\mathcal{U}$(0.0, 1.0) & $0.405^{+0.053}_{-0.049}$\\ 
\textit{Spitzer} Quadratic limb darkening & $q_\mathrm{1,S}$ &  & $\mathcal{U}$(0.0, 1.0) & $0.00077^{+0.0018}_{-0.0006}$\\ 
\textit{Spitzer} Quadratic limb darkening & $q_\mathrm{2,S}$ &  & $\mathcal{U}$(0.0, 1.0) & $0.46^{+0.35}_{-0.32}$\\ 
Stellar Radius & $R_\star$ & $R_\odot$ & $\mathcal{N}$(2.072, 0.022) & $2.082^{+0.022}_{-0.024}$\\ 
Stellar Inclination & $i_\mathrm{\star}$ & deg & $\mathcal{U}$(0, 90) & $55.5^{+2.3}_{-2.9}$\\ 
Sky-projected spin orbit angle & $\lambda$ & deg & $\mathcal{N}$(69.5, 3.0) & $-69.2^{+3.1}_{-3.4}$\\ 
Projected stellar rotational velocity & $v\sin{i_\mathrm{\star}}$ & km s$^{-1}$ & $\mathcal{N}$(109.0, 4.0) & $101.7^{+3.5}_{-4.2}$\\ 
Gravity darkening coefficient & $\beta$ &  & fixed & 0.199\\ 
Stellar effective/polar temperature\tablefoottext{a} & $T_\mathrm{eff/pole}$ & K & fixed & 7490\\ 
Hotspot offset & $\Delta\phi$ & deg & $\mathcal{U}$(-180, 180) & $2.0^{+8.9}_{-9.4}$\\ 
$C_{m\ell}$ power coefficient & $\ln{C_\mathrm{1,1}}$ &  & $\mathcal{U}$(-2.75, 1.25) & $-1.25^{+0.35}_{-0.38}$\\ 
Mean background temperature & $\bar{T}$ & K & $\mathcal{U}$(1000, 4000) & $2350^{+130}_{-150}$\\ 
Fluid number & $\alpha$ &  & fixed & 0.6\\ 
Drag frequency & $\omega_\mathrm{drag}$ &  & fixed & 4.5\\ 
Single-scattering albedo & $\omega_\mathrm{scat}$ &  & $\mathcal{U}$(0.0, 1.0) & $0.71^{+0.11}_{-0.18}$\\ 
Scattering asymmetry factor & $g$ &  & $\mathcal{N}$(0.0, 0.07) & $-0.010^{+0.070}_{-0.061}$\\ 
\hline 
\multicolumn{5}{l}{\textit{Derived parameters}} \\ 
Stellar density & $\rho_\star$ & $\rho_\odot$ &  & $0.2104^{+0.0019}_{-0.0017}$\\
Stellar mass & $M_\star$ & $M_\odot$ &  & $1.900^{+0.063}_{-0.068}$\\ 
Transit duration & $t_\mathrm{14}$ & hr &  & $4.226^{+0.010}_{-0.011}$\\ 
Radius ratio & $R_\mathrm{p}/R_\star$ &  &  & $0.07884^{+0.00022}_{-0.00021}$\\ 
Semi major axis & $a$ & R$_\star$ &  & $4.1676^{+0.0047}_{-0.0051}$\\ 
Semi major axis & $a$ & au &  & $0.040352^{+0.000046}_{-0.000049}$\\ 
Orbital inclination & $i_\mathrm{p}$ & deg &  & $88.45\pm0.17$\\ 
Eccentricity & $e$ &  &  & $0.00034^{+0.00034}_{-0.00023}$\\ 
Argument of periastron & $\omega_\mathrm{peri}$ & deg &  & $-16^{+67}_{-52}$\\ 
Planetary radius & $R_\mathrm{p}$ & $R_\mathrm{J}$ &  & $1.597^{+0.018}_{-0.019}$\\ 
Planetary equilibrium temperature & $T_\mathrm{equil}$ & K &  & $2594.3^{+1.6}_{-1.5}$\\ 
Stellar rotational period & $P_\mathrm{rot}$ & d &  & $0.853^{+0.017}_{-0.016}$\\ 
Oblateness & $f_\mathrm{obl}$ &  &  & $0.0439\pm0.0018$\\ 
Stellar polar gravity & $g_\mathrm{pole}$ & m s$^{-2}$ &  & $125.6^{+1.7}_{-1.8}$\\ 
Spin-orbit angle & $\Psi$ & deg &  & $72.1^{+2.5}_{-2.4}$\\ 
\textit{CHEOPS} occultation depth & $\delta_\mathrm{occ,C}$ & ppm &  & $133^{+19}_{-18}$\\ 
\textit{Spitzer} occultation depth & $\delta_\mathrm{occ,S}$ & ppm &  & $2116^{+53}_{-52}$\\ 
Integrated dayside temperature & $T_\mathrm{d}$ & K &  & $3062^{+66}_{-68}$\\ 
\textit{CHEOPS} nightside flux & $F_\mathrm{night,C}$ & ppm &  & $0.96^{+2.60}_{-0.73}$\\ 
\textit{Spitzer} nightside flux & $F_\mathrm{night,S}$ & ppm &  & $670^{+330}_{-310}$\\ 
Integrated nightside temperature & $T_\mathrm{n}$ & K &  & $1720\pm330$\\ 
\textit{CHEOPS} phase curve amplitude & $F_\mathrm{amp,C}$ & ppm &  & $131\pm19$\\ 
\textit{Spitzer} phase curve amplitude & $F_\mathrm{amp,S}$ & ppm &  & $1470^{+330}_{-340}$\\ 
Geometric albedo & $A_\mathrm{g}$ &  &  & $\agkelp$ \\ 
Phase integral & $q$ &  &  & $1.559^{+0.095}_{-0.100}$\\ 
Spherical albedo & $A_\mathrm{s}$ &  &  & $\asphere$\\ 
\hline 
\end{tabular}
\tablefoottext{a}{One parameter is used for both the $T_\mathrm{eff}$ (used in the phase-curve model) and $T_\mathrm{pole}$ (used in the asymmetric transit model).}
\end{table*} 

\subsection{Spin-orbit alignment}\label{sec:spinorbit}

\begin{figure}
    \centering
    \includegraphics{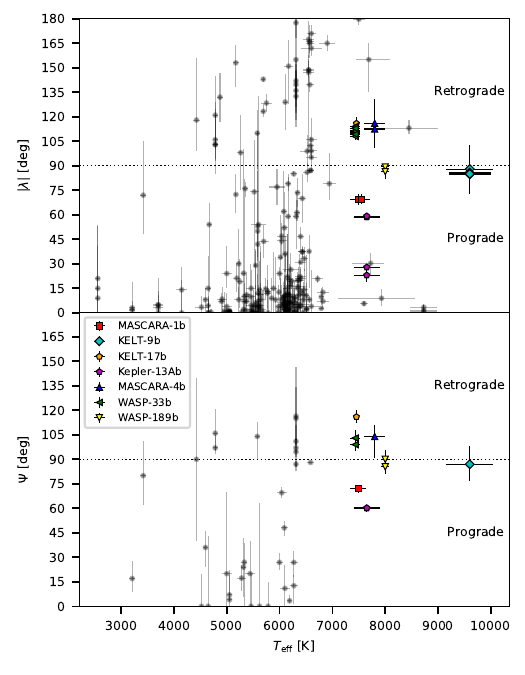}
    \caption{All published measurements of the absolute value of the sky projected spin-orbit angle $|\lambda|$ (top panel) and true spin-orbit angle $\Psi$ (bottom panel) as a function of $T_\mathrm{eff}$, according to TEPCat. All planets with measurements of $\Psi$ with $T_\mathrm{eff}>7000$ K are marked specifically. In cases where more than one measurement exists for a given planet, they are all shown. Some values of $\lambda$ were converted to fall in the range $-180<\lambda<180$ deg.}
    \label{fig:spinorbits}
\end{figure}

As detailed in \cref{sec:transit_fit}, in our transit only fits we obtained values of $v\sin{i_\star}$ and $\lambda$ consistent with Doppler tomography presented in T17 when they were allowed to vary freely. However, in the phase curve model we chose to place Gaussian priors on both parameters, due to the higher significance and the relative insensitivity to correlations with other parameters for the Doppler tomography measurement. Unsurprisingly, our measurements of  $v\sin{i_\star}=101.7^{+3.5}_{-4.2}$ km\,s$^{-1}$ and $\lambda=-69.2^{+3.1}_{-3.4}$ deg are both within 2$\sigma$ of the Doppler tomography\footnote{Although the result of $\lambda=69.5±3.0$ deg given by T17 is approximately the negative of our result, the difference arises due to their study using the $\beta$ definition of $\lambda$, while we adopted the opposite $\alpha$ definiton (see \cref{sec:stellar_model}).}. Unlike the Doppler tomography, the transit light curve encodes information about the stellar inclination $i_\star$, which we measured to be $55.5^{+2.3}_{-2.9}$ deg. As this breaks the degeneracy between $v$ and $i_\star$, we measured the stellar rotation period $P_\mathrm{rot}$ to be $0.853^{+0.017}_{-0.016}$ d. This rapid rotation causes the stellar equator to bulge to a radius $104.39\pm0.18 \%$ the size of the polar radius. By calculating the local surface gravity at both the equator and pole and utilising Von Zeipel's theorem \citep{vonZeipel}, we measured a stellar equatorial effective temperature of $7357.8^{+5.8}_{-4.9}$ K, over 100 K lower than the temperature at the poles.

Using \cref{eq:spinorbit} we derived a true spin-orbit angle of $72.1^{+2.5}_{-2.4}$ deg. Our measurements of $\lambda$ (top panel) and $\Psi$ (bottom panel) are shown in \cref{fig:spinorbits} in the context of similar measurements collected in the TEPCat orbital obliquity catalogue\footnote{\url{https://www.astro.keele.ac.uk/jkt/tepcat/obliquity.html}} \citep{TEPCat}. Spectroscopic methods have facilitated many more measurements of $\lambda$ than $\Psi$, with stars where $T_\mathrm{eff}<7000$\,K most likely to occupy aligned, prograde orbits. Stars with $T_\mathrm{eff}>7000$\,K are less amenable to RM effect measurement so the sample is much smaller, but following the trend proposed by \citet{WinnHJs}, the results appear less clustered and more evenly distributed across the full range of prograde and retrograde orbits. MASCARA-1\,b occupies a fairly sparsely populated section of the axes with a prograde near-polar orbit with misalignment and $T_\mathrm{eff}$ comparable to measurements for Kepler-13A\,b that agree with the Doppler tomography \citep{Johnsonk13,Masudak13,Howarthk13}. WASP-33\,b, for example, appears at the edge of a larger cluster of planets with near-polar orbits of similar magnitudes, albeit with retrograde orbits.

Recently, \citet{PerpendicularPlanets} note that the majority of planets with misaligned orbits occupy nearly polar orbits in the range $\Psi=80 - 120$ deg. Like Kepler-13A\,b, while relatively close to polar the $\Psi$ of MASCARA-1\,b falls outside this particularly clustered range. Although not as massive as Kepler-13A\,b, with $M_\mathrm{p}=3.7\pm0.9 \mathrm{M_J}$ measured by T17, MASCARA-1\,b has a relatively large mass compared to other hot Jupiters, which may support a tentative trend of more massive misaligned planets being less likely to occupy polar orbits. Measurements for a much larger sample of planets transiting stars with $T_\mathrm{eff}>7000$\,K will be necessary to reveal more detailed trends.

\subsection{Phase curve analysis}

\begin{figure}
    \centering
    \includegraphics{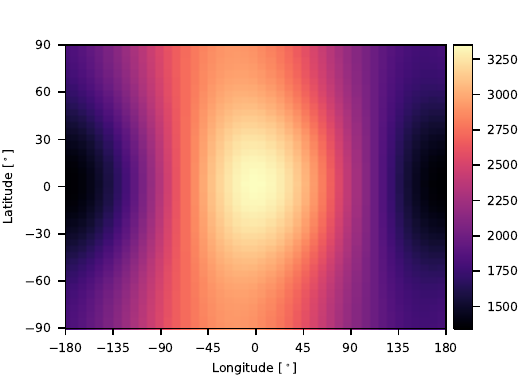}
    \caption{Temperature map as a function of planetary latitude and longitude derived from the best fitting phase curve parameters. The colour bar on the right shows the corresponding temperatures. The longitude zero-point is defined with respect to the sub-stellar point.}
    \label{fig:temperature_map}
\end{figure}

As described in \cref{sec:planetary_model}, the thermal component of the phase curve model is evaluated by generating a 2D temperature map, and integrating the blackbody flux between the respective filter response functions. The temperature map associated with our median result is displayed in \cref{fig:temperature_map}. The chevron-shaped patterns of common local temperatures are the trademark signature of the $h_{m,\ell}$ basis functions, which replicate the shape of the global temperature distributions produced by many GCMs \citep[e.g.][]{ShowmanGCM}. As $\omega_\mathrm{scat}$ is fixed to 4.5 in the fit, the hotspot offset $\Delta \phi$ is roughly equivalent to the offset between the phase of superior conjunction and maximum flux, which with $\Delta \phi=2.0^{+8.9}_{-9.4}$ deg is consistent with zero.

\begin{figure}
    \centering
    \includegraphics[width=\columnwidth]{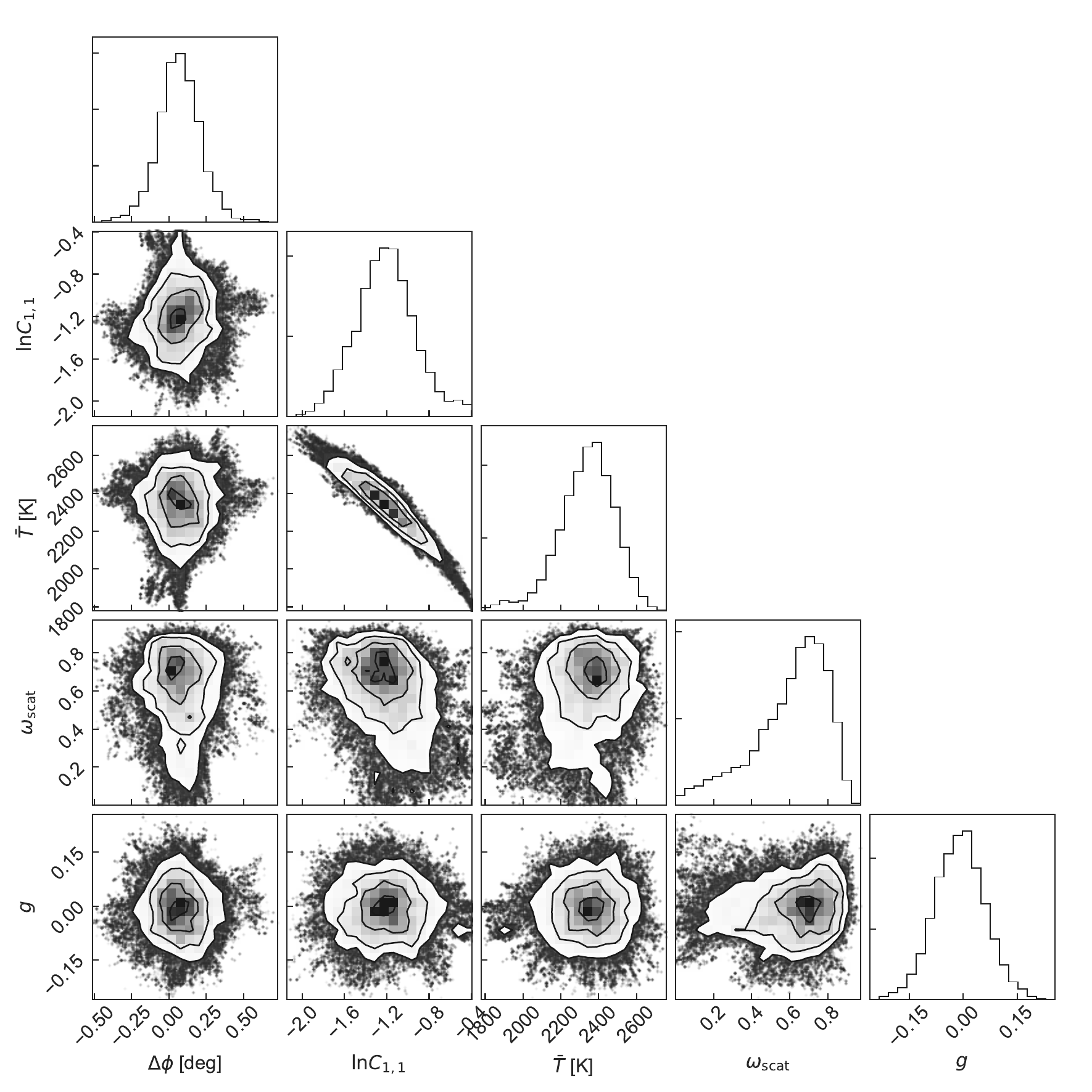}
    \caption{Corner plot demonstrating the correlations between the five free parameters describing the phase variations. Increasing densities of data points are marked with contour lines and darker background shades. The top row displays a histogram showing the posterior distribution for the parameter associated with that column. Generated using the \texttt{corner} package \citep{corner}.}
    \label{fig:corner_phase}
\end{figure}

A corner plot showing the correlations between the phase curve parameters is shown in \cref{fig:corner_phase}. The negative correlation between $\ln{c_\mathrm{1,1}}$ and $\bar{T}$ is analogous to the relationship between the phase curve amplitude and the minimum flux from sinusoidal phase curve parameterisations. While the thermal flux variations are wavelength-dependent, the reflected light component contributes the same phase-dependent flux in all bandpasses. The nonzero values favoured in the sampling of $\omega_\mathrm{scat}$, which scales the amplitude of the reflected flux, suggests that the \textit{CHEOPS} occultation is too deep with respect to the \textit{Spitzer} occultation to be described by blackbody flux alone.

\Cref{tab:results} also lists a range of parameters derived from the phase curves. Although we list the \textit{CHEOPS} phase curve parameters, we caution against interpreting any significance from the \textit{CHEOPS} nightside flux value of $0.96^{+2.60}_{-0.73}$ ppm. This is derived from the posterior distributions of phase curve parameters fitting both the \textit{Spitzer} and \textit{CHEOPS} light curves simultaneously, but is predominantly informed by the \textit{Spitzer} data.


\citet{Heng_reflected} describe how the geometric albedo $A_\mathrm{g}$, spherical albedo $A_\mathrm{s}$, and phase integral $q$ are derived from our parameterisation of the reflected light phase curve (see \cref{sec:planetary_model}). Our measurement of $A_g=0.171^{+0.066}_{-0.068}$ constitutes a $2.5\sigma$ detection of reflected light from the dayside (see \cref{sec:emission_spectrum} for a wider discussion). Our measurement of $q=1.559^{+0.095}_{-0.100}$ is consistent with isotropic scattering ($q=3/2$). Although the reflected light is only present in significant quantities in the \textit{CHEOPS} bandpass and full-phase coverage is preferable when measuring $q$, the 43.5\% full-phase coverage that the \textit{CHEOPS} observations achieve are sufficient to meaningfully constrain this value.

\begin{figure}
    \centering
    \includegraphics{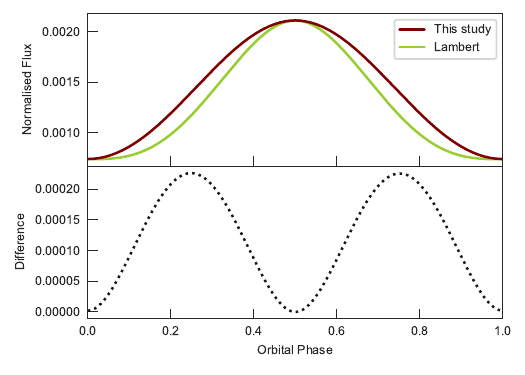}
    \caption{Top panel: phase curve model associated with the median result for the \textit{Spitzer} bandpass (maroon), which is the sum of the thermal and reflected light components described in \cref{sec:planetary_model}. A Lambertian phase curve model with the same amplitude, minimum flux, and hotspot offset (green) is also shown. Bottom panel: flux difference between the two models.}
    \label{fig:kelp_vs_lambert}
\end{figure}

These derived parameters facilitate a more direct comparison with the fit to the \textit{Spitzer} phase curve previously presented in \citet{BellM1}. Our measurement of occultation depth and nightside flux -- with the corresponding dayside and nightside integrated temperatures -- are each 1-2$\sigma$ higher than the \citet{BellM1} equivalents. Although different approaches were employed between the two studies for light curve detrending, another likely provenance is the different phase curve parameterisations used, which is demonstrated in \cref{fig:kelp_vs_lambert}.

\subsection{Energy budget: Bond albedo and recirculation efficiency}\label{sec:budget}

Given the hemispherically averaged dayside and nightside temperature, one may derive the Bond albedo $A_{\rm B}$ and redistribution factor $f$ using zero-dimensional (0D) `box models' as stated in equations (4) and (5) of \citet{Cowanphasecurve}.  However, this depends on the choice of boundary condition associated with the nightside temperature, for example demanding that $f=2/3$ in the limit of no heat redistribution.  
This issue may be avoided by deriving the Bond albedo directly from the 2D temperature map, as demonstrated in  \citet{Keating2D} and M21.
\begin{equation}
    A_{\rm B} = 1 - \left( \frac{a}{R_\star} \right)^2 \frac{\int^{\pi}_{-\pi} \int^{\pi}_0 F_\mathrm{p} \sin\theta ~d\theta ~d\phi}{\pi \sigma_\mathrm{SB} T_\mathrm{eff}^4},
\end{equation}
where $\sigma_\mathrm{SB}$ is the Stefan-Boltzmann constant. The redistribution factor $\varepsilon$ is simply the ratio of the flux from the integrated nightside and dayside \citep{GCM}, 
\begin{equation}
\varepsilon = \frac{\int^{\pi}_{\pi/2} \int^{\pi}_{0} F_\mathrm{p} \sin\theta ~d\theta ~d\phi + \int^{-\pi/2}_{-\pi} \int^{\pi}_{0} F_\mathrm{p} \sin\theta ~d\theta ~d\phi}{\int^{\pi/2}_{-\pi/2} \int^{\pi}_0 F_\mathrm{p} \sin\theta ~d\theta ~d\phi}.
\end{equation}
In the 2D approach, the redistribution factor $f$ is undefined.

The derived values of $\varepsilon$ and $A_\mathrm{B}$ using both the 0D approach from \citet{Cowanphasecurve} and 2D approach from M21 are displayed in \cref{tab:fAB}, and in both cases they are within 1$\sigma$ of each other. Both methods yield a large proportion of samples with negative values of $A_\mathrm{B}$, which suggests that it could be underestimated due to spectral features in the emission spectrum of MASCARA-1\,b within the \textit{Spitzer}/IRAC \SI{4.5}{\micro\meter} bandpass causing a significant deviation from the flux predicted by a Planck function. 


Formally, the Bond albedo is the spherical albedo weighted by the stellar spectral energy distribution $F_\star$, integrated over all wavelengths \citep{MarleyAlbedo},
\begin{equation}\label{eq:ABAs}
    A_{\rm B} = \frac{\int^{\infty}_{0} A_\mathrm{s}\left (\lambda \right ) ~F_\star ~d\lambda}{\int^{\infty}_{0} F_\star ~d\lambda}.
\end{equation}
In this study, each quantity is derived separately: $A_\mathrm{B}$ as described in the previous paragraphs using the thermal component of the light curve model and $A_\mathrm{s}$ as the product of $A_\mathrm{g}$ and the phase integral $q$ from the reflected component of the light curve model. Formally, the spherical albedo is defined at a single wavelength, as is the case for the geometric albedo; both quantities are intrinsic properties of the atmosphere that do not formally involve the star. In practice, spherical and geometric albedos are commonly reported as bandpass-integrated quantities, even for the planets of the Solar System \citep[e.g. Table 7 of][]{PearlAlbedos}; an exception is for Cassini data of Jupiter (e.g. Figure 3 of \citealt{LiCassini}).  Our CHEOPS bandpass-integrated\footnote{Although practically speaking we have actually measured $A_\mathrm{s}$ collectively across the \textit{CHEOPS} and \textit{Spitzer}/IRAC \SI{4.5}{\micro\meter} bandpasses, in reality the fact that the stellar flux in the \SI{4.5}{\micro\meter} bandpass constitutes $\sim0.3\%$ of the combined total means that it effectively approximates the $A_\mathrm{s}$ for the \textit{CHEOPS} integrated bandpass only.} measurement of $A_\mathrm{s}=\asphere$ is larger than our measurement of $A_\mathrm{B}$ from the 2D temperature map. 

\begin{figure}
    \centering
    \includegraphics{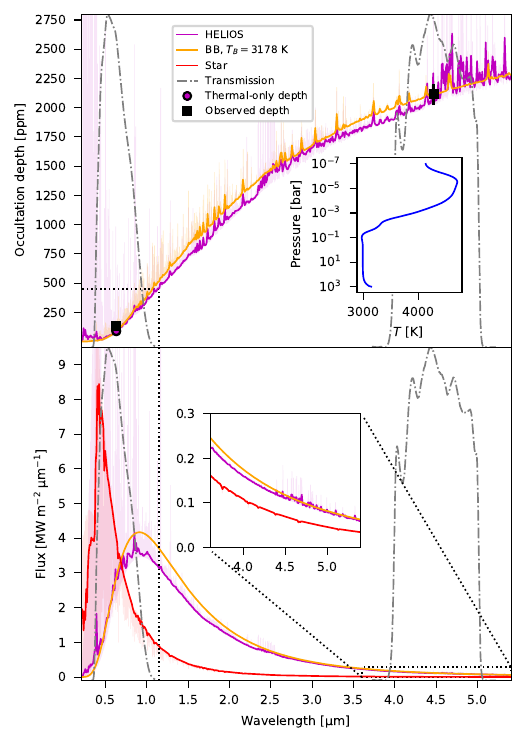}
    \caption{Top: emission spectra generated to fit the \textit{Spitzer} occultation depth, as described in \cref{sec:emission_spectrum}. The spectrum from the \texttt{HELIOS} grid is shown in magenta and the best fitting blackbody is shown in orange. The unbinned spectra are shown with thin, lighter lines. Binned spectra are also shown in darker, thicker lines for display purposes. The measured \textit{CHEOPS} and \textit{Spitzer} occultation depths are shown with black squares with error bars, and the implied depths according to the spectra are shown with coloured circles (obscured by the measured depth in the case of \textit{Spitzer}). The \textit{CHEOPS} and \textit{Spitzer} response functions are shown in grey. Inset top: the temperature-pressure profile of the best fitting \texttt{HELIOS} spectrum. Bottom: the same spectra, but shown in terms of flux instead of occultation depth. The PHOENIX spectrum used for the star is shown in red, shrunk by a factor of 50 for display purposes. Right inset: a closer look at the flux at wavelengths covered by the \textit{Spitzer}/IRAC \SI{4.5}{\micro\meter} bandpass. Other dotted lines in both panels are regions that are displayed in \cref{fig:spectra_cheops}.}
    \label{fig:spectra_full}
\end{figure}

\begin{figure}
    \centering
    \includegraphics{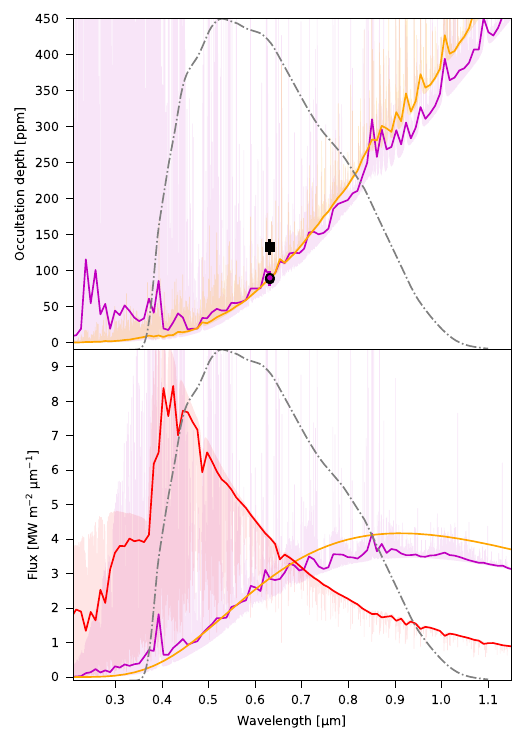}
    \caption{Closer look at the emission spectra at wavelengths covered by the \textit{CHEOPS} bandpass, enclosed by dotted lines in \cref{fig:spectra_full}.}
    \label{fig:spectra_cheops}
\end{figure}

A possible cause of the relatively poor agreement between these two values could arise due to the \textit{CHEOPS} and \textit{Spitzer}/IRAC \SI{4.5}{\micro\meter} bandpasses only covering broad optical and part of the mid-infrared, respectively. Consulting the PHOENIX spectrum that modelled the stellar flux in the light curve fit (shown in red in the bottom panels of \cref{fig:spectra_full,fig:spectra_cheops}), these bandpasses only capture $\sim 69\%$ of this flux. In the event that MASCARA-1\,b reflected none of the light incident upon it at wavelengths outside of the \textit{CHEOPS} and \SI{4.5}{\micro\meter} bandpasses (i.e. $A_\mathrm{s,out}=0$), and assuming that optical wavelengths and longer are well-described by the $A_\mathrm{s}$ derived in our light curve model (i.e. $A_\mathrm{s,in}=\asphere$), using \cref{eq:ABAs} we computed $A_\mathrm{B}=0.184^{+0.067}_{-0.069}$. Although still over double the $A_\mathrm{B}$ that we derived for the temperature maps, they are only discrepant at $1.2\sigma$. The stellar flux that falls outside of these bandpasses is split relatively evenly between the ultraviolet and near-infrared. If MASCARA-1\,b reflects a much lower proportion of incident light from its host in the ultraviolet than in the optical, this would be a reversal of reflected properties measured on the basis of a \textit{Hubble}/STIS secondary eclipse observation of HD 189733\,b \citep{evanshd189}. Reflection from hot Jupiters in the near-infrared has not been well-studied to date, as the observed flux in this range is generally dominated by thermal emission.



\begin{table}
\centering
\caption[]{Heat redistribution efficiency $\varepsilon$ and Bond albedo $A_\mathrm{B}$ derived using both the 0D method from \citet{Cowanphasecurve} and the 2D method using temperature maps from M21.}
\label{tab:fAB}
\begin{tabular}{ccc}
\hline \hline
Parameter & 0D & 2D  \\
\hline 
$\varepsilon$ & $0.23^{+0.13}_{-0.10}$ & $0.204^{+0.089}_{-0.068}$\\
$A_\mathrm{B}$ & $\cowanAB$ & $\morrisAB$\\
\hline 
\end{tabular} 
\end{table} 

When assuming that the thermal emission from the dayside follows a blackbody distribution (as our light curve model does), the dayside integrated temperature of $T_\mathrm{d}=3062^{+66}_{-68}$ K confirms the position of the MASCARA-1 dayside amongst the hottest of any known exoplanet. We report a 3$\sigma$ detection of flux from the nightside and a corresponding nightside integrated temperature of $T_\mathrm{n}=1720\pm330$ K. When taken together with the hotspot offset $\Delta\phi$ consistent with zero and the derived $\varepsilon$, this indicates the relatively poor redistribution of heat from dayside to nightside that is typical for the majority of hot Jupiters. However, a comparison with the hot Jupiters analysed in \citet{WongNorth} places the redistribution efficiency of MASCARA-1\,b for both the 0D and 2D methods above that of the majority of the sample. Correspondingly, the extreme levels of irradiation MASCARA-1\,b receives from its host place the derived $T_\mathrm{n}$ as the second hottest in the sample, with only HAT-P-7\,b hotter, although likely also cooler than KELT-9b \citep{MansfieldK9,BellM1} which did not form part of the sample. We add the caveat that different analyses of \textit{Spitzer}/IRAC phase curves have yielded dramatically different conclusions about the energy budgets of the corresponding exoplanets \citep[see e.g.][and references therein]{Mayw43}. 


\subsection{Emission spectrum retrieval}
\label{sec:emission_spectrum}

Although the previous section presented a self-consistent analysis of the thermal and reflective properties, for the sake of computational efficiency it necessarily included some major simplifications. In particular, we assumed that the thermal emission from each point in the 2D temperature map could be described by a Planck function, and reduced the spectra for planet, star, and filter transmission functions down to only 15 bins per bandpass. To perform a more detailed spectral analysis of the dayside atmosphere, we generated a grid of emission spectra using \texttt{HELIOS}\footnote{\url{https://github.com/exoclime/HELIOS}} \citep{HELIOS1,HELIOS2} to jointly interpret the \textit{CHEOPS} and \textit{Spitzer} occultation depths. 

We used a planetary radius of 1.597 $R_\mathrm{J}$ and an orbital distance of 0.040352 au derived from our light curve fitting, and a surface gravity of $\log g = 3.55$ in cgs units informed by the planetary mass reported in T17. Since the metallicity of the host star is compatible with solar element abundances, for simplicity we correspondingly adopted these to describe the chemical composition of the planet. The considered opacity sources and corresponding references can be found in \citet{Wongk9}.

We generated a grid of self-consistent \texttt{HELIOS} atmosphere models for MASCARA-1\,b as a function of the heat redistribution efficiency $\varepsilon$ (see \citealt{Lendlw189} for an example of such a grid constructed for WASP-189b). In a post-processing step, we then generated a high-resolution planetary emission spectrum without contributions by scattering for each point in the grid. These spectra, thus, contain only the thermal emission part of the planet's spectrum. To calculate the theoretical occultation depths implied by these emission spectra, we used the same PHOENIX spectrum as was used for the computation of the thermal flux modulation (see \cref{sec:planetary_model}).

The high-resolution spectra are averaged over the \textit{Spitzer} and \textit{CHEOPS} bandpasses to obtain the corresponding, theoretical occultation depths as seen by these instruments. Since the bandpass-averaged occultation depths vary smoothly with the heat redistribution factor $\varepsilon$, we parameterised them with a spline function for each filter.

We then used an MC algorithm to find the distribution of $\varepsilon$ that best fit the \textit{Spitzer} occultation depth within its error bars, and also taking into account the uncertainty on the transit depth and the $R_p/R_\star$ value reported in \cref{tab:results}. Although flux originating from both thermal emission and reflected light comprised the model fitting the \textit{Spitzer} occultation, this step implicitly assumes that all of the flux from the dayside in \textit{Spitzer} bandpass arises due to thermal emission. As thermal emission is expected to dominate reflected light in the mid-infrared for ultra-hot Jupiters such as MASCARA-1\,b (which is indeed true in our light curve model), we deem this to be an acceptable assumption.

This analysis yields a result of $\varepsilon=0.307\pm0.088$ ($f=0.538\pm0.038$). The spectra for the median value of $\varepsilon$ from this process are displayed in magenta in \cref{fig:spectra_full}. As can be seen in the zoom of the spectra for the \textit{CHEOPS} bandpass shown in \cref{fig:spectra_cheops}, the \textit{CHEOPS} occultation depth predicted using this method is $88\pm10$ ppm---significantly less than the observed value of $133^{+19}_{-18}$ ppm from our light curve model. Since we neglected scattering in the post-process emission spectra, the difference between the predicted occultation depth and the measured one can be translated into the geometric albedo $A_\mathrm{g}$ in the \textit{CHEOPS} bandpass, for which we obtained a value of $A_\mathrm{g}=\aghelios$. 

We also tested the effect of replacing the emission spectrum described above with a blackbody (shown in orange in \cref{fig:spectra_full,fig:spectra_cheops}), and fitting to the \textit{Spitzer} occultation depth as a function of the brightness temperature $T_\mathrm{B}$. This was optimised with $T_B=3178\pm53$ K, which resulted in a predicted \textit{CHEOPS} occultation depth of 90 ppm and an $A_\mathrm{g}$ of $\agbb$. The blackbody spectrum is broadly very similar to the \texttt{HELIOS} spectrum, with the main differences arising due to metal emission lines close to the \textit{Spitzer} bandpass.

Given that the dayside emission spectrum produced by \texttt{HELIOS} is so closely approximated by a blackbody, the retrieval should return results in good agreement with those resulting from the light curve model, as is reported in the previous section. From this, we indeed see that the measurement of $A_g=\agkelp$ is in good agreement with both the retrieved \texttt{HELIOS} and blackbody values, and $T_\mathrm{d}=3062^{+66}_{-68}$ K $\sim 1.5\sigma$ away from the retrieved dayside $T_\mathrm{B}$.

Although these values of $A_\mathrm{g}$ translate into the dayside MASCARA-1 reflecting a small fraction of the radiation incident upon it, this is a trend seen across the vast majority of hot Jupiters for which equivalent observations exist. Depending on which derivation of $A_\mathrm{g}$ that we adopt (light curve fit, \texttt{HELIOS} retrieval or blackbody retrieval), the significance of the detection of reflected light ranges from 2-2.5$\sigma$. We select the $A_\mathrm{g}$ of $\agkelp$ derived from the light curve fit as our headline figure for this study, as the thermal and reflected components of flux are derived self-consistently using information from the full light curves, rather than the occultation depths alone. \citet{WongSouth} and \citet{WongNorth} report that an observed positive correlation between $A_\mathrm{g}$ and $T_d$ breaks down at $T_d\sim 3000$ K. Although the detection of reflected light in this study is relatively marginal, it is significantly higher than the estimated $A_\mathrm{g}$ for other hot Jupiters with $T_d>3000$ K such as WASP-18\,b \citep[$A_g<0.03$]{Maxtedw18} and WASP-33\,b \citep[$A_g<0.08$]{Zhangw33,vonessenw33}\footnote{The upper limits are reported at 2$\sigma$ by \citet{WongNorth}.} This may indicate that the MASCARA-1\,b dayside is not completely dominated by \ce{H-} absorption \citep{Arcangeli_H-,Lothringer_H-}, and that some highly reflecting species may not have dissociated close to the terminator.


\section{Conclusions}
\label{sec:conclusion}

We have presented a joint analysis of \textit{Spitzer} and \textit{CHEOPS} light curves of MASCARA-1\,b, which has yielded the most precise constraints to date on a range of parameters including the period, epoch, eccentricity, stellar density, planetary radius, and orbital separation. The light curves of planets transiting fast rotating stars simultaneously encode information about the stellar inclination $i_\star$, the planetary inclination $i_p$, and the sky-projected spin-orbit angle $\lambda$, which in turn can be used to derive the true spin-orbit angle $\Psi$. However, the large degeneracies that exist between many of the transit parameters makes their accurate determination challenging. Doppler tomography observations do not resolve $\Psi$, but do generally allow $v\sin{i_\star}$ and $\lambda$ to be robustly measured. The use of priors informed by the tomography when fitting the transit light curves is a powerful tool to facilitate an accurate measurement of $\Psi$. \citet{Masudak13} recommend the placement of a Gaussian prior on both $v\sin{i_\star}$ and $\lambda$ from tomography when fitting such transits. As tomography also enables accurate measurements of $b$ that are less subject to degeneracies than the transits, we go one step further and suggest that Gaussian priors should be placed on $b$ in a similar way.

Our strategy of fitting both the \textit{CHEOPS} and \textit{Spitzer} transits with a common set of transit parameters also greatly reduced the effect of these degeneracies, measuring values of $v\sin{i_\star}$ and $\lambda$ consistent with the Doppler tomography even without the use of Gaussian priors, and allowed an independent measurement of $\Psi=70.8^{+10.8}_{-12.1}$ deg. We propose that the acquisition of high precision transit light curves of fast rotators at multiple wavelengths may routinely provide a more robust determination of the transit parameters where Doppler tomography is not available, though this should be tested with upcoming observations. The use of \textit{CHEOPS} to target hot Jupiters transiting fast rotating stars that have already been observed by \textit{TESS} will provide a way to test whether observation in two overlapping optical bandpasses produces the same effect. As previous studies do not search for asymmetry in the \textit{Spitzer} transits of fast rotators such as KELT-9 \citep{MansfieldK9}, Kepler-13A\,b \citep{Shporerk13}, and WASP-33 \citep{Zhangw33}, we suggest that the combined reanalysis of these light curves in conjunction with their corresponding \textit{TESS} light curves may facilitate similar studies to the one we have performed for MASCARA-1.

When applying Gaussian priors on $b$, $v\sin{i_\star}$, and $\lambda$ informed by the Doppler tomography, we measured a more precise value of $\Psi=72.1^{+2.5}_{-2.4}$ deg, which is the value we report. This result follows the trend of observed polar or near-polar orbits planets orbiting stars with $T_\mathrm{eff}>7000$ K, though measurements for a larger number of systems are needed to properly test theories describing the underlying physics.

In \cref{sec:precession}, we reported that the transit light curves show no evidence for significant orbital precession. As a result, we were not able to constrain the stellar gravitational quadrupole moment $J_\mathrm{2}$: the term primarily dictating the rate of precession. The combination of the observational strategy and photometric precision afforded by the \textit{Kepler} mission made it uniquely well-suited to search for orbital precession in hot Jupiters orbiting oblate stars---unfortunately the only target in the \textit{Kepler} field for which this was possible was Kepler-13A\,b. The ability of \textit{CHEOPS} to replicate such a study is also hampered by the gaps in the light curve, often limiting the ability of the impact parameter $b$ to be accurately determined from a single transit. 

A strategy that future \textit{CHEOPS} observations could adopt to mitigate against these issues could be to observe multiple transits of a suitable target across numerous years, ensuring that observations in a given observing season collectively sample all phases of the transit. When phase-folding all of the transits for each observing season, this strategy would facilitate search for orbital precession from year to year. Adopting this strategy for a sector of \textit{TESS} observations allows \citet{Szabo20} to confirm the continuing decrease in $b$ well after the \textit{Kepler} mission had concluded. Although it would not come close to matching the time resolution achieved by \textit{Kepler}, \textit{CHEOPS} could observe optimal targets across a far wider area of the sky. Alongside MASCARA-1\,b, this would include optimal targets such as KELT-9\,b and WASP-189\,b where the signatures of precession are expected to be larger and more easily detectable than for Kepler-13A\,b. \textit{TESS} observations will supplement these searches, and for MASCARA-1 they are provisionally scheduled to occur in sector 52 (Aug 2022).

We also studied the atmosphere of MASCARA-1\,b by measuring the occultation depths and phase-dependent flux modulation for both \textit{CHEOPS} and \textit{Spitzer}. Our light curve fitting suggests that MASCARA-1\,b is within a small sample of hot Jupiters with an integrated dayside temperature $T_\mathrm{d}>3000$ K. The corresponding nightside temperature of $1720\pm330$ K and the derived heat redistribution factor of $0.571^{+0.054}_{-0.078}$ suggests inefficient but existent recirculation of heat from the dayside. Our retrieval confirmed  the assumption of our light curve model that the dayside emission spectrum is well described by a blackbody. While low, the measurement of $A_\mathrm{g}=\agkelp$ from our light curve model stands out compared to other ultra-hot Jupiters where \ce{H-} absorption is expected to dominate in the dayside. The fact that our derived value of $A_\mathrm{s}$ was somewhat larger than that of $A_\mathrm{B}$ may suggest that MASCARA-1\,b is less reflective at ultraviolet and near-infrared wavelengths than in the optical. The precision achieved for the \textit{CHEOPS} occultations suggest that future \textit{CHEOPS} observations of the MASCARA-1\,b full-phase curve should significantly detect the planetary signal, which would allow the reflective properties at both optical and IR wavelengths to be separately determined.

To model the phase-dependent flux variations in the light curve, we used a novel, physically motivated phase curve parameterisation presented in M21 that generates a 2D temperature map to model the thermal emission and can describe any reflection law \citep{Heng_reflected}. This novel approach enabled us to separately derive the $A_\mathrm{g}$, spherical albedo $A_\mathrm{s}$, Bond albedo $A_\mathrm{B}$, and phase integral $q$ of MASCARA-1. This approach to fitting multi-wavelength occultation and phase curve data in the optical and IR also provides a template for the analysis of spectrally resolved \textit{James Webb Space Telescope} (\textit{JWST}) data. Due to the limited phase coverage of the \textit{CHEOPS} data, we fitted the data from both bandpasses with a single set of phase curve parameters, which implicitly assumed that they share a common albedo, and that the spectral energy distribution of the planet is well represented by a blackbody. Although a similar strategy may still be necessary when analysing occultation data, the use of a unique set of parameters per wavelength bin in the exquisite-precision spectrally resolved phase curves that \textit{JWST} will acquire will allow for variations in thermal emission and reflection to be taken into account in the light curve fitting.

\begin{acknowledgements}\\



CHEOPS is an ESA mission in partnership with Switzerland with important contributions to the payload and the ground segment from Austria, Belgium, France, Germany, Hungary, Italy, Portugal, Spain, Sweden, and the United Kingdom. The CHEOPS Consortium would like to gratefully acknowledge the support received by all the agencies, offices, universities, and industries involved. Their flexibility and willingness to explore new approaches were essential to the success of this mission. 
MJH and YA acknowledge the support of the Swiss National Fund under grant 200020\_172746. 
SH gratefully acknowledges CNES funding through the grant 837319. 
D.K. acknowledges partial financial support from the Center for Space and Habitability (CSH), the PlanetS National Center of Competence in Research (NCCR), and the Swiss National Science Foundation and the Swiss-based MERAC Foundation. 
ACC and TGW acknowledge support from STFC consolidated grant number ST/M001296/1. 
PM acknowledges support from STFC research grant number ST/M001040/1. 
This work was also partially supported by a grant from the Simons Foundation (PI Queloz, grant number 327127). 
B.-O.D. acknowledges support from the Swiss National Science Foundation (PP00P2-190080).
S.S. has received funding from the European Research Council (ERC) under the European Union’s Horizon 2020 research and innovation programme (grant agreement No 833925, project STAREX).
ABr was supported by the SNSA. 
This work was supported by FCT - Fundação para a Ciência e a Tecnologia through national funds and by FEDER through COMPETE2020 - Programa Operacional Competitividade e Internacionalizacão by these grants: UID/FIS/04434/2019, UIDB/04434/2020, UIDP/04434/2020, PTDC/FIS-AST/32113/2017 \& POCI-01-0145-FEDER- 032113, PTDC/FIS-AST/28953/2017 \& POCI-01-0145-FEDER-028953, PTDC/FIS-AST/28987/2017 \& POCI-01-0145-FEDER-028987, O.D.S.D. is supported in the form of work contract (DL 57/2016/CP1364/CT0004) funded by national funds through FCT. 
We acknowledge support from the Spanish Ministry of Science and Innovation and the European Regional Development Fund through grants ESP2016-80435-C2-1-R, ESP2016-80435-C2-2-R, PGC2018-098153-B-C33, PGC2018-098153-B-C31, ESP2017-87676-C5-1-R, MDM-2017-0737 Unidad de Excelencia Maria de Maeztu-Centro de Astrobiologí­a (INTA-CSIC), as well as the support of the Generalitat de Catalunya/CERCA programme. The MOC activities have been supported by the ESA contract No. 4000124370. 
S.C.C.B. acknowledges support from FCT through FCT contracts nr. IF/01312/2014/CP1215/CT0004. 
XB, SC, DG, MF, and JL acknowledge their roles as ESA-appointed CHEOPS science team members. 
This project was supported by the CNES. 
The Belgian participation to CHEOPS has been supported by the Belgian Federal Science Policy Office (BELSPO) in the framework of the PRODEX Program, and by the University of Liège through an ARC grant for Concerted Research Actions financed by the Wallonia-Brussels Federation. 
This project has received funding from the European Research Council (ERC) under the European Union’s Horizon 2020 research and innovation programme (project {\sc Four Aces}. 
grant agreement No 724427). 
DE acknowledges financial support from the Swiss National Science Foundation for project 200021\_200726.
CMP and MF gratefully acknowledge the support of the Swedish National Space Agency (DNR 65/19, 174/18). 
DG gratefully acknowledges financial support from the CRT foundation under Grant No. 2018.2323 `Gaseousor rocky? Unveiling the nature of small worlds'. 
M.G. is an F.R.S.-FNRS Senior Research Associate. 
KGI is the ESA CHEOPS Project Scientist and is responsible for the ESA CHEOPS Guest Observers Programme. She does not participate in, or contribute to, the definition of the Guaranteed Time Programme of the CHEOPS mission through which observations described in this paper have been taken, nor to any aspect of target selection for the programme. 
This work was granted access to the HPC resources of MesoPSL financed by the Region Ile de France and the project Equip@Meso (reference ANR-10-EQPX-29-01) of the programme Investissements d'Avenir supervised by the Agence Nationale pour la Recherche. 
Acknowledges support from the Spanish Ministry of Science and Innovation and the European Regional Development Fund through grant PGC2018-098153-B- C33, as well as the support of the Generalitat de Catalunya/CERCA programme. 
S.G.S. acknowledge support from FCT through FCT contract nr. CEECIND/00826/2018 and POPH/FSE (EC). 
This project has been supported by the Hungarian National Research, Development and Innovation Office (NKFIH) grants GINOP-2.3.2-15-2016-00003, K-119517, K-125015, and the City of Szombathely under Agreement No.\ 67.177-21/2016. 
V.V.G. is an F.R.S-FNRS Research Associate. 
We thank Taylor J. Bell for supplying and discussing the analysis of the \textit{Spitzer} phase curve presented in \citet{BellM1}.

\end{acknowledgements}

\bibliographystyle{aa}
\bibliography{biblio}

\begin{thebibliography}{135}
\expandafter\ifx\csname natexlab\endcsname\relax\def\natexlab#1{#1}\fi

\bibitem[{{Ahlers} {et~al.}(2015){Ahlers}, {Barnes}, \& {Barnes}}]{Ahlersk89}
{Ahlers}, J.~P., {Barnes}, J.~W., \& {Barnes}, R. 2015, \apj, 814, 67

\bibitem[{{Ahlers} {et~al.}(2019){Ahlers}, {Barnes}, \& {Myers}}]{Ahlersk976}
{Ahlers}, J.~P., {Barnes}, J.~W., \& {Myers}, S.~A. 2019, \aj, 158, 88

\bibitem[{{Ahlers} {et~al.}(2020{\natexlab{a}}){Ahlers}, {Johnson}, {Stassun},
  {Col{\'o}n}, {Barnes}, {Stevens}, {Beatty}, {Gaudi}, {Collins}, {Rodriguez},
  {Ricker}, {Vanderspek}, {Latham}, {Seager}, {Winn}, {Jenkins}, {Caldwell},
  {Goeke}, {Osborn}, {Paegert}, {Rowden}, \& {Tenenbaum}}]{Ahlersk9}
{Ahlers}, J.~P., {Johnson}, M.~C., {Stassun}, K.~G., {et~al.}
  2020{\natexlab{a}}, \aj, 160, 4

\bibitem[{{Ahlers} {et~al.}(2020{\natexlab{b}}){Ahlers}, {Kruse}, {Col{\'o}n},
  {Dorval}, {Talens}, {Snellen}, {Albrecht}, {Otten}, {Ricker}, {Vanderspek},
  {Latham}, {Seager}, {Winn}, {Jenkins}, {Haworth}, {Cartwright}, {Morris},
  {Rowden}, {Tenenbaum}, \& {Ting}}]{AhlersM4}
{Ahlers}, J.~P., {Kruse}, E., {Col{\'o}n}, K.~D., {et~al.} 2020{\natexlab{b}},
  \apj, 888, 63

\bibitem[{{Ahlers} {et~al.}(2014){Ahlers}, {Seubert}, \& {Barnes}}]{Ahlersk368}
{Ahlers}, J.~P., {Seubert}, S.~A., \& {Barnes}, J.~W. 2014, \apj, 786, 131

\bibitem[{{Albrecht} {et~al.}(2021){Albrecht}, {Marcussen}, {Winn}, {Dawson},
  \& {Knudstrup}}]{PerpendicularPlanets}
{Albrecht}, S.~H., {Marcussen}, M.~L., {Winn}, J.~N., {Dawson}, R.~I., \&
  {Knudstrup}, E. 2021, \apjl, 916, L1

\bibitem[{{Anderson} {et~al.}(2018){Anderson}, {Temple}, {Nielsen}, {Burdanov},
  {Hellier}, {Bouchy}, {Brown}, {Collier Cameron}, {Gillon}, {Jehin}, {Maxted},
  {Pepe}, {Pollacco}, {Pozuelos}, {Queloz}, {S{\'e}gransan}, {Smalley},
  {Triaud}, {Turner}, {Udry}, \& {West}}]{w189}
{Anderson}, D.~R., {Temple}, L.~Y., {Nielsen}, L.~D., {et~al.} 2018, arXiv
  e-prints, arXiv:1809.04897

\bibitem[{{Angerhausen} {et~al.}(2015){Angerhausen}, {DeLarme}, \&
  {Morse}}]{Angerhausen}
{Angerhausen}, D., {DeLarme}, E., \& {Morse}, J.~A. 2015, \pasp, 127, 1113

\bibitem[{{Arcangeli} {et~al.}(2018){Arcangeli}, {D{\'e}sert}, {Line}, {Bean},
  {Parmentier}, {Stevenson}, {Kreidberg}, {Fortney}, {Mansfield}, \&
  {Showman}}]{Arcangeli_H-}
{Arcangeli}, J., {D{\'e}sert}, J.-M., {Line}, M.~R., {et~al.} 2018, \apjl, 855,
  L30

\bibitem[{{Barnes}(2009)}]{Barnes09}
{Barnes}, J.~W. 2009, \apj, 705, 683

\bibitem[{{Barnes} {et~al.}(2015){Barnes}, {Ahlers}, {Seubert}, \&
  {Relles}}]{Barnesk2138}
{Barnes}, J.~W., {Ahlers}, J.~P., {Seubert}, S.~A., \& {Relles}, H.~M. 2015,
  \apjl, 808, L38

\bibitem[{{Barnes} {et~al.}(2011){Barnes}, {Linscott}, \&
  {Shporer}}]{Barnesk13}
{Barnes}, J.~W., {Linscott}, E., \& {Shporer}, A. 2011, \apjs, 197, 10

\bibitem[{{Bell} {et~al.}(2021){Bell}, {Dang}, {Cowan}, {Bean}, {D{\'e}sert},
  {Fortney}, {Keating}, {Kempton}, {Kreidberg}, {Line}, {Mansfield},
  {Parmentier}, {Stevenson}, {Swain}, \& {Zellem}}]{BellM1}
{Bell}, T.~J., {Dang}, L., {Cowan}, N.~B., {et~al.} 2021, \mnras, 504, 3316

\bibitem[{{Benomar} {et~al.}(2014){Benomar}, {Masuda}, {Shibahashi}, \&
  {Suto}}]{Asteroseis_hp7}
{Benomar}, O., {Masuda}, K., {Shibahashi}, H., \& {Suto}, Y. 2014, \pasj, 66,
  94

\bibitem[{{Benz} {et~al.}(2021){Benz}, {Broeg}, {Fortier}, {Rando}, {Beck},
  {Beck}, {Queloz}, {Ehrenreich}, {Maxted}, {Isaak}, {Billot}, {Alibert},
  {Alonso}, {Ant{\'o}nio}, {Asquier}, {Bandy}, {B{\'a}rczy}, {Barrado},
  {Barros}, {Baumjohann}, {Bekkelien}, {Bergomi}, {Biondi}, {Bonfils},
  {Borsato}, {Brandeker}, {Busch}, {Cabrera}, {Cessa}, {Charnoz}, {Chazelas},
  {Collier Cameron}, {Corral Van Damme}, {Cortes}, {Davies}, {Deleuil},
  {Deline}, {Delrez}, {Demangeon}, {Demory}, {Erikson}, {Farinato}, {Fossati},
  {Fridlund}, {Futyan}, {Gandolfi}, {Garcia Munoz}, {Gillon}, {Guterman},
  {Gutierrez}, {Hasiba}, {Heng}, {Hernandez}, {Hoyer}, {Kiss}, {Kovacs},
  {Kuntzer}, {Laskar}, {Lecavelier des Etangs}, {Lendl}, {L{\'o}pez}, {Lora},
  {Lovis}, {L{\"u}ftinger}, {Magrin}, {Malvasio}, {Marafatto}, {Michaelis}, {de
  Miguel}, {Modrego}, {Munari}, {Nascimbeni}, {Olofsson}, {Ottacher},
  {Ottensamer}, {Pagano}, {Palacios}, {Pall{\'e}}, {Peter}, {Piazza}, {Piotto},
  {Pizarro}, {Pollaco}, {Ragazzoni}, {Ratti}, {Rauer}, {Ribas}, {Rieder},
  {Rohlfs}, {Safa}, {Salatti}, {Santos}, {Scandariato}, {S{\'e}gransan},
  {Simon}, {Smith}, {Sordet}, {Sousa}, {Steller}, {Szab{\'o}}, {Szoke},
  {Thomas}, {Tschentscher}, {Udry}, {Van Grootel}, {Viotto}, {Walter},
  {Walton}, {Wildi}, \& {Wolter}}]{BenzCHEOPS}
{Benz}, W., {Broeg}, C., {Fortier}, A., {et~al.} 2021, Experimental Astronomy,
  51, 109

\bibitem[{{Blackwell} \& {Shallis}(1977)}]{Blackwell1977}
{Blackwell}, D.~E. \& {Shallis}, M.~J. 1977, \mnras, 180, 177

\bibitem[{{Bonfanti} {et~al.}(2021){Bonfanti}, {Delrez}, {Hooton}, {Wilson},
  {Fossati}, {Alibert}, {Hoyer}, {Mustill}, {Osborn}, {Adibekyan}, {Gandolfi},
  {Salmon}, {Sousa}, {Tuson}, {Van Grootel}, {Cabrera}, {Nascimbeni}, {Maxted},
  {Barros}, {Billot}, {Bonfils}, {Borsato}, {Broeg}, {Davies}, {Deleuil},
  {Demangeon}, {Fridlund}, {Lacedelli}, {Lendl}, {Persson}, {Santos},
  {Scandariato}, {Szab{\'o}}, {Collier Cameron}, {Udry}, {Benz}, {Beck},
  {Ehrenreich}, {Fortier}, {Isaak}, {Queloz}, {Alonso}, {Asquier}, {Bandy},
  {B{\'a}rczy}, {Barrado}, {Barrag{\'a}n}, {Baumjohann}, {Beck}, {Bekkelien},
  {Bergomi}, {Brandeker}, {Busch}, {Cessa}, {Charnoz}, {Chazelas}, {Corral Van
  Damme}, {Demory}, {Erikson}, {Farinato}, {Futyan}, {Garcia Mu{\~n}oz},
  {Gillon}, {Guedel}, {Guterman}, {Hasiba}, {Heng}, {Hernandez}, {Kiss},
  {Kuntzer}, {Laskar}, {Lecavelier des Etangs}, {Lovis}, {Magrin}, {Malvasio},
  {Marafatto}, {Michaelis}, {Munari}, {Olofsson}, {Ottacher}, {Ottensamer},
  {Pagano}, {Pall{\'e}}, {Peter}, {Piazza}, {Piotto}, {Pollacco}, {Ragazzoni},
  {Rando}, {Ratti}, {Rauer}, {Ribas}, {Rieder}, {Rohlfs}, {Safa}, {Salatti},
  {S{\'e}gransan}, {Simon}, {Smith}, {Sordet}, {Steller}, {Thomas},
  {Tschentscher}, {Van Eylen}, {Viotto}, {Walter}, {Walton}, {Wildi}, \&
  {Wolter}}]{bonfanti21}
{Bonfanti}, A., {Delrez}, L., {Hooton}, M.~J., {et~al.} 2021, \aap, 646, A157

\bibitem[{{Bonfanti} {et~al.}(2016){Bonfanti}, {Ortolani}, \&
  {Nascimbeni}}]{bonfanti16}
{Bonfanti}, A., {Ortolani}, S., \& {Nascimbeni}, V. 2016, \aap, 585, A5

\bibitem[{{Bonfanti} {et~al.}(2015){Bonfanti}, {Ortolani}, {Piotto}, \&
  {Nascimbeni}}]{bonfanti15}
{Bonfanti}, A., {Ortolani}, S., {Piotto}, G., \& {Nascimbeni}, V. 2015, \aap,
  575, A18

\bibitem[{{Borsa} {et~al.}(2021){Borsa}, {Lanza}, {Raspantini}, {Rainer},
  {Fossati}, {Brogi}, {Di Mauro}, {Gratton}, {Pino}, {Benatti}, {Bignamini},
  {Bonomo}, {Claudi}, {Esposito}, {Frustagli}, {Maggio}, {Maldonado},
  {Mancini}, {Micela}, {Nascimbeni}, {Poretti}, {Scandariato}, {Sicilia},
  {Sozzetti}, {Boschin}, {Cosentino}, {Covino}, {Desidera}, {Di Fabrizio},
  {Fiorenzano}, {Harutyunyan}, {Knapic}, {Molinari}, {Pagano}, {Pedani}, \&
  {Piotto}}]{Borsaw33}
{Borsa}, F., {Lanza}, A.~F., {Raspantini}, I., {et~al.} 2021, \aap, 653, A104

\bibitem[{{Borsato} {et~al.}(2021){Borsato}, {Piotto}, {Gandolfi},
  {Nascimbeni}, {Lacedelli}, {Marzari}, {Billot}, {Maxted}, {Sousa}, {Cameron},
  {Bonfanti}, {Wilson}, {Serrano}, {Garai}, {Alibert}, {Alonso}, {Asquier},
  {B{\'a}rczy}, {Bandy}, {Barrado}, {Barros}, {Baumjohann}, {Beck}, {Beck},
  {Benz}, {Bonfils}, {Brandeker}, {Broeg}, {Cabrera}, {Charnoz}, {Csizmadia},
  {Davies}, {Deleuil}, {Delrez}, {Demangeon}, {Demory}, {des Etangs},
  {Ehrenreich}, {Erikson}, {Escud{\'e}}, {Fortier}, {Fossati}, {Fridlund},
  {Gillon}, {Guedel}, {Hasiba}, {Heng}, {Hoyer}, {Isaak}, {Kiss}, {Kopp},
  {Laskar}, {Lendl}, {Lovis}, {Magrin}, {Munari}, {Olofsson}, {Ottensamer},
  {Pagano}, {Pall{\'e}}, {Peter}, {Pollacco}, {Queloz}, {Ragazzoni}, {Rando},
  {Rauer}, {Ribas}, {S{\'e}gransan}, {Santos}, {Scandariato}, {Simon}, {Smith},
  {Steller}, {Szab{\'o}}, {Thomas}, {Udry}, {Van Grootel}, \&
  {Walton}}]{BorsatoTTV}
{Borsato}, L., {Piotto}, G., {Gandolfi}, D., {et~al.} 2021, \mnras, 506, 3810

\bibitem[{{Bourrier} {et~al.}(2018){Bourrier}, {Lovis}, {Beust}, {Ehrenreich},
  {Henry}, {Astudillo-Defru}, {Allart}, {Bonfils}, {S{\'e}gransan}, {Delfosse},
  {Cegla}, {Wyttenbach}, {Heng}, {Lavie}, \& {Pepe}}]{Reloaded_GJ436}
{Bourrier}, V., {Lovis}, C., {Beust}, H., {et~al.} 2018, \nat, 553, 477

\bibitem[{{Brothwell} {et~al.}(2014){Brothwell}, {Watson}, {H{\'e}brard},
  {Triaud}, {Cegla}, {Santerne}, {H{\'e}brard}, {Anderson}, {Pollacco},
  {Simpson}, {Bouchy}, {Brown}, {Chew}, {Collier Cameron}, {Armstrong},
  {Barros}, {Bento}, {Bochinski}, {Burwitz}, {Busuttil}, {Delrez}, {Doyle},
  {Faedi}, {Fumel}, {Gillon}, {Haswell}, {Hellier}, {Jehin}, {Kolb}, {Lendl},
  {Liebig}, {Maxted}, {McCormac}, {Miller}, {Norton}, {Pepe}, {Queloz},
  {Rodr{\'\i}guez}, {S{\'e}gransan}, {Skillen}, {Smalley}, {Stassun}, {Udry},
  {West}, \& {Wheatley}}]{BrothwellRM}
{Brothwell}, R.~D., {Watson}, C.~A., {H{\'e}brard}, G., {et~al.} 2014, \mnras,
  440, 3392

\bibitem[{{Casasayas-Barris} {et~al.}(2021){Casasayas-Barris}, {Palle},
  {Stangret}, {Bourrier}, {Tabernero}, {Yan}, {Borsa}, {Allart}, {Zapatero
  Osorio}, {Lovis}, {Sousa}, {Chen}, {Oshagh}, {Santos}, {Pepe}, {Rebolo},
  {Molaro}, {Cristiani}, {Adibekyan}, {Alibert}, {Allende Prieto}, {Bouchy},
  {Demangeon}, {Di Marcantonio}, {D'Odorico}, {Ehrenreich}, {Figueira},
  {G{\'e}nova Santos}, {Gonz{\'a}lez Hern{\'a}ndez}, {Lavie}, {Lillo-Box}, {Lo
  Curto}, {Martins}, {Mehner}, {Micela}, {Nunes}, {Poretti}, {Sozzetti},
  {Su{\'a}rez Mascare{\~n}o}, \& {Udry}}]{Reloaded_HD209}
{Casasayas-Barris}, N., {Palle}, E., {Stangret}, M., {et~al.} 2021, \aap, 647,
  A26

\bibitem[{{Castelli} \& {Kurucz}(2003)}]{Castelli2003}
{Castelli}, F. \& {Kurucz}, R.~L. 2003, in IAU Symposium, Vol. 210, Modelling
  of Stellar Atmospheres, ed. N.~{Piskunov}, W.~W. {Weiss}, \& D.~F. {Gray},
  A20

\bibitem[{{Cegla} {et~al.}(2016){Cegla}, {Lovis}, {Bourrier}, {Beeck},
  {Watson}, \& {Pepe}}]{Reloaded_def}
{Cegla}, H.~M., {Lovis}, C., {Bourrier}, V., {et~al.} 2016, \aap, 588, A127

\bibitem[{{Chaplin} {et~al.}(2013){Chaplin}, {Sanchis-Ojeda}, {Campante},
  {Handberg}, {Stello}, {Winn}, {Basu}, {Christensen-Dalsgaard}, {Davies},
  {Metcalfe}, {Buchhave}, {Fischer}, {Bedding}, {Cochran}, {Elsworth},
  {Gilliland}, {Hekker}, {Huber}, {Isaacson}, {Karoff}, {Kawaler}, {Kjeldsen},
  {Latham}, {Lund}, {Lundkvist}, {Marcy}, {Miglio}, {Barclay}, \&
  {Lissauer}}]{Asteroseis_k50}
{Chaplin}, W.~J., {Sanchis-Ojeda}, R., {Campante}, T.~L., {et~al.} 2013, \apj,
  766, 101

\bibitem[{{Chen} {et~al.}(2014){Chen}, {van Boekel}, {Wang}, {Nikolov},
  {Seemann}, \& {Henning}}]{Chenw46}
{Chen}, G., {van Boekel}, R., {Wang}, H., {et~al.} 2014, \aap, 567, A8

\bibitem[{{Claret}(2016)}]{Claret_GD}
{Claret}, A. 2016, \aap, 588, A15

\bibitem[{{Claret}(2021)}]{Claret_CHEOPS}
{Claret}, A. 2021, Research Notes of the American Astronomical Society, 5, 13

\bibitem[{{Claret} \& {Bloemen}(2011)}]{Claret_Spitzer}
{Claret}, A. \& {Bloemen}, S. 2011, \aap, 529, A75

\bibitem[{{Collier Cameron} {et~al.}(2010){Collier Cameron}, {Guenther},
  {Smalley}, {McDonald}, {Hebb}, {Andersen}, {Augusteijn}, {Barros}, {Brown},
  {Cochran}, {Endl}, {Fossey}, {Hartmann}, {Maxted}, {Pollacco}, {Skillen},
  {Telting}, {Waldmann}, \& {West}}]{WASP33disc}
{Collier Cameron}, A., {Guenther}, E., {Smalley}, B., {et~al.} 2010, \mnras,
  407, 507

\bibitem[{{Cowan} \& {Agol}(2011)}]{Cowanphasecurve}
{Cowan}, N.~B. \& {Agol}, E. 2011, \apj, 726, 82

\bibitem[{{Csizmadia}(2020)}]{TLCM}
{Csizmadia}, S. 2020, \mnras, 496, 4442

\bibitem[{{Dalal} {et~al.}(2019){Dalal}, {H{\'e}brard}, {Lecavelier des
  {\'E}tangs}, {Petit}, {Bourrier}, {Laskar}, {K{\"o}nig}, \&
  {Correia}}]{Reloaded_HD3167}
{Dalal}, S., {H{\'e}brard}, G., {Lecavelier des {\'E}tangs}, A., {et~al.} 2019,
  \aap, 631, A28

\bibitem[{{Deline} {et~al.}(2022){Deline}, {Hooton}, {Lendl}, {Morris},
  {Salmon}, {Olofsson}, {Broeg}, {Ehrenreich}, {Beck}, {Brandeker}, {Hoyer},
  {Sulis}, {Van Grootel}, {Bourrier}, {Demangeon}, {Demory}, {Heng},
  {Parviainen}, {Serrano}, {Singh}, {Bonfanti}, {Fossati}, {Kitzmann}, {Sousa},
  {Wilson}, {Alibert}, {Alonso}, {Anglada}, {B{\'a}rczy}, {Barrado Navascues},
  {Barros}, {Baumjohann}, {Beck}, {Bekkelien}, {Benz}, {Billot}, {Bonfils},
  {Cabrera}, {Charnoz}, {Collier Cameron}, {Corral van Damme}, {Csizmadia},
  {Davies}, {Deleuil}, {Delrez}, {de Roche}, {Erikson}, {Fortier}, {Fridlund},
  {Futyan}, {Gandolfi}, {Gillon}, {G{\"u}del}, {Gutermann}, {Hasiba}, {Isaak},
  {Kiss}, {Laskar}, {Lecavelier des Etangs}, {Lovis}, {Magrin}, {Maxted},
  {Munari}, {Nascimbeni}, {Ottensamer}, {Pagano}, {Pall{\'e}}, {Peter},
  {Piotto}, {Pollacco}, {Queloz}, {Ragazzoni}, {Rando}, {Rauer}, {Ribas},
  {Santos}, {Scandariato}, {S{\'e}gransan}, {Simon}, {Smith}, {Steller},
  {Szab{\'o}}, {Thomas}, {Udry}, {Walter}, \& {Walton}}]{2022arXiv220104518D}
{Deline}, A., {Hooton}, M.~J., {Lendl}, M., {et~al.} 2022, arXiv e-prints,
  arXiv:2201.04518

\bibitem[{{Delrez} {et~al.}(2021){Delrez}, {Ehrenreich}, {Alibert}, {Bonfanti},
  {Borsato}, {Fossati}, {Hooton}, {Hoyer}, {Pozuelos}, {Salmon}, {Sulis},
  {Wilson}, {Adibekyan}, {Bourrier}, {Brandeker}, {Charnoz}, {Deline},
  {Guterman}, {Haldemann}, {Hara}, {Oshagh}, {Sousa}, {Van Grootel}, {Alonso},
  {Anglada-Escud{\'e}}, {B{\'a}rczy}, {Barrado}, {Barros}, {Baumjohann},
  {Beck}, {Bekkelien}, {Benz}, {Billot}, {Bonfils}, {Broeg}, {Cabrera},
  {Collier Cameron}, {Davies}, {Deleuil}, {Delisle}, {Demangeon}, {Demory},
  {Erikson}, {Fortier}, {Fridlund}, {Futyan}, {Gandolfi}, {Garcia Mu{\~n}oz},
  {Gillon}, {Guedel}, {Heng}, {Kiss}, {Laskar}, {Lecavelier des Etangs},
  {Lendl}, {Lovis}, {Maxted}, {Nascimbeni}, {Olofsson}, {Osborn}, {Pagano},
  {Pall{\'e}}, {Piotto}, {Pollacco}, {Queloz}, {Rauer}, {Ragazzoni}, {Ribas},
  {Santos}, {Scandariato}, {S{\'e}gransan}, {Simon}, {Smith}, {Steller},
  {Szab{\'o}}, {Thomas}, {Udry}, \& {Walton}}]{Delrez_nu2lupi}
{Delrez}, L., {Ehrenreich}, D., {Alibert}, Y., {et~al.} 2021, Nature Astronomy,
  5, 775

\bibitem[{{Demory} {et~al.}(2013){Demory}, {de Wit}, {Lewis}, {Fortney},
  {Zsom}, {Seager}, {Knutson}, {Heng}, {Madhusudhan}, {Gillon}, {Barclay},
  {Desert}, {Parmentier}, \& {Cowan}}]{Demoryk7}
{Demory}, B.-O., {de Wit}, J., {Lewis}, N., {et~al.} 2013, \apjl, 776, L25

\bibitem[{{Demory} {et~al.}(2016){Demory}, {Gillon}, {de Wit}, {Madhusudhan},
  {Bolmont}, {Heng}, {Kataria}, {Lewis}, {Hu}, {Krick}, {Stamenkovi{\'c}},
  {Benneke}, {Kane}, \& {Queloz}}]{Demory55cnce}
{Demory}, B.-O., {Gillon}, M., {de Wit}, J., {et~al.} 2016, \nat, 532, 207

\bibitem[{{Dorval} {et~al.}(2020){Dorval}, {Talens}, {Otten}, {Brahm},
  {Jord{\'a}n}, {Torres}, {Vanzi}, {Zapata}, {Henry}, {Paredes}, {Jao},
  {James}, {Hinojosa}, {Bakos}, {Csubry}, {Bhatti}, {Suc}, {Osip}, {Mamajek},
  {Mellon}, {Wyttenbach}, {Stuik}, {Kenworthy}, {Bailey}, {Ireland},
  {Crawford}, {Lomberg}, {Kuhn}, \& {Snellen}}]{MASCARA4disc}
{Dorval}, P., {Talens}, G.~J.~J., {Otten}, G.~P.~P.~L., {et~al.} 2020, \aap,
  635, A60

\bibitem[{{Dupret}(2001)}]{Dupret01}
{Dupret}, M.~A. 2001, \aap, 366, 166

\bibitem[{{Dyudina} {et~al.}(2016){Dyudina}, {Zhang}, {Li}, {Kopparla},
  {Ingersoll}, {Dones}, {Verbiscer}, \& {Yung}}]{Dyudina_phasecurves}
{Dyudina}, U., {Zhang}, X., {Li}, L., {et~al.} 2016, \apj, 822, 76

\bibitem[{{Esteves} {et~al.}(2015){Esteves}, {De Mooij}, \&
  {Jayawardhana}}]{Esteves}
{Esteves}, L.~J., {De Mooij}, E. J.~W., \& {Jayawardhana}, R. 2015, \apj, 804,
  150

\bibitem[{{Evans} {et~al.}(2013){Evans}, {Pont}, {Sing}, {Aigrain}, {Barstow},
  {D{\'e}sert}, {Gibson}, {Heng}, {Knutson}, \& {Lecavelier des
  Etangs}}]{evanshd189}
{Evans}, T.~M., {Pont}, F., {Sing}, D.~K., {et~al.} 2013, \apjl, 772, L16

\bibitem[{{Foreman-Mackey}(2016)}]{corner}
{Foreman-Mackey}, D. 2016, The Journal of Open Source Software, 1, 24

\bibitem[{{Foreman-Mackey}(2018)}]{celerite2}
{Foreman-Mackey}, D. 2018, Research Notes of the American Astronomical Society,
  2, 31

\bibitem[{{Foreman-Mackey} {et~al.}(2017){Foreman-Mackey}, {Agol},
  {Ambikasaran}, \& {Angus}}]{celerite1}
{Foreman-Mackey}, D., {Agol}, E., {Ambikasaran}, S., \& {Angus}, R. 2017, \aj,
  154, 220

\bibitem[{{Foreman-Mackey} {et~al.}(2013){Foreman-Mackey}, {Hogg}, {Lang}, \&
  {Goodman}}]{emcee}
{Foreman-Mackey}, D., {Hogg}, D.~W., {Lang}, D., \& {Goodman}, J. 2013, \pasp,
  125, 306

\bibitem[{{Gaia Collaboration} {et~al.}(2021){Gaia Collaboration}, {Brown},
  {Vallenari}, {Prusti}, {de Bruijne}, {Babusiaux}, {Biermann}, {Creevey},
  {Evans}, {Eyer}, {Hutton}, {Jansen}, {Jordi}, {Klioner}, {Lammers},
  {Lindegren}, {Luri}, {Mignard}, {Panem}, {Pourbaix}, {Randich}, {Sartoretti},
  {Soubiran}, {Walton}, {Arenou}, {Bailer-Jones}, {Bastian}, {Cropper},
  {Drimmel}, {Katz}, {Lattanzi}, {van Leeuwen}, {Bakker}, {Cacciari},
  {Casta{\~n}eda}, {De Angeli}, {Ducourant}, {Fabricius}, {Fouesneau},
  {Fr{\'e}mat}, {Guerra}, {Guerrier}, {Guiraud}, {Jean-Antoine Piccolo},
  {Masana}, {Messineo}, {Mowlavi}, {Nicolas}, {Nienartowicz}, {Pailler},
  {Panuzzo}, {Riclet}, {Roux}, {Seabroke}, {Sordo}, {Tanga}, {Th{\'e}venin},
  {Gracia-Abril}, {Portell}, {Teyssier}, {Altmann}, {Andrae}, {Bellas-Velidis},
  {Benson}, {Berthier}, {Blomme}, {Brugaletta}, {Burgess}, {Busso}, {Carry},
  {Cellino}, {Cheek}, {Clementini}, {Damerdji}, {Davidson}, {Delchambre},
  {Dell'Oro}, {Fern{\'a}ndez-Hern{\'a}ndez}, {Galluccio}, {Garc{\'\i}a-Lario},
  {Garcia-Reinaldos}, {Gonz{\'a}lez-N{\'u}{\~n}ez}, {Gosset}, {Haigron},
  {Halbwachs}, {Hambly}, {Harrison}, {Hatzidimitriou}, {Heiter},
  {Hern{\'a}ndez}, {Hestroffer}, {Hodgkin}, {Holl}, {Jan{\ss}en}, {Jevardat de
  Fombelle}, {Jordan}, {Krone-Martins}, {Lanzafame}, {L{\"o}ffler}, {Lorca},
  {Manteiga}, {Marchal}, {Marrese}, {Moitinho}, {Mora}, {Muinonen}, {Osborne},
  {Pancino}, {Pauwels}, {Petit}, {Recio-Blanco}, {Richards}, {Riello},
  {Rimoldini}, {Robin}, {Roegiers}, {Rybizki}, {Sarro}, {Siopis}, {Smith},
  {Sozzetti}, {Ulla}, {Utrilla}, {van Leeuwen}, {van Reeven}, {Abbas}, {Abreu
  Aramburu}, {Accart}, {Aerts}, {Aguado}, {Ajaj}, {Altavilla}, {{\'A}lvarez},
  {{\'A}lvarez Cid-Fuentes}, {Alves}, {Anderson}, {Anglada Varela}, {Antoja},
  {Audard}, {Baines}, {Baker}, {Balaguer-N{\'u}{\~n}ez}, {Balbinot}, {Balog},
  {Barache}, {Barbato}, {Barros}, {Barstow}, {Bartolom{\'e}}, {Bassilana},
  {Bauchet}, {Baudesson-Stella}, {Becciani}, {Bellazzini}, {Bernet}, {Bertone},
  {Bianchi}, {Blanco-Cuaresma}, {Boch}, {Bombrun}, {Bossini}, {Bouquillon},
  {Bragaglia}, {Bramante}, {Breedt}, {Bressan}, {Brouillet}, {Bucciarelli},
  {Burlacu}, {Busonero}, {Butkevich}, {Buzzi}, {Caffau}, {Cancelliere},
  {C{\'a}novas}, {Cantat-Gaudin}, {Carballo}, {Carlucci}, {Carnerero},
  {Carrasco}, {Casamiquela}, {Castellani}, {Castro-Ginard}, {Castro Sampol},
  {Chaoul}, {Charlot}, {Chemin}, {Chiavassa}, {Cioni}, {Comoretto}, {Cooper},
  {Cornez}, {Cowell}, {Crifo}, {Crosta}, {Crowley}, {Dafonte}, {Dapergolas},
  {David}, {David}, {de Laverny}, {De Luise}, {De March}, {De Ridder}, {de
  Souza}, {de Teodoro}, {de Torres}, {del Peloso}, {del Pozo}, {Delbo},
  {Delgado}, {Delgado}, {Delisle}, {Di Matteo}, {Diakite}, {Diener},
  {Distefano}, {Dolding}, {Eappachen}, {Edvardsson}, {Enke}, {Esquej}, {Fabre},
  {Fabrizio}, {Faigler}, {Fedorets}, {Fernique}, {Fienga}, {Figueras},
  {Fouron}, {Fragkoudi}, {Fraile}, {Franke}, {Gai}, {Garabato},
  {Garcia-Gutierrez}, {Garc{\'\i}a-Torres}, {Garofalo}, {Gavras}, {Gerlach},
  {Geyer}, {Giacobbe}, {Gilmore}, {Girona}, {Giuffrida}, {Gomel}, {Gomez},
  {Gonzalez-Santamaria}, {Gonz{\'a}lez-Vidal}, {Granvik},
  {Guti{\'e}rrez-S{\'a}nchez}, {Guy}, {Hauser}, {Haywood}, {Helmi}, {Hidalgo},
  {Hilger}, {H{\l}adczuk}, {Hobbs}, {Holland}, {Huckle}, {Jasniewicz},
  {Jonker}, {Juaristi Campillo}, {Julbe}, {Karbevska}, {Kervella}, {Khanna},
  {Kochoska}, {Kontizas}, {Kordopatis}, {Korn}, {Kostrzewa-Rutkowska},
  {Kruszy{\'n}ska}, {Lambert}, {Lanza}, {Lasne}, {Le Campion}, {Le Fustec},
  {Lebreton}, {Lebzelter}, {Leccia}, {Leclerc}, {Lecoeur-Taibi}, {Liao},
  {Licata}, {Lindstr{\o}m}, {Lister}, {Livanou}, {Lobel}, {Madrero Pardo},
  {Managau}, {Mann}, {Marchant}, {Marconi}, {Marcos Santos}, {Marinoni},
  {Marocco}, {Marshall}, {Martin Polo}, {Mart{\'\i}n-Fleitas}, {Masip},
  {Massari}, {Mastrobuono-Battisti}, {Mazeh}, {McMillan}, {Messina},
  {Michalik}, {Millar}, {Mints}, {Molina}, {Molinaro}, {Moln{\'a}r},
  {Montegriffo}, {Mor}, {Morbidelli}, {Morel}, {Morris}, {Mulone}, {Munoz},
  {Muraveva}, {Murphy}, {Musella}, {Noval}, {Ord{\'e}novic}, {Orr{\`u}},
  {Osinde}, {Pagani}, {Pagano}, {Palaversa}, {Palicio}, {Panahi}, {Pawlak},
  {Pe{\~n}alosa Esteller}, {Penttil{\"a}}, {Piersimoni}, {Pineau}, {Plachy},
  {Plum}, {Poggio}, {Poretti}, {Poujoulet}, {Pr{\v{s}}a}, {Pulone}, {Racero},
  {Ragaini}, {Rainer}, {Raiteri}, {Rambaux}, {Ramos}, {Ramos-Lerate}, {Re
  Fiorentin}, {Regibo}, {Reyl{\'e}}, {Ripepi}, {Riva}, {Rixon}, {Robichon},
  {Robin}, {Roelens}, {Rohrbasser}, {Romero-G{\'o}mez}, {Rowell}, {Royer},
  {Rybicki}, {Sadowski}, {Sagrist{\`a} Sell{\'e}s}, {Sahlmann}, {Salgado},
  {Salguero}, {Samaras}, {Sanchez Gimenez}, {Sanna}, {Santove{\~n}a},
  {Sarasso}, {Schultheis}, {Sciacca}, {Segol}, {Segovia}, {S{\'e}gransan},
  {Semeux}, {Shahaf}, {Siddiqui}, {Siebert}, {Siltala}, {Slezak}, {Smart},
  {Solano}, {Solitro}, {Souami}, {Souchay}, {Spagna}, {Spoto}, {Steele},
  {Steidelm{\"u}ller}, {Stephenson}, {S{\"u}veges}, {Szabados}, {Szegedi-Elek},
  {Taris}, {Tauran}, {Taylor}, {Teixeira}, {Thuillot}, {Tonello}, {Torra},
  {Torra}, {Turon}, {Unger}, {Vaillant}, {van Dillen}, {Vanel}, {Vecchiato},
  {Viala}, {Vicente}, {Voutsinas}, {Weiler}, {Wevers}, {Wyrzykowski}, {Yoldas},
  {Yvard}, {Zhao}, {Zorec}, {Zucker}, {Zurbach}, \&
  {Zwitter}}]{GaiaCollaboration2020}
{Gaia Collaboration}, {Brown}, A.~G.~A., {Vallenari}, A., {et~al.} 2021, \aap,
  649, A1

\bibitem[{{Gaudi} {et~al.}(2017){Gaudi}, {Stassun}, {Collins}, {Beatty},
  {Zhou}, {Latham}, {Bieryla}, {Eastman}, {Siverd}, {Crepp}, {Gonzales},
  {Stevens}, {Buchhave}, {Pepper}, {Johnson}, {Colon}, {Jensen}, {Rodriguez},
  {Bozza}, {Novati}, {D'Ago}, {Dumont}, {Ellis}, {Gaillard}, {Jang-Condell},
  {Kasper}, {Fukui}, {Gregorio}, {Ito}, {Kielkopf}, {Manner}, {Matt}, {Narita},
  {Oberst}, {Reed}, {Scarpetta}, {Stephens}, {Yeigh}, {Zambelli}, {Fulton},
  {Howard}, {James}, {Penny}, {Bayliss}, {Curtis}, {Depoy}, {Esquerdo},
  {Gould}, {Joner}, {Kuhn}, {Labadie-Bartz}, {Lund}, {Marshall}, {McLeod},
  {Pogge}, {Relles}, {Stockdale}, {Tan}, {Trueblood}, \&
  {Trueblood}}]{KELT9Gaudi}
{Gaudi}, B.~S., {Stassun}, K.~G., {Collins}, K.~A., {et~al.} 2017, \nat, 546,
  514

\bibitem[{{Gibson}(2014)}]{GibsonGP2}
{Gibson}, N.~P. 2014, \mnras, 445, 3401

\bibitem[{{Gibson} {et~al.}(2012){Gibson}, {Aigrain}, {Roberts}, {Evans},
  {Osborne}, \& {Pont}}]{GibsonGP}
{Gibson}, N.~P., {Aigrain}, S., {Roberts}, S., {et~al.} 2012, \mnras, 419, 2683

\bibitem[{{Heng} \& {Li}(2021)}]{HengCassini}
{Heng}, K. \& {Li}, L. 2021, \apjl, 909, L20

\bibitem[{{Heng} {et~al.}(2021){Heng}, {Morris}, \&
  {Kitzmann}}]{Heng_reflected}
{Heng}, K., {Morris}, B.~M., \& {Kitzmann}, D. 2021, Nature Astronomy, 5, 1001

\bibitem[{{Heng} \& {Workman}(2014)}]{HengWorkman}
{Heng}, K. \& {Workman}, J. 2014, \apjs, 213, 27

\bibitem[{{Henyey} \& {Greenstein}(1941)}]{HenyeyGreenstein}
{Henyey}, L.~G. \& {Greenstein}, J.~L. 1941, \apj, 93, 70

\bibitem[{{Herman} {et~al.}(2018){Herman}, {de Mooij}, {Huang}, \&
  {Jayawardhana}}]{Hermank13}
{Herman}, M.~K., {de Mooij}, E. J.~W., {Huang}, C.~X., \& {Jayawardhana}, R.
  2018, \aj, 155, 13

\bibitem[{{Higson} {et~al.}(2019){Higson}, {Handley}, {Hobson}, \&
  {Lasenby}}]{dns1}
{Higson}, E., {Handley}, W., {Hobson}, M., \& {Lasenby}, A. 2019, Statistics
  and Computing, 29, 891

\bibitem[{{Hirano} {et~al.}(2020){Hirano}, {Gaidos}, {Winn}, {Dai}, {Fukui},
  {Kuzuhara}, {Kotani}, {Tamura}, {Hjorth}, {Albrecht}, {Huber}, {Bolmont},
  {Harakawa}, {Hodapp}, {Ishizuka}, {Jacobson}, {Konishi}, {Kudo}, {Kurokawa},
  {Nishikawa}, {Omiya}, {Serizawa}, {Ueda}, \& {Weiss}}]{T1RM}
{Hirano}, T., {Gaidos}, E., {Winn}, J.~N., {et~al.} 2020, \apjl, 890, L27

\bibitem[{{Hooton} {et~al.}(2019){Hooton}, {de Mooij}, {Watson}, {Gibson},
  {Galindo-Guil}, {Clavero}, \& {Merritt}}]{Hootonw12}
{Hooton}, M.~J., {de Mooij}, E. J.~W., {Watson}, C.~A., {et~al.} 2019, \mnras,
  486, 2397

\bibitem[{{Hooton} {et~al.}(2018){Hooton}, {Watson}, {de Mooij}, {Gibson}, \&
  {Kitzmann}}]{Hootonk9}
{Hooton}, M.~J., {Watson}, C.~A., {de Mooij}, E. J.~W., {Gibson}, N.~P., \&
  {Kitzmann}, D. 2018, \apjl, 869, L25

\bibitem[{{Howarth} \& {Morello}(2017)}]{Howarthk13}
{Howarth}, I.~D. \& {Morello}, G. 2017, \mnras, 470, 932

\bibitem[{{Hoyer} {et~al.}(2020){Hoyer}, {Guterman}, {Demangeon}, {Sousa},
  {Deleuil}, {Meunier}, \& {Benz}}]{DRPHoyer}
{Hoyer}, S., {Guterman}, P., {Demangeon}, O., {et~al.} 2020, \aap, 635, A24

\bibitem[{{Husser} {et~al.}(2013){Husser}, {Wende-von Berg}, {Dreizler},
  {Homeier}, {Reiners}, {Barman}, \& {Hauschildt}}]{PHOENIX}
{Husser}, T.~O., {Wende-von Berg}, S., {Dreizler}, S., {et~al.} 2013, \aap,
  553, A6

\bibitem[{{Johnson} {et~al.}(2014){Johnson}, {Cochran}, {Albrecht},
  {Dodson-Robinson}, {Winn}, \& {Gullikson}}]{Johnsonk13}
{Johnson}, M.~C., {Cochran}, W.~D., {Albrecht}, S., {et~al.} 2014, \apj, 790,
  30

\bibitem[{{Kamiaka} {et~al.}(2018){Kamiaka}, {Benomar}, \&
  {Suto}}]{Asteroseis_inc}
{Kamiaka}, S., {Benomar}, O., \& {Suto}, Y. 2018, \mnras, 479, 391

\bibitem[{{Kass} \& {Raftery}(1995)}]{BayesFactor}
{Kass}, R.~E. \& {Raftery}, A.~E. 1995, Journal of the American Statistical
  Association, 90, 773

\bibitem[{{Keating} {et~al.}(2019){Keating}, {Cowan}, \& {Dang}}]{Keating2D}
{Keating}, D., {Cowan}, N.~B., \& {Dang}, L. 2019, Nature Astronomy, 3, 1092

\bibitem[{{Kipping}(2013)}]{limbdark}
{Kipping}, D.~M. 2013, \mnras, 435, 2152

\bibitem[{{Kurucz}(2013)}]{2013ascl.soft03024K}
{Kurucz}, R.~L. 2013, {ATLAS12: Opacity sampling model atmosphere program}

\bibitem[{{Leleu} {et~al.}(2021){Leleu}, {Alibert}, {Hara}, {Hooton}, {Wilson},
  {Robutel}, {Delisle}, {Laskar}, {Hoyer}, {Lovis}, {Bryant}, {Ducrot},
  {Cabrera}, {Delrez}, {Acton}, {Adibekyan}, {Allart}, {Allende Prieto},
  {Alonso}, {Alves}, {Anderson}, {Angerhausen}, {Anglada Escud{\'e}},
  {Asquier}, {Barrado}, {Barros}, {Baumjohann}, {Bayliss}, {Beck}, {Beck},
  {Bekkelien}, {Benz}, {Billot}, {Bonfanti}, {Bonfils}, {Bouchy}, {Bourrier},
  {Bou{\'e}}, {Brandeker}, {Broeg}, {Buder}, {Burdanov}, {Burleigh},
  {B{\'a}rczy}, {Cameron}, {Chamberlain}, {Charnoz}, {Cooke}, {Corral Van
  Damme}, {Correia}, {Cristiani}, {Damasso}, {Davies}, {Deleuil}, {Demangeon},
  {Demory}, {Di Marcantonio}, {Di Persio}, {Dumusque}, {Ehrenreich}, {Erikson},
  {Figueira}, {Fortier}, {Fossati}, {Fridlund}, {Futyan}, {Gandolfi},
  {Garc{\'\i}a Mu{\~n}oz}, {Garcia}, {Gill}, {Gillen}, {Gillon}, {Goad},
  {Gonz{\'a}lez Hern{\'a}ndez}, {Guedel}, {G{\"u}nther}, {Haldemann},
  {Henderson}, {Heng}, {Hogan}, {Isaak}, {Jehin}, {Jenkins}, {Jord{\'a}n},
  {Kiss}, {Kristiansen}, {Lam}, {Lavie}, {Lecavelier des Etangs}, {Lendl},
  {Lillo-Box}, {Lo Curto}, {Magrin}, {Martins}, {Maxted}, {McCormac}, {Mehner},
  {Micela}, {Molaro}, {Moyano}, {Murray}, {Nascimbeni}, {Nunes}, {Olofsson},
  {Osborn}, {Oshagh}, {Ottensamer}, {Pagano}, {Pall{\'e}}, {Pedersen}, {Pepe},
  {Persson}, {Peter}, {Piotto}, {Polenta}, {Pollacco}, {Poretti}, {Pozuelos},
  {Queloz}, {Ragazzoni}, {Rando}, {Ratti}, {Rauer}, {Raynard}, {Rebolo},
  {Reimers}, {Ribas}, {Santos}, {Scandariato}, {Schneider}, {Sebastian},
  {Sestovic}, {Simon}, {Smith}, {Sousa}, {Sozzetti}, {Steller}, {Su{\'a}rez
  Mascare{\~n}o}, {Szab{\'o}}, {S{\'e}gransan}, {Thomas}, {Thompson},
  {Tilbrook}, {Triaud}, {Turner}, {Udry}, {Van Grootel}, {Venus}, {Verrecchia},
  {Vines}, {Walton}, {West}, {Wheatley}, {Wolter}, \& {Zapatero
  Osorio}}]{TOI178}
{Leleu}, A., {Alibert}, Y., {Hara}, N.~C., {et~al.} 2021, \aap, 649, A26

\bibitem[{{Lendl} {et~al.}(2020){Lendl}, {Csizmadia}, {Deline}, {Fossati},
  {Kitzmann}, {Heng}, {Hoyer}, {Salmon}, {Benz}, {Broeg}, {Ehrenreich},
  {Fortier}, {Queloz}, {Bonfanti}, {Brandeker}, {Collier Cameron}, {Delrez},
  {Garcia Mu{\~n}oz}, {Hooton}, {Maxted}, {Morris}, {Van Grootel}, {Wilson},
  {Alibert}, {Alonso}, {Asquier}, {Bandy}, {B{\'a}rczy}, {Barrado}, {Barros},
  {Baumjohann}, {Beck}, {Beck}, {Bekkelien}, {Bergomi}, {Billot}, {Biondi},
  {Bonfils}, {Bourrier}, {Busch}, {Cabrera}, {Cessa}, {Charnoz}, {Chazelas},
  {Corral Van Damme}, {Davies}, {Deleuil}, {Demangeon}, {Demory}, {Erikson},
  {Farinato}, {Fridlund}, {Futyan}, {Gandolfi}, {Gillon}, {Guterman}, {Hasiba},
  {Hernandez}, {Isaak}, {Kiss}, {Kuntzer}, {Lecavelier des Etangs},
  {L{\"u}ftinger}, {Laskar}, {Lovis}, {Magrin}, {Malvasio}, {Marafatto},
  {Michaelis}, {Munari}, {Nascimbeni}, {Olofsson}, {Ottacher}, {Ottensamer},
  {Pagano}, {Pall{\'e}}, {Peter}, {Piazza}, {Piotto}, {Pollacco}, {Ratti},
  {Rauer}, {Ragazzoni}, {Rando}, {Ribas}, {Rieder}, {Rohlfs}, {Safa}, {Santos},
  {Scandariato}, {S{\'e}gransan}, {Simon}, {Singh}, {Smith}, {Sordet}, {Sousa},
  {Steller}, {Szab{\'o}}, {Thomas}, {Tschentscher}, {Udry}, {Viotto}, {Walter},
  {Walton}, {Wildi}, \& {Wolter}}]{Lendlw189}
{Lendl}, M., {Csizmadia}, S., {Deline}, A., {et~al.} 2020, \aap, 643, A94

\bibitem[{{Li} {et~al.}(2018){Li}, {Jiang}, {West}, {Gierasch}, {Perez-Hoyos},
  {Sanchez-Lavega}, {Fletcher}, {Fortney}, {Knowles}, {Porco}, {Baines}, {Fry},
  {Mallama}, {Achterberg}, {Simon}, {Nixon}, {Orton}, {Dyudina}, {Ewald}, \&
  {Schmude}}]{LiCassini}
{Li}, L., {Jiang}, X., {West}, R.~A., {et~al.} 2018, Nature Communications, 9,
  3709

\bibitem[{{Lindegren} {et~al.}(2021){Lindegren}, {Bastian}, {Biermann},
  {Bombrun}, {de Torres}, {Gerlach}, {Geyer}, {Hern{\'a}ndez}, {Hilger},
  {Hobbs}, {Klioner}, {Lammers}, {McMillan}, {Ramos-Lerate},
  {Steidelm{\"u}ller}, {Stephenson}, \& {van Leeuwen}}]{Lindegren2020}
{Lindegren}, L., {Bastian}, U., {Biermann}, M., {et~al.} 2021, \aap, 649, A4

\bibitem[{{Lothringer} {et~al.}(2018){Lothringer}, {Barman}, \&
  {Koskinen}}]{Lothringer_H-}
{Lothringer}, J.~D., {Barman}, T., \& {Koskinen}, T. 2018, \apj, 866, 27

\bibitem[{{Maeder}(2009)}]{Maeder}
{Maeder}, A. 2009, {Physics, Formation and Evolution of Rotating Stars}

\bibitem[{{Malik} {et~al.}(2017){Malik}, {Grosheintz}, {Mendon{\c{c}}a},
  {Grimm}, {Lavie}, {Kitzmann}, {Tsai}, {Burrows}, {Kreidberg}, {Bedell},
  {Bean}, {Stevenson}, \& {Heng}}]{HELIOS1}
{Malik}, M., {Grosheintz}, L., {Mendon{\c{c}}a}, J.~M., {et~al.} 2017, \aj,
  153, 56

\bibitem[{{Malik} {et~al.}(2019){Malik}, {Kitzmann}, {Mendon{\c{c}}a}, {Grimm},
  {Marleau}, {Linder}, {Tsai}, \& {Heng}}]{HELIOS2}
{Malik}, M., {Kitzmann}, D., {Mendon{\c{c}}a}, J.~M., {et~al.} 2019, \aj, 157,
  170

\bibitem[{{Mandel} \& {Agol}(2002)}]{mandelagol}
{Mandel}, K. \& {Agol}, E. 2002, \apjl, 580, L171

\bibitem[{{Mansfield} {et~al.}(2020){Mansfield}, {Bean}, {Stevenson},
  {Komacek}, {Bell}, {Tan}, {Malik}, {Beatty}, {Wong}, {Cowan}, {Dang},
  {D{\'e}sert}, {Fortney}, {Gaudi}, {Keating}, {Kempton}, {Kreidberg}, {Line},
  {Parmentier}, {Stassun}, {Swain}, \& {Zellem}}]{MansfieldK9}
{Mansfield}, M., {Bean}, J.~L., {Stevenson}, K.~B., {et~al.} 2020, \apjl, 888,
  L15

\bibitem[{{Marigo} {et~al.}(2017){Marigo}, {Girardi}, {Bressan}, {Rosenfield},
  {Aringer}, {Chen}, {Dussin}, {Nanni}, {Pastorelli}, {Rodrigues}, {Trabucchi},
  {Bladh}, {Dalcanton}, {Groenewegen}, {Montalb{\'a}n}, \& {Wood}}]{marigo17}
{Marigo}, P., {Girardi}, L., {Bressan}, A., {et~al.} 2017, \apj, 835, 77

\bibitem[{{Marley} {et~al.}(1999){Marley}, {Gelino}, {Stephens}, {Lunine}, \&
  {Freedman}}]{MarleyAlbedo}
{Marley}, M.~S., {Gelino}, C., {Stephens}, D., {Lunine}, J.~I., \& {Freedman},
  R. 1999, \apj, 513, 879

\bibitem[{{Masuda}(2015)}]{Masudak13}
{Masuda}, K. 2015, \apj, 805, 28

\bibitem[{{Maxted} {et~al.}(2013){Maxted}, {Anderson}, {Doyle}, {Gillon},
  {Harrington}, {Iro}, {Jehin}, {Lafreni{\`e}re}, {Smalley}, \&
  {Southworth}}]{Maxtedw18}
{Maxted}, P.~F.~L., {Anderson}, D.~R., {Doyle}, A.~P., {et~al.} 2013, \mnras,
  428, 2645

\bibitem[{{Maxted} {et~al.}(2021){Maxted}, {Ehrenreich}, {Wilson}, {Jones},
  {Piaulet}, {Kitzmann}, \& {Hoeijmakers}}]{Maxted_classics}
{Maxted}, P.~F.~L., {Ehrenreich}, D., {Wilson}, T.~G., {et~al.} 2021, MNRAS
  submitted

\bibitem[{{May} \& {Stevenson}(2020)}]{Mayw43}
{May}, E.~M. \& {Stevenson}, K.~B. 2020, \aj, 160, 140

\bibitem[{{McDonald} {et~al.}(2012){McDonald}, {Zijlstra}, \&
  {Boyer}}]{2012MNRAS.427..343M}
{McDonald}, I., {Zijlstra}, A.~A., \& {Boyer}, M.~L. 2012, \mnras, 427, 343

\bibitem[{{McLaughlin}(1924)}]{McLaughlin}
{McLaughlin}, D.~B. 1924, \apj, 60, 22

\bibitem[{{Morris} {et~al.}(2021{\natexlab{a}}){Morris}, {Delrez}, {Brandeker},
  {Cameron}, {Simon}, {Futyan}, {Olofsson}, {Hoyer}, {Fortier}, {Demory},
  {Lendl}, {Wilson}, {Oshagh}, {Heng}, {Ehrenreich}, {Sulis}, {Alibert},
  {Alonso}, {Anglada Escud{\'e}}, {Barrado}, {Barros}, {Baumjohann}, {Beck},
  {Beck}, {Bekkelien}, {Benz}, {Bergomi}, {Billot}, {Bonfils}, {Bourrier},
  {Broeg}, {B{\'a}rczy}, {Cabrera}, {Charnoz}, {Davies}, {De Miguel Ferreras},
  {Deleuil}, {Deline}, {Demangeon}, {Erikson}, {Floren}, {Fossati}, {Fridlund},
  {Gandolfi}, {Garc{\'\i}a Mu{\~n}oz}, {Gillon}, {Guedel}, {Guterman}, {Isaak},
  {Kiss}, {Laskar}, {Lecavelier des Etangs}, {Lieder}, {Lovis}, {Magrin},
  {Maxted}, {Nascimbeni}, {Ottensamer}, {Pagano}, {Pall{\'e}}, {Peter},
  {Piotto}, {Pizarro Rubio}, {Pollacco}, {Pozuelos}, {Queloz}, {Ragazzoni},
  {Rando}, {Rauer}, {Ribas}, {Santos}, {Scandariato}, {Smith}, {Sousa},
  {Steller}, {Szab{\'o}}, {S{\'e}gransan}, {Thomas}, {Udry}, {Ulmer}, {Van
  Grootel}, \& {Walton}}]{Morris55cnce}
{Morris}, B.~M., {Delrez}, L., {Brandeker}, A., {et~al.} 2021{\natexlab{a}},
  \aap, 653, A173

\bibitem[{{Morris} {et~al.}(2021{\natexlab{b}}){Morris}, {Heng}, {Jones},
  {Piaulet}, {Demory}, {Kitzmann}, \& {Hoeijmakers}}]{Morris2021_hml}
{Morris}, B.~M., {Heng}, K., {Jones}, K., {et~al.} 2021{\natexlab{b}}, arXiv
  e-prints, arXiv:2110.11837

\bibitem[{{Morris}(1985)}]{Morris_ellipsoid1}
{Morris}, S.~L. 1985, \apj, 295, 143

\bibitem[{{Morris} \& {Naftilan}(1993)}]{Morris_ellipsoid2}
{Morris}, S.~L. \& {Naftilan}, S.~A. 1993, \apj, 419, 344

\bibitem[{{Owens} {et~al.}(2021){Owens}, {de Mooij}, {Watson}, \&
  {Hooton}}]{Owens_w12}
{Owens}, N., {de Mooij}, E.~J.~W., {Watson}, C.~A., \& {Hooton}, M.~J. 2021,
  \mnras, 503, L38

\bibitem[{{Pamyatnykh}(1999)}]{Kappa}
{Pamyatnykh}, A.~A. 1999, \actaa, 49, 119

\bibitem[{{Parviainen}(2015)}]{pytransit}
{Parviainen}, H. 2015, \mnras, 450, 3233

\bibitem[{{Pearl} \& {Conrath}(1991)}]{PearlAlbedos}
{Pearl}, J.~C. \& {Conrath}, B.~J. 1991, \jgr, 96, 18921

\bibitem[{{Perna} {et~al.}(2012){Perna}, {Heng}, \& {Pont}}]{GCM}
{Perna}, R., {Heng}, K., \& {Pont}, F. 2012, \apj, 751, 59

\bibitem[{{Piskunov} \& {Valenti}(2017)}]{pv2017}
{Piskunov}, N. \& {Valenti}, J.~A. 2017, \aap, 597, A16

\bibitem[{{Queloz} {et~al.}(2000){Queloz}, {Eggenberger}, {Mayor}, {Perrier},
  {Beuzit}, {Naef}, {Sivan}, \& {Udry}}]{QuelozRM}
{Queloz}, D., {Eggenberger}, A., {Mayor}, M., {et~al.} 2000, \aap, 359, L13

\bibitem[{{Ram{\'\i}rez-Agudelo} {et~al.}(2013){Ram{\'\i}rez-Agudelo},
  {Sim{\'o}n-D{\'\i}az}, {Sana}, {de Koter}, {Sab{\'\i}n-Sanjul{\'\i}an}, {de
  Mink}, {Dufton}, {Gr{\"a}fener}, {Evans}, {Herrero}, {Langer}, {Lennon},
  {Ma{\'\i}z Apell{\'a}niz}, {Markova}, {Najarro}, {Puls}, {Taylor}, \&
  {Vink}}]{VFTS102}
{Ram{\'\i}rez-Agudelo}, O.~H., {Sim{\'o}n-D{\'\i}az}, S., {Sana}, H., {et~al.}
  2013, \aap, 560, A29

\bibitem[{{Ricker} {et~al.}(2015){Ricker}, {Winn}, {Vanderspek}, {Latham},
  {Bakos}, {Bean}, {Berta-Thompson}, {Brown}, {Buchhave}, {Butler}, {Butler},
  {Chaplin}, {Charbonneau}, {Christensen-Dalsgaard}, {Clampin}, {Deming},
  {Doty}, {De Lee}, {Dressing}, {Dunham}, {Endl}, {Fressin}, {Ge}, {Henning},
  {Holman}, {Howard}, {Ida}, {Jenkins}, {Jernigan}, {Johnson}, {Kaltenegger},
  {Kawai}, {Kjeldsen}, {Laughlin}, {Levine}, {Lin}, {Lissauer}, {MacQueen},
  {Marcy}, {McCullough}, {Morton}, {Narita}, {Paegert}, {Palle}, {Pepe},
  {Pepper}, {Quirrenbach}, {Rinehart}, {Sasselov}, {Sato}, {Seager},
  {Sozzetti}, {Stassun}, {Sullivan}, {Szentgyorgyi}, {Torres}, {Udry}, \&
  {Villasenor}}]{TESS}
{Ricker}, G.~R., {Winn}, J.~N., {Vanderspek}, R., {et~al.} 2015, Journal of
  Astronomical Telescopes, Instruments, and Systems, 1, 014003

\bibitem[{{Rossiter}(1924)}]{Rossiter}
{Rossiter}, R.~A. 1924, \apj, 60, 15

\bibitem[{{Russell}(1916)}]{PhaseIntegral}
{Russell}, H.~N. 1916, \apj, 43, 173

\bibitem[{{Ryabchikova} {et~al.}(2015){Ryabchikova}, {Piskunov}, {Kurucz},
  {Stempels}, {Heiter}, {Pakhomov}, \& {Barklem}}]{Ryabchikova2015}
{Ryabchikova}, T., {Piskunov}, N., {Kurucz}, R.~L., {et~al.} 2015, \physscr,
  90, 054005

\bibitem[{{Scuflaire} {et~al.}(2008){Scuflaire}, {Th{\'e}ado}, {Montalb{\'a}n},
  {Miglio}, {Bourge}, {Godart}, {Thoul}, \& {Noels}}]{scuflaire08}
{Scuflaire}, R., {Th{\'e}ado}, S., {Montalb{\'a}n}, J., {et~al.} 2008, \apss,
  316, 83

\bibitem[{{Showman} \& {Polvani}(2011)}]{ShowmanGCM}
{Showman}, A.~P. \& {Polvani}, L.~M. 2011, \apj, 738, 71

\bibitem[{{Shporer} {et~al.}(2014){Shporer}, {O'Rourke}, {Knutson},
  {Szab{\'o}}, {Zhao}, {Burrows}, {Fortney}, {Agol}, {Cowan}, {Desert},
  {Howard}, {Isaacson}, {Lewis}, {Showman}, \& {Todorov}}]{Shporerk13}
{Shporer}, A., {O'Rourke}, J.~G., {Knutson}, H.~A., {et~al.} 2014, \apj, 788,
  92

\bibitem[{{Shporer} {et~al.}(2019){Shporer}, {Wong}, {Huang}, {Line},
  {Stassun}, {Fetherolf}, {Kane}, {Bouma}, {Daylan}, {G{\"u}enther}, {Ricker},
  {Latham}, {Vanderspek}, {Seager}, {Winn}, {Jenkins}, {Glidden},
  {Berta-Thompson}, {Ting}, {Li}, \& {Haworth}}]{Shporer_w18}
{Shporer}, A., {Wong}, I., {Huang}, C.~X., {et~al.} 2019, \aj, 157, 178

\bibitem[{{Skilling}(2004)}]{ns1}
{Skilling}, J. 2004, in American Institute of Physics Conference Series, Vol.
  735, Bayesian Inference and Maximum Entropy Methods in Science and
  Engineering: 24th International Workshop on Bayesian Inference and Maximum
  Entropy Methods in Science and Engineering, ed. R.~{Fischer}, R.~{Preuss}, \&
  U.~V. {Toussaint}, 395--405

\bibitem[{Skilling(2006)}]{ns2}
Skilling, J. 2006, Bayesian Anal., 1, 833

\bibitem[{{Skrutskie} {et~al.}(2006){Skrutskie}, {Cutri}, {Stiening},
  {Weinberg}, {Schneider}, {Carpenter}, {Beichman}, {Capps}, {Chester},
  {Elias}, {Huchra}, {Liebert}, {Lonsdale}, {Monet}, {Price}, {Seitzer},
  {Jarrett}, {Kirkpatrick}, {Gizis}, {Howard}, {Evans}, {Fowler}, {Fullmer},
  {Hurt}, {Light}, {Kopan}, {Marsh}, {McCallon}, {Tam}, {Van Dyk}, \&
  {Wheelock}}]{Skrutskie2006}
{Skrutskie}, M.~F., {Cutri}, R.~M., {Stiening}, R., {et~al.} 2006, \aj, 131,
  1163

\bibitem[{{Southworth}(2011)}]{TEPCat}
{Southworth}, J. 2011, \mnras, 417, 2166

\bibitem[{{Speagle}(2020)}]{dynesty}
{Speagle}, J.~S. 2020, \mnras, 493, 3132

\bibitem[{{Stefansson} {et~al.}(2020){Stefansson}, {Ca{\~n}as}, {Wisniewski},
  {Robertson}, {Mahadevan}, {Maney}, {Kanodia}, {Beard}, {Bender}, {Brunt},
  {Clemens}, {Cochran}, {Diddams}, {Endl}, {Ford}, {Fredrick}, {Halverson},
  {Hearty}, {Hebb}, {Huehnerhoff}, {Jennings}, {Kaplan}, {Levi}, {Lubar},
  {Metcalf}, {Monson}, {Morris}, {Ninan}, {Nitroy}, {Ramsey}, {Roy}, {Schwab},
  {Sigurdsson}, {Terrien}, \& {Wright}}]{Asteroseis_G9}
{Stefansson}, G., {Ca{\~n}as}, C., {Wisniewski}, J., {et~al.} 2020, \aj, 159,
  100

\bibitem[{{Swayne} {et~al.}(2021){Swayne}, {Maxted}, {Triaud}, {Sousa},
  {Broeg}, {Flor{\'e}n}, {Guterman}, {Simon}, {Boisse}, {Bonfanti}, {Martin},
  {Santerne}, {Salmon}, {Standing}, {Van Grootel}, {Wilson}, {Alibert},
  {Alonso}, {Anglada Escud{\'e}}, {Asquier}, {B{\'a}rczy}, {Barrado}, {Barros},
  {Battley}, {Baumjohann}, {Beck}, {Beck}, {Bekkelien}, {Benz}, {Billot},
  {Bonfils}, {Brandeker}, {Busch}, {Cabrera}, {Charnoz}, {Collier Cameron},
  {Csizmadia}, {Davies}, {Deleuil}, {Deline}, {Delrez}, {Demangeon}, {Demory},
  {Dransfield}, {Ehrenreich}, {Erikson}, {Fortier}, {Fossati}, {Fridlund},
  {Futyan}, {Gandolfi}, {Gillon}, {Guedel}, {H{\'e}brard}, {Heidari},
  {Hellier}, {Heng}, {Hobson}, {Hoyer}, {Isaak}, {Kiss}, {Kunovac
  Hod{\v{z}}i{\'c}}, {Lalitha}, {Laskar}, {Lecavelier des Etangs}, {Lendl},
  {Lovis}, {Magrin}, {Marafatto}, {McCormac}, {Miller}, {Nascimbeni},
  {Olofsson}, {Ottensamer}, {Pagano}, {Pall{\'e}}, {Peter}, {Piotto},
  {Pollacco}, {Queloz}, {Ragazzoni}, {Rando}, {Rauer}, {Ribas}, {Santos},
  {Scandariato}, {S{\'e}gransan}, {Smith}, {Steinberger}, {Steller},
  {Szab{\'o}}, {Thomas}, {Udry}, {Walter}, {Walton}, \& {Willett}}]{SwayneEBLM}
{Swayne}, M.~I., {Maxted}, P.~F.~L., {Triaud}, A.~H.~M.~J., {et~al.} 2021,
  \mnras, 506, 306

\bibitem[{{Szab{\'o}} {et~al.}(2021){Szab{\'o}}, {Gandolfi}, {Brandeker},
  {Csizmadia}, {Garai}, {Billot}, {Broeg}, {Ehrenreich}, {Fortier}, {Fossati},
  {Hoyer}, {Kiss}, {Lecavelier des Etangs}, {Maxted}, {Ribas}, {Alibert},
  {Alonso}, {Anglada Escud{\'e}}, {B{\'a}rczy}, {Barros}, {Barrado},
  {Baumjohann}, {Beck}, {Beck}, {Bekkelien}, {Bonfils}, {Benz}, {Borsato},
  {Busch}, {Cabrera}, {Charnoz}, {Collier Cameron}, {Van Damme}, {Davies},
  {Delrez}, {Deleuil}, {Demangeon}, {Demory}, {Erikson}, {Fridlund}, {Futyan},
  {Garc{\'\i}a Mu{\~n}oz}, {Gillon}, {Guedel}, {Guterman}, {Heng}, {Isaak},
  {Lacedelli}, {Laskar}, {Lendl}, {Lovis}, {Luntzer}, {Magrin}, {Nascimbeni},
  {Olofsson}, {Osborn}, {Ottensamer}, {Pagano}, {Pall{\'e}}, {Peter}, {Piazza},
  {Piotto}, {Pollacco}, {Queloz}, {Ragazzoni}, {Rando}, {Rauer}, {Santos},
  {Scandariato}, {S{\'e}gransan}, {Serrano}, {Sicilia}, {Simon}, {Smith},
  {Sousa}, {Steller}, {Thomas}, {Udry}, {Van Grootel}, {Walton}, \&
  {Wilson}}]{Szabo_aumic}
{Szab{\'o}}, G.~M., {Gandolfi}, D., {Brandeker}, A., {et~al.} 2021, \aap, 654,
  A159

\bibitem[{{Szab{\'o}} {et~al.}(2012){Szab{\'o}}, {P{\'a}l}, {Derekas}, {Simon},
  {Szalai}, \& {Kiss}}]{Szabo12}
{Szab{\'o}}, G.~M., {P{\'a}l}, A., {Derekas}, A., {et~al.} 2012, \mnras, 421,
  L122

\bibitem[{{Szab{\'o}} {et~al.}(2020){Szab{\'o}}, {Pribulla}, {P{\'a}l},
  {B{\'o}di}, {Kiss}, \& {Derekas}}]{Szabo20}
{Szab{\'o}}, G.~M., {Pribulla}, T., {P{\'a}l}, A., {et~al.} 2020, \mnras, 492,
  L17

\bibitem[{{Szab{\'o}} {et~al.}(2011){Szab{\'o}}, {Szab{\'o}}, {Benk{\H{o}}},
  {Lehmann}, {Mez{\H{o}}}, {Simon}, {K{\H{o}}v{\'a}ri}, {Hodos{\'a}n},
  {Reg{\'a}ly}, \& {Kiss}}]{Szabo11}
{Szab{\'o}}, G.~M., {Szab{\'o}}, R., {Benk{\H{o}}}, J.~M., {et~al.} 2011,
  \apjl, 736, L4

\bibitem[{{Talens} {et~al.}(2017){Talens}, {Albrecht}, {Spronck}, {Lesage},
  {Otten}, {Stuik}, {Van Eylen}, {Van Winckel}, {Pollacco}, {McCormac},
  {Grundahl}, {Fredslund Andersen}, {Antoci}, \& {Snellen}}]{Talens}
{Talens}, G.~J.~J., {Albrecht}, S., {Spronck}, J.~F.~P., {et~al.} 2017, \aap,
  606, A73

\bibitem[{{Talens} {et~al.}(2018){Talens}, {Albrecht}, {Spronck}, {Lesage},
  {Otten}, {Stuik}, {Van Eylen}, {Van Winckel}, {Pollacco}, {McCormac},
  {Grundahl}, {Fredslund Andersen}, {Antoci}, \& {Snellen}}]{Talens_corr}
{Talens}, G.~J.~J., {Albrecht}, S., {Spronck}, J.~F.~P., {et~al.} 2018, \aap,
  613, C2

\bibitem[{{Tsai} {et~al.}(2014){Tsai}, {Dobbs-Dixon}, \& {Gu}}]{Shami2014}
{Tsai}, S.-M., {Dobbs-Dixon}, I., \& {Gu}, P.-G. 2014, \apj, 793, 141

\bibitem[{{Valenti} \& {Piskunov}(1996)}]{vp96}
{Valenti}, J.~A. \& {Piskunov}, N. 1996, \aaps, 118, 595

\bibitem[{{Van Grootel} {et~al.}(2021){Van Grootel}, {Pozuelos}, {Thuillier},
  {Charpinet}, {Delrez}, {Beck}, {Fortier}, {Hoyer}, {Sousa}, {Barlow},
  {Billot}, {D{\'e}vora-Pajares}, {{\O}stensen}, {Alibert}, {Alonso}, {Anglada
  Escud{\'e}}, {Asquier}, {Barrado}, {Barros}, {Baumjohann}, {Beck},
  {Bekkelien}, {Benz}, {Bonfils}, {Brandeker}, {Broeg}, {Bruno}, {B{\'a}rczy},
  {Cabrera}, {Cameron}, {Charnoz}, {Davies}, {Deleuil}, {Demangeon}, {Demory},
  {Ehrenreich}, {Erikson}, {Fossati}, {Fridlund}, {Futyan}, {Gandolfi},
  {Gillon}, {Guedel}, {Heng}, {Isaak}, {Kiss}, {Laskar}, {Lecavelier des
  Etangs}, {Lendl}, {Lovis}, {Magrin}, {Maxted}, {Mecina}, {Mustill},
  {Nascimbeni}, {Olofsson}, {Ottensamer}, {Pagano}, {Pall{\'e}}, {Peter},
  {Piotto}, {Plesseria}, {Pollacco}, {Queloz}, {Ragazzoni}, {Rando}, {Rauer},
  {Ribas}, {Santos}, {Scandariato}, {S{\'e}gransan}, {Silvotti}, {Simon},
  {Smith}, {Steller}, {Szab{\'o}}, {Thomas}, {Udry}, {Viotto}, {Walton},
  {Westerdorff}, \& {Wilson}}]{VanGrootelCHEOPS}
{Van Grootel}, V., {Pozuelos}, F.~J., {Thuillier}, A., {et~al.} 2021, \aap,
  650, A205

\bibitem[{{von Essen} {et~al.}(2020){von Essen}, {Mallonn}, {Borre}, {Antoci},
  {Stassun}, {Khalafinejad}, \& {Tautvai{\v{s}}ien{\.{e}}}}]{vonessenw33}
{von Essen}, C., {Mallonn}, M., {Borre}, C.~C., {et~al.} 2020, \aap, 639, A34

\bibitem[{{von Zeipel}(1924)}]{vonZeipel}
{von Zeipel}, H. 1924, \mnras, 84, 665

\bibitem[{{Watanabe} {et~al.}(2020){Watanabe}, {Narita}, \&
  {Johnson}}]{Watanabew33}
{Watanabe}, N., {Narita}, N., \& {Johnson}, M.~C. 2020, \pasj, 72, 19

\bibitem[{{Winn} {et~al.}(2010){Winn}, {Fabrycky}, {Albrecht}, \&
  {Johnson}}]{WinnHJs}
{Winn}, J.~N., {Fabrycky}, D., {Albrecht}, S., \& {Johnson}, J.~A. 2010, \apjl,
  718, L145

\bibitem[{{Winn} {et~al.}(2007){Winn}, {Holman}, {Bakos}, {P{\'a}l}, {Johnson},
  {Williams}, {Shporer}, {Mazeh}, {Fernandez}, {Latham}, \&
  {Gillon}}]{Winn_timeave}
{Winn}, J.~N., {Holman}, M.~J., {Bakos}, G.~{\'A}., {et~al.} 2007, \aj, 134,
  1707

\bibitem[{{Wong} {et~al.}(2021){Wong}, {Kitzmann}, {Shporer}, {Heng},
  {Fetherolf}, {Benneke}, {Daylan}, {Kane}, {Vanderspek}, {Seager}, {Winn},
  {Jenkins}, \& {Ting}}]{WongNorth}
{Wong}, I., {Kitzmann}, D., {Shporer}, A., {et~al.} 2021, \aj, 162, 127

\bibitem[{{Wong} {et~al.}(2020{\natexlab{a}}){Wong}, {Shporer}, {Daylan},
  {Benneke}, {Fetherolf}, {Kane}, {Ricker}, {Vanderspek}, {Latham}, {Winn},
  {Jenkins}, {Boyd}, {Glidden}, {Goeke}, {Sha}, {Ting}, \&
  {Yahalomi}}]{WongSouth}
{Wong}, I., {Shporer}, A., {Daylan}, T., {et~al.} 2020{\natexlab{a}}, \aj, 160,
  155

\bibitem[{{Wong} {et~al.}(2020{\natexlab{b}}){Wong}, {Shporer}, {Kitzmann},
  {Morris}, {Heng}, {Hoeijmakers}, {Demory}, {Ahlers}, {Mansfield}, {Bean},
  {Daylan}, {Fetherolf}, {Rodriguez}, {Benneke}, {Ricker}, {Latham},
  {Vanderspek}, {Seager}, {Winn}, {Jenkins}, {Burke}, {Christiansen}, {Essack},
  {Rose}, {Smith}, {Tenenbaum}, \& {Yahalomi}}]{Wongk9}
{Wong}, I., {Shporer}, A., {Kitzmann}, D., {et~al.} 2020{\natexlab{b}}, \aj,
  160, 88

\bibitem[{{Wright} {et~al.}(2010){Wright}, {Eisenhardt}, {Mainzer}, {Ressler},
  {Cutri}, {Jarrett}, {Kirkpatrick}, {Padgett}, {McMillan}, {Skrutskie},
  {Stanford}, {Cohen}, {Walker}, {Mather}, {Leisawitz}, {Gautier}, {McLean},
  {Benford}, {Lonsdale}, {Blain}, {Mendez}, {Irace}, {Duval}, {Liu}, {Royer},
  {Heinrichsen}, {Howard}, {Shannon}, {Kendall}, {Walsh}, {Larsen}, {Cardon},
  {Schick}, {Schwalm}, {Abid}, {Fabinsky}, {Naes}, \& {Tsai}}]{Wright2010}
{Wright}, E.~L., {Eisenhardt}, P. R.~M., {Mainzer}, A.~K., {et~al.} 2010, \aj,
  140, 1868

\bibitem[{{Zhang} {et~al.}(2018){Zhang}, {Knutson}, {Kataria}, {Schwartz},
  {Cowan}, {Showman}, {Burrows}, {Fortney}, {Todorov}, {Desert}, {Agol}, \&
  {Deming}}]{Zhangw33}
{Zhang}, M., {Knutson}, H.~A., {Kataria}, T., {et~al.} 2018, \aj, 155, 83

\bibitem[{{Zhou} \& {Huang}(2013)}]{ZhouHuang13}
{Zhou}, G. \& {Huang}, C.~X. 2013, \apjl, 776, L35

\end{thebibliography}

\appendix

\section{Table of noise parameters}

\begin{table}[h]
\caption[]{Similar to \cref{tab:results}, but displaying the baseline and noise parameters. All parameters are dimensionless.}
\label{tab:noise}
\begin{tabular}{lccc}
\hline \hline
\multicolumn{2}{c}{Parameter} & Prior & Best fit \\ 
\hline 
White noise \textit{Spitzer} & ln$\sigma_\mathrm{w,S}$ & $\mathcal{U}$(-12, -6) & $-10.94^{+0.66}_{-0.92}$\\ 
White noise \textit{CHEOPS} & ln$\sigma_\mathrm{w,C}$ & $\mathcal{U}$(-12, -6) & $-9.348^{+0.066}_{-0.072}$\\ 
GP hyperparameter & ln$S_\mathrm{0}$ & $\mathcal{U}$(-30, 0) & $-23.0^{+1.6}_{-1.0}$\\ 
GP hyperparameter & ln$\omega_\mathrm{0}$ & $\mathcal{U}$(-15, 15) & $1.88^{+0.31}_{-0.45}$\\ 
GP hyperparameter & ln$Q$ & fixed & $\ln{10}$\\ 
Constant 1 & $c_\mathrm{1}$ & $\mathcal{U}$(0, 2) & $1.00673\pm0.00021$\\ 
$y$ centroid 1 & $y_\mathrm{1}$ & $\mathcal{U}$(-1, 1) & $0.0155^{+0.0015}_{-0.0014}$\\ 
$x^2$ centroid 1 & $x^2_\mathrm{1}$ & $\mathcal{U}$(-1, 1) & $-0.00116^{+0.00049}_{-0.00048}$\\ 
$y^2$ centroid 1 & $y^2_\mathrm{1}$ & $\mathcal{U}$(-1, 1) & $0.0063^{+0.0031}_{-0.0028}$\\ 
$xy$ centroid 1 & $xy_\mathrm{1}$ & $\mathcal{U}$(-1, 1) & $-0.0\pm0.001$\\ 
$x$ FWHM 1 & $xw_\mathrm{1}$ & $\mathcal{U}$(-1, 1) & $-0.00426^{+0.00069}_{-0.00065}$\\ 
$y$ centroid 1 & $y_\mathrm{1}$ & $\mathcal{U}$(-1, 1) & $-0.0083^{+0.0012}_{-0.0013}$\\ 
$y^2$ FWHM 1 & $y^w2_\mathrm{1}$ & $\mathcal{U}$(-1, 1) & $-0.00044\pm0.00058$\\ 
Constant 2 & $c_\mathrm{2}$ & $\mathcal{U}$(0, 2) & $1.00105^{+0.00054}_{-0.00051}$\\ 
$x$ centroid 2 & $x_\mathrm{2}$ & $\mathcal{U}$(-1, 1) & $-0.0027^{+0.00037}_{-0.00036}$\\ 
$y$ centroid 2 & $y_\mathrm{2}$ & $\mathcal{U}$(-1, 1) & $0.01109^{+0.0009}_{-0.00086}$\\ 
$x$ FWHM 2 & $xw_\mathrm{2}$ & $\mathcal{U}$(-1, 1) & $-0.00519^{+0.00063}_{-0.00056}$\\ 
$y$ centroid 2 & $y_\mathrm{2}$ & $\mathcal{U}$(-1, 1) & $0.0101^{+0.0013}_{-0.0012}$\\ 
$xy$ FWHM 2 & $xwyw_\mathrm{2}$ & $\mathcal{U}$(-1, 1) & $-0.00221^{+0.00078}_{-0.00082}$\\ 
Constant 3 & $c_\mathrm{3}$ & $\mathcal{U}$(0, 2) & $0.99927^{+0.00039}_{-0.00041}$\\ 
$x$ centroid 3 & $x_\mathrm{3}$ & $\mathcal{U}$(-1, 1) & $-0.00399^{+0.00028}_{-0.00029}$\\ 
$y$ centroid 3 & $y_\mathrm{3}$ & $\mathcal{U}$(-1, 1) & $-0.00285^{+0.00043}_{-0.00044}$\\ 
$y^2$ centroid 3 & $y^2_\mathrm{3}$ & $\mathcal{U}$(-1, 1) & $0.0096^{+0.0013}_{-0.0012}$\\ 
$xy$ centroid 3 & $xy_\mathrm{3}$ & $\mathcal{U}$(-1, 1) & $-0.00175^{+0.00083}_{-0.00089}$\\ 
$x$ FWHM 3 & $xw_\mathrm{3}$ & $\mathcal{U}$(-1, 1) & $-0.00279^{+0.00081}_{-0.00079}$\\ 
$x^2$ FWHM 3 & $xw^2_\mathrm{3}$ & $\mathcal{U}$(-1, 1) & $0.00107^{+0.00083}_{-0.00087}$\\ 
$y^2$ FWHM 3 & $y^w2_\mathrm{3}$ & $\mathcal{U}$(-1, 1) & $0.0002^{+0.00057}_{-0.00062}$\\ 
$xy$ FWHM 3 & $xwyw_\mathrm{3}$ & $\mathcal{U}$(-1, 1) & $-0.0007^{+0.0013}_{-0.0011}$\\ 
Constant 4 & $c_\mathrm{4}$ & $\mathcal{U}$(0, 2) & $1.000073^{+0.000028}_{-0.000055}$\\ 
Glint 4 & $g_\mathrm{4}$ & $\mathcal{U}$(0, 2) & $1.01\pm0.12$\\ 
Constant 5 & $c_\mathrm{5}$ & $\mathcal{U}$(0, 2) & $0.999919^{+0.000048}_{-0.000053}$\\ 
Glint 5 & $g_\mathrm{5}$ & $\mathcal{U}$(0, 2) & $0.97^{+0.12}_{-0.13}$\\ 
Constant 6 & $c_\mathrm{6}$ & $\mathcal{U}$(0, 2) & $0.999938^{+0.000056}_{-0.00004}$\\ 
Sky background 6 & bg${6}$ & $\mathcal{U}$(-1, 1) & $0.000415^{+0.000087}_{-0.000079}$\\ 
Glint 6 & $g_\mathrm{6}$ & $\mathcal{U}$(0, 2) & $1.05^{+0.18}_{-0.17}$\\ 
Constant 7 & $c_\mathrm{7}$ & $\mathcal{U}$(0, 2) & $1.000174^{+0.00011}_{-0.00005}$\\ 
Glint 7 & $g_\mathrm{7}$ & $\mathcal{U}$(0, 2) & $1.108^{+0.089}_{-0.088}$\\ 
\hline 
\end{tabular}
\end{table}

\section{Corner plots of asymmetric transit fits}


\begin{figure*}[h]
    \centering
    \includegraphics[width=0.8\linewidth]{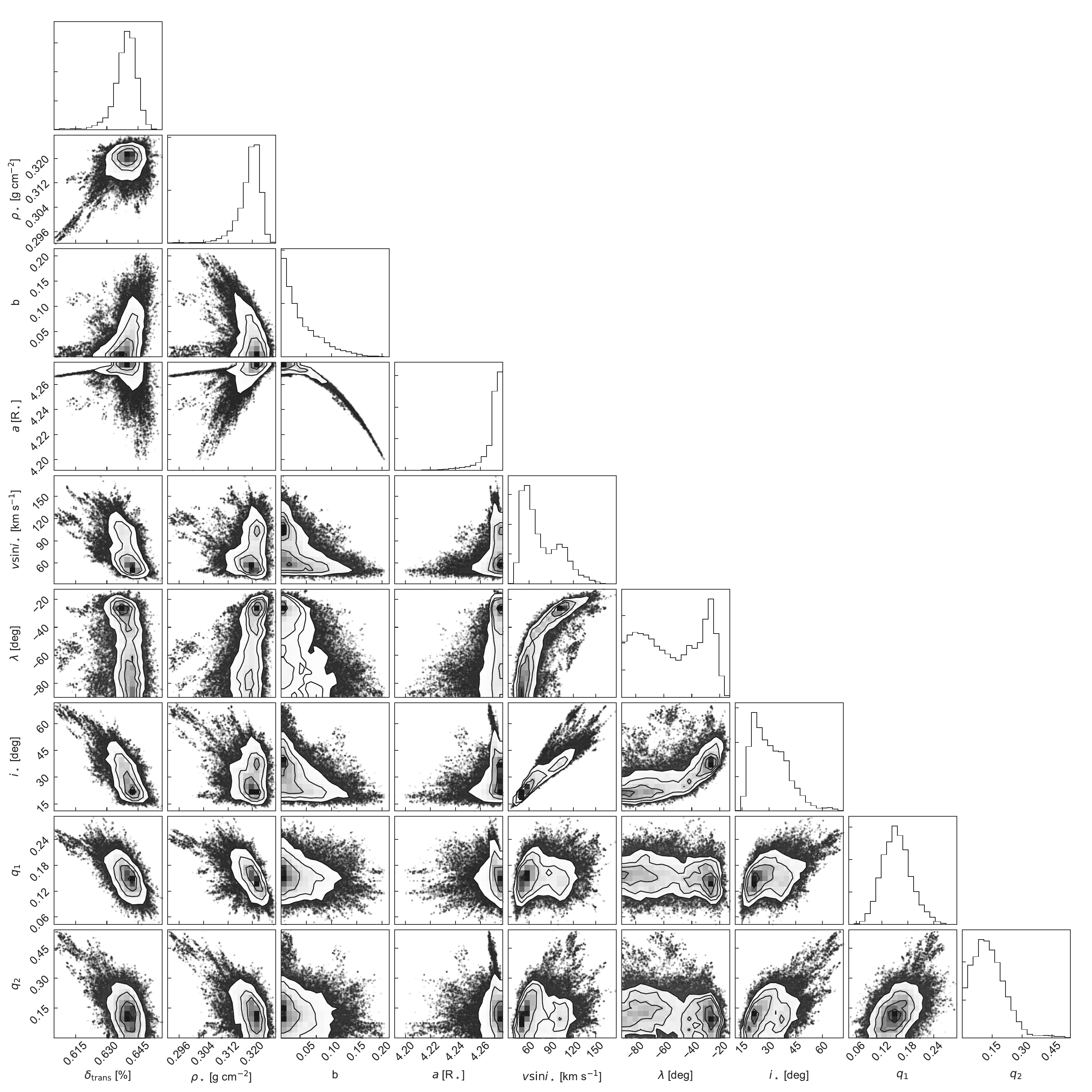}
    \caption{Corner plot of the \textit{CHEOPS} transit-only fit without the placement of Gaussian priors on $b$, $v\sin{i_\star}$, and $\lambda$ informed by the Doppler tomography. The strategy is described in full in \cref{sec:transit_fit} and the corresponding results are displayed in \cref{tab:GDtransit}.} 
    \label{fig:corner_cheops_noprior}
\end{figure*}

\begin{figure*}[h]
    \centering
    \includegraphics[width=0.8\linewidth]{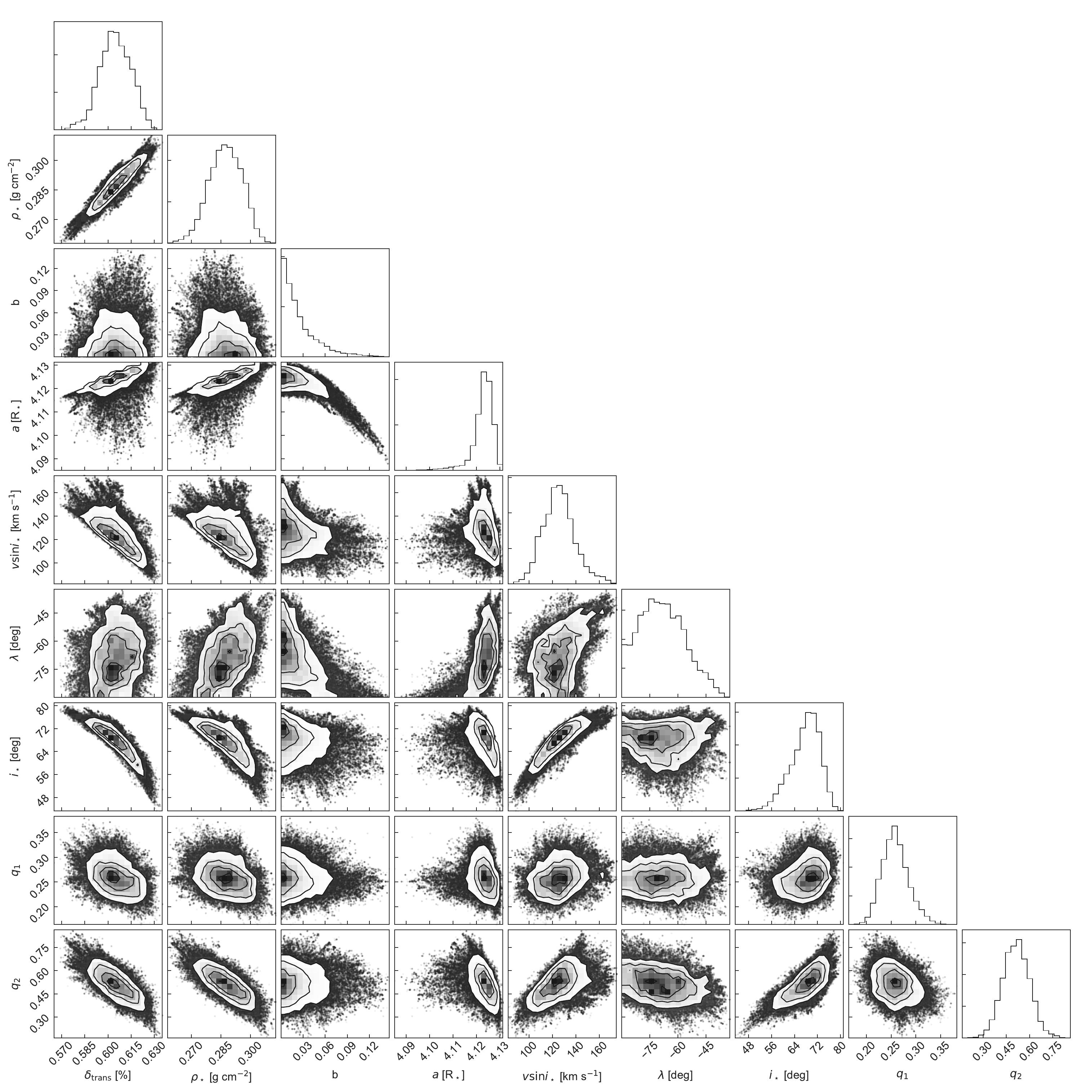}
    \caption{Corner plot of the \textit{CHEOPS}+\textit{Spitzer} transit-only fit without the placement of Gaussian priors on $b$, $v\sin{i_\star}$, and $\lambda$ informed by the Doppler tomography. The strategy is described in full in \cref{sec:transit_fit} and the corresponding results are displayed in \cref{tab:GDtransit}.} 
    \label{fig:corner_both_noprior}
\end{figure*}

\begin{figure*}[h]
    \centering
    \includegraphics[width=0.8\linewidth]{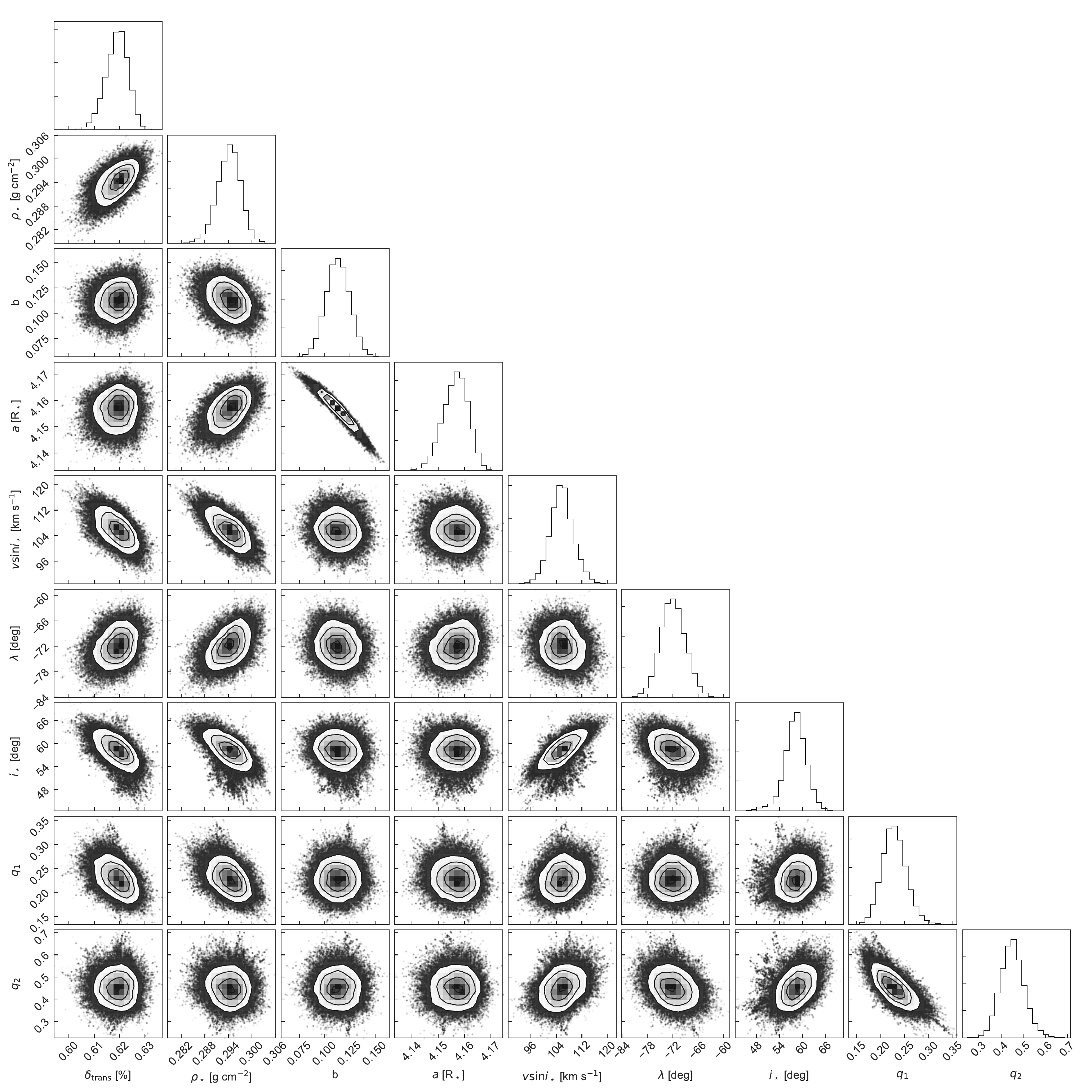}
    \caption{Corner plot of the \textit{CHEOPS} transit-only fit with the placement of Gaussian priors on $b$, $v\sin{i_\star}$, and $\lambda$ informed by the Doppler tomography. The strategy is described in full in \cref{sec:transit_fit} and the corresponding results are displayed in \cref{tab:GDtransit}.} 
    \label{fig:corner_cheops_prior}
\end{figure*}

\begin{figure*}[h]
    \centering
    \includegraphics[width=0.8\linewidth]{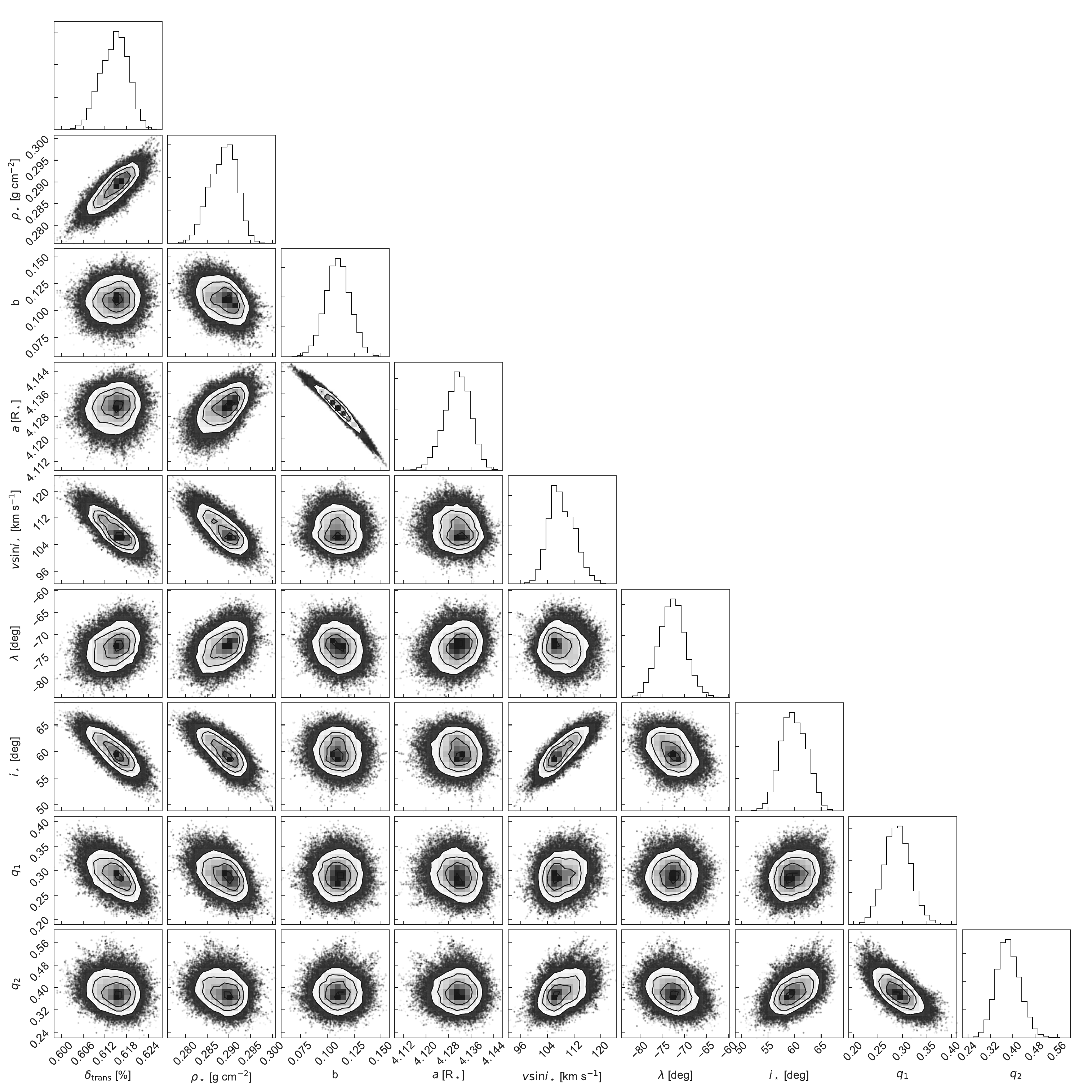}
    \caption{Corner plot of the \textit{CHEOPS}+\textit{Spitzer} transit-only fit with the placement of Gaussian priors on $b$, $v\sin{i_\star}$, and $\lambda$ informed by the Doppler tomography. The strategy is described in full in \cref{sec:transit_fit} and the corresponding results are displayed in \cref{tab:GDtransit}. This approach of using both the \textit{CHEOPS} and \textit{Spitzer} light curves and using priors from tomography is the one adopted for the fit to the full set of light curves presented in \cref{sec:discussion}.} 
    \label{fig:corner_both_prior}
\end{figure*}

\end{document}